\documentclass[a4paper,11pt]{article}
\pdfoutput=1 

\usepackage{jheppub} 
\usepackage[utf8]{inputenc}
\usepackage{setspace}
\usepackage{subfigure}
\usepackage{hyperref}
\definecolor{refkey}{gray}{0.45}
\definecolor{labelkey}{RGB}{155,48,48}

\usepackage{mathtools}
\usepackage{physics}
\usepackage{simplewick}
\usepackage[obeyFinal]{todonotes}
\usepackage[normalem]{ulem}

\def\Tr{\text{Tr}}

\def\beq{\begin{eqnarray}}\def\eeq{\end{eqnarray}}
\def\be{\begin{equation}}\def\ee{\end{equation}}

\def\mes[#1]{d^{3}{#1}}

\def\del{\partial}

\newcommand{\half}{\frac{1}{2}}

\def\del{\partial}

\def\order{\ensuremath{\mathcal{O}}}

\reversemarginpar

\definecolor{UI_blue}{RGB}{32, 64, 151}
\definecolor{UI_red}{RGB}{187, 62, 24}
\definecolor{UI_blue2}{RGB}{0, 84, 147}
\definecolor{UI_red2}{RGB}{159, 32, 66}
\definecolor{UI_gray}{RGB}{169, 169, 169}
\definecolor{UI_sepia}{RGB}{112, 66, 20}
\definecolor{UI_bittersweet}{RGB}{254, 111, 94}
\definecolor{UI_emerald}{RGB}{80, 200, 120}
\definecolor{UI_olivegreen}{RGB}{181, 179, 92}
\definecolor{UI_cadetblue}{RGB}{95, 158, 160}
\definecolor{UI_fuchsia}{RGB}{255, 0, 255}
\definecolor{UI_midnightblue}{RGB}{25, 25, 112}
\definecolor{UI_royalblue}{RGB}{0,35, 102}
\definecolor{UI_periwinkle}{RGB}{204, 204, 255}
\definecolor{UI_redorange}{RGB}{255, 83, 73}
\definecolor{UI_brickred}{RGB}{203,65,84}	
\definecolor{UI_forestgreen}{RGB}{34, 139, 34}
\definecolor{UI_tan}{RGB}{210,180,140}	
\definecolor{UI_burlywood}{RGB}{222,184,135}
\definecolor{UI_burlywood}{RGB}{192,64,0}
\definecolor{UI_darkorchid}{RGB}{153,50,204}

\usepackage{titlesec}
\usepackage{verbatim}
\usepackage{fancyhdr}
\usepackage{enumerate}
\hypersetup{colorlinks=true,citecolor=blue,urlcolor=black}
\usepackage{slashed, enumitem}
\usepackage{array}
\newcolumntype{P}[1]{>{\centering\arraybackslash}p{#1}}
\usepackage{anyfontsize}
	\vspace{-11cm}
	
	\author[a]{Kanhu Kishore Nanda,}
	\author[b]{Sunil Kumar Sake,}
	\author[a]{Sandip P. Trivedi,}
	
	\affiliation[a]{\it Department of Theoretical Physics,
		Tata Institute of Fundamental Research,\\  Colaba, Mumbai, India, 400005\\}
	\affiliation[b]{\it Department of Physics, Osaka University,\\  Toyonaka, Osaka, Japan, 560-0043\\}

	\emailAdd{kanhu.nanda@tifr.res.in }
	\emailAdd{sunilsake1@gmail.com}
	\emailAdd{sandip@theory.tifr.res.in}

	\vspace{1cm}

	\abstract{We discuss the canonical quantisation of JT gravity in de Sitter space, following earlier work by Henneaux, with particular attention to the problem of time. 
	Choosing the  dilaton as the physical clock, we define a norm and operator expectation values for states  and explore  the classical limit.
	We find that requiring a conserved and finite norm and well-defined expectation values for operators imposes significant restrictions on states, as does the   requirement of  a   classical limit. However, these requirements can all be met, with the dilaton providing a satisfactory physical clock. We construct several examples and analyse them in detail. We find that in fact  an infinite number of states exist which  meet the various conditions mentioned above.  }

	\title{JT Gravity in de Sitter Space and the Problem of Time}
	
	\preprint{\parbox{3cm}{TIFR/TH/23-17\\OU-HET-1192}}
	
\begin{document}
		\maketitle
		\flushbottom
	
		\vskip 10pt

		\newpage
		\section{Introduction}
		Any attempt to apply  quantum mechanics to the universe as a whole immediately leads to a host of  conceptual questions. 
		For example, what is the meaning and interpretation of wave function of the universe? Who can observe its consequences, i.e., who are well defined observables? And what are the acceptable wave functions?  
		
		This paper will not attempt to provide an answer to all such questions thereby providing a conceptually sound framework for quantum cosmology. 
		Rather, our aim is much more modest. In one simple example of two-dimensional gravity we will attempt to address some of these questions.
		Our hope is that such attempts will prove relevant for four  and higher  dimensions too, eventually leading to a conceptually complete  understanding of the subject. Before proceeding let us note that there is of course considerable literature already on Quantum Cosmology, in particular pertaining to the  conceptual issues  we mentioned above, for an incomplete list of reviews and references see, \cite{Spradlin:2001pw,Halliwell:1989myn,Hartle:1992as,Isham:1992ms,Witten:2001kn,Linde:1990flp,Gell-Mann:1989lly,Hartletasi,Vilenkin:1986cy,PhysRevD.28.2960,Banks:1984cw}.
		
		 One reason why $2$ dimensions is much simpler  than $4$ is that there are no propagating gravitons in $2$ dimensions.  Nevertheless, some of the key  issues alluded to above are present  even in this simplified setting  and one can hope to address them here, shorn of the additional complications that arise in $4$ dimensions, some of which are tied to the lack of renormalisability once gravity becomes dynamical. 
		 
		 The model we will study is Jackiw-Teitelboim (JT) gravity \cite{JACKIW1985343}, \cite{Teitelboim:1983ux} in deSitter space. Its action is 
		 \be
		 \label{JTact1}
		I= {1\over {16\pi G} }\left(\int d^2 x\,\sqrt{-g}\, (\phi R-U(\phi))             -2\int_{bdy}\sqrt{-\gamma}\phi K \right)
		 \ee
		 and it contains the metric and a scalar called the dilaton. 
		 A more complete description is given in section \ref{canquJT}. 
		 This model is so simple that in fact it has no dynamical degrees of freedom\footnote{Towards the end of the paper we will briefly comment on the  behaviour of JT gravity once extra matter fields with propagating degrees of freedom are added. }. 
		 
		 One of the  questions which we will seek to address is the well known problem of time. This problem refers to the fact that since  the  time coordinate is only  a way to label points in spacetime, 
		 gauge invariant  states in quantum gravity must  be invariant with respect to time reparametrisations. This then leads to a question: how does a physical notion of time arise?		 The problem is particularly acute in deSitter space and more generally cosmologies with closed spatial slices. There are no ``large" coordinate transformations in such cases and the Hamiltonian must vanish. 
		 On the other hand in a physical sense these cosmologies clearly evolve with time, and in fact this is their  most striking feature. It is generally understood that 
		 a physical notion of time  must  arise  from    some physical degree of freedom internal to the system. The problem of time refers to  finding such a  degree of freedom  which can satisfactorily play the role  of physical time, allowing one to define at each value for this variable, a Hilbert space, or at least a set of acceptable physical states. Observables  which are well defined in these states would then tell us about the properties of the universe as it evolves.

		 		 Following the important early work of \cite{Hennauxjt,Maldacena:2019cbz,Verlinde:2020zld}, we  canonically quantise JT gravity in this paper.
		 Our analysis will also be connected to  path integral quantisation of JT gravity in dS  space, and in particular to the understanding of the Hartle-Hawking wave function which has been obtained in this manner \cite{Moitra:2021uiv,Moitra:2022glw}.  In our investigation, we solve the gauge constraints, after canonically quantising the theory,  and are able to obtain the most general gauge invariant states. As we will see, these  are  infinite  in number.

		 		 This then sets the stage for our study of the problem of time. We explore the idea that  the    dilaton  plays the role of the physical clock.
				  	The spatial geometry of the universe at a given instant of time is a circle in this theory,  and the only  coordinate invariant  degree of freedom,   arising from the $2$ dimensional metric characterizing the spatial geometry, is the length $l$ of this circle. The Wheeler De-Witt equation (WDW)   then  turns out to be  quite simple, taking the form of a massive Klein-Gordon  (KG) equation in a space of  signature $(-1,1)$ with the dilaton, $\phi$, and $l$  being null directions. All solutions to this equation are physical states.
					
						 We  find that these   physical states exhibit a host of behaviours. The Klein-Gordon equation suggests a natural norm for states which we call the Klein-Gordon norm (KG norm)   whose consequences we  explore. 
				 We  find that the KG  norm is  not  conserved for all states. Among those with a conserved norm, we  find some which lead to a positive probability density  $p(l)$ for the universe having size $l$, others for which $p(l)$ is not positive for all values of $l$, and yet others for which $p(l)$ becomes positive, for all $l$, only at late times, i.e. at large values of $\phi$.  
				 
				 The problem of time is also related to whether a good   classical limit arises. With an eye towards reproducing some aspects of our universe one is specially interested in situations where 
				 at late time, and at large size,  the universe becomes approximately classical. It is clear from eq.(\ref{JTact1}) that in  JT theory   the $G\rightarrow 0$ limit can  be related to a classical limit, and this limit should also be  related to the $\phi\rightarrow \infty$ limit,  i.e., to the late time behaviour. 
				 Thus we might expect, having chosen the dilaton as our clock, that  late times in this model coincide with a classical domain arising. 
				
				 In fact, we  find  that many, but not all,  states do become classical when $\phi\rightarrow \infty$.  
				And in the classical limit we often find that   the KG norm  reduces to an $L_2$ norm at large $\phi$, with a manifestly positive probability  density, $p(l)$. 
				Obstructions to achieving the classical limit can often arise though and are  of different types. 
				In some cases,  fluctuations about the mean for moments of the length, $\langle l^n\rangle $,  continue to remain large at large $\phi$. In other cases there are multiple  branches in the  wave function, each of which would have become classical at large $\phi$ corresponding  to a different  classical expanding universe,  but the interference between these different branches does not die away. And finally, there are situations where there are multiple branches corresponding to  both expanding and contracting universes which continue to interfere at late times. In short, our investigation reveals that the requirement that a good classical limit is obtained imposes a significant restriction on the allowed wave functions. 
				
				We should also mention that in contrast, there are also examples where   the wave function contains various  expanding and contracting branches which interfere at early times but  which, at   large $\phi$, stop interfering and decohere from each other. In such cases,  a well defined norm and probability density $p(l)$ can then be defined for each branch  separately,  and in many cases the fluctuations in $l$ about its mean  value  can becomes small, as one would expect for a classical universe.

			The general picture which then emerges from our investigation is that the dilaton can satisfactorily  play the role of a physical clock in this theory, but the restrictions imposed by this requirement are not automatically met by  physical states and in fact impose a significant restriction on them.

			The absence of local propagating degrees, as mentioned above,  for this model helps in its solution, but is a drawback in physical terms. It leaves us with very few physical observables.  As mentioned above the only physical geometric property for a circle is its length, and in the quantum theory this gives rise to the length operator, whose various moments $\langle l^n\rangle $ are physical properties of a state.  However, even these moments are best thought of as ``calculables" -  gauge invariant quantities that can be calculated from a (well defined) wave function, rather than observables. Whether they can actually be measured by observers given the exponential expansion of the universe,  is not  clear and requires a more involved discussion.

			Let us end this introduction with a few  more comments. As mentioned, we find that there are several gauge invariant states - which meet the Hamiltonian and Momentum constraints - that are not normalisable with respect to the Klein-Gordon norm, or for which the probability density $p(l)$ is not positive for all values of $l$. 
			Should such states be regarded as unphysical and discarded?
			In other words what criterion should be used to select physically acceptable states?
			In thinking about this question one  is led  back to confronting some of the other conceptual issues mentioned at the start above, tied to the meaning and operational significance of the wave function. Given our incomplete understanding, we believe,  it  is  best to keep an open mind in this regard. It could be that once additional degrees of freedom are included some of the problems with these states are alleviated.  E.g.,  gauge invariant observables could then exist which can be calculated  in such states, or it could be that some of the issues tied  to  the existence of the norm and positivity of $p(l)$ become less of a concern. 
			
			One example is provided by the Hartle-Hawking state. This state has a diverging norm, as has been noticed before, \cite{Maldacena:2019cbz,Verlinde:2020zld,Vilenkin:2021awm,Moitra:2022glw},  and as  we will discuss in section \ref{exmp}. From the path integral point of view the divergence is tied to the presence of conformal killing vectors and one expects that it can be dealt with after additional matter fields are introduced, for correlation functions with enough matter insertions.  This example illustrates how physical observables could arise once additional degrees of freedom are included.
			
		This paper is structured as follows. In section \ref{canquJT}, we discuss the canonical quantization procedure and construct the physical states that satisfy the constraint equation. In section \ref{pbtime}, we discuss the problem of time and propose dilaton a s a physical clock. With this choice of clock we analyze the properties of the states constructed such the conservation and positivity of the norm, classical limit etc. In section \ref{exsec} we study some examples to illustrate more concretely the various properties of states discussed in earlier sections. In section \ref{dtprop}, we extend our discussion to the case of multiple boundaries (universes). In section \ref{concs}, we end with some concluding remarks and open questions. Some additional related calculations are presented in appendices. 

			 For related work in JT gravity in dS space, see the following references \cite{Vilenkin:2021awm, Narayan:2020pyj, Witten:2020ert,  Blommaert:2020tht, Chakraborty:2023los, Chakraborty:2023yed, Constantinidis:2008ty,  Stanford:2019vob,Stanford:2020qhm, Strobl:1993yn, Anegawa:2023wrk,Aguilar-Gutierrez:2021bns,Jensen:2023eza,Baek:2022ozg,Castro:2022cuo,Aalsma:2022swk,Rahman:2022jsf,Svesko:2022txo,Teresi:2021qff,Kames-King:2021etp,Addazi:2021vvy,Niermann:2021wco,Balasubramanian:2020xqf,Hartman:2020khs, Chen:2020tes,Mirbabayi:2020grb,Cotler:2019dcj,witten2020matrix,eberhardt20232d,mertens2023solvable}.

		
		\section{Canonical Quantization of JT gravity in dS space}
		\label{canquJT}
		
		The Jackiw-Teitelboim (JT) theory   consists of 2D gravity coupled to a scalar, $\phi$, called the dilaton,
		with an action, 
		\be
		I_{\text{JT}}= \frac{1}{16\pi G}\pqty{\int d^2 x\,\sqrt{-g}\, \phi ( R-\Lambda)-2\int_{bdy}\sqrt{-\gamma}\phi K }\label{jtacta}.
		\ee
 
		Note that this theory has no dimensionless constant. $\Lambda$ is the only dimensionful parameter - it is positive and we will set it equal to $2$, giving,
		\be
		I_{\text{JT}}= \frac{1}{16\pi G}\pqty{\int d^2 x\,\sqrt{-g}\, \phi ( R-2)-2\int_{bdy}\sqrt{-\gamma}\phi K }
		   \label{jtact}
		 \ee
		$G$ - the two dimensional Newton constant - is dimensionless and can be absorbed by rescaling $\phi$,
		${\phi \over 16 \pi G} \rightarrow \phi$. 
		
		There is actually an additional term which is purely topological:
		\be
		\label{top}
		I_{\text{top}}={\phi_0\over 16 \pi G}  \left(\int d^2x \, \sqrt{-g} R -2 \int_{bdy}\sqrt{-\gamma} K\right)
		\ee
		with 
		 $\phi_0$ being a constant.  This term will not play much of a role in our discussion since we will not consider topology changing processes and mostly consider  spacetimes with one boundary  and no handles, for which it    equals\footnote{More generally, when we are dealing with higher topologies involving multi-boundaries or extra handles, the wave function or amplitude will be proportional to $e^{S_0 \chi}$, where $\chi=2-B-2H$.}		
		\be
		\label{vals0}
		S_0={\phi_0\over 4 G}. 
		\ee
		JT gravity can be obtained from dimensional reduction of the near-extremal black hole in $4$ dim dS space, \cite{Maldacena:2019cbz}. In this case $S_0$ is half the entropy 
		of the $4$ dim solution, obtained by adding the contributions of the black hole and cosmological horizons.
		
		One more comment is in order. Sometimes we will find it worthwhile to work with an   action that has a more general potential $U(\phi)$ for the dilaton,
		\be
		I_{\text{JT}}= \frac{1}{16\pi G}\pqty{\int d^2 x\,\sqrt{-g}\, (\phi  R-U(\phi))-2\int_{bdy}\sqrt{-\gamma}\phi K }
		   \label{jtactb}
		 \ee
		
		We will discuss the canonical quantisation of this theory below. Before doing so let us  review some facts about the classical behaviour of   the theory. 
		
		\subsection{Classical Solutions}
\label{cllimiwfa}
The classical limit corresponds to take $G\rightarrow 0$.
In this limit  the solutions to the equations of motion  from the action eq.(\ref{jtact}) are given by, 
\be	R=2,\,\nabla_\mu\nabla_\nu \phi-g_{\mu\nu}\nabla^2\phi-g_{\mu\nu}\phi=0
\ee
The first condition tells us that all solutions are  locally de Sitter space. 

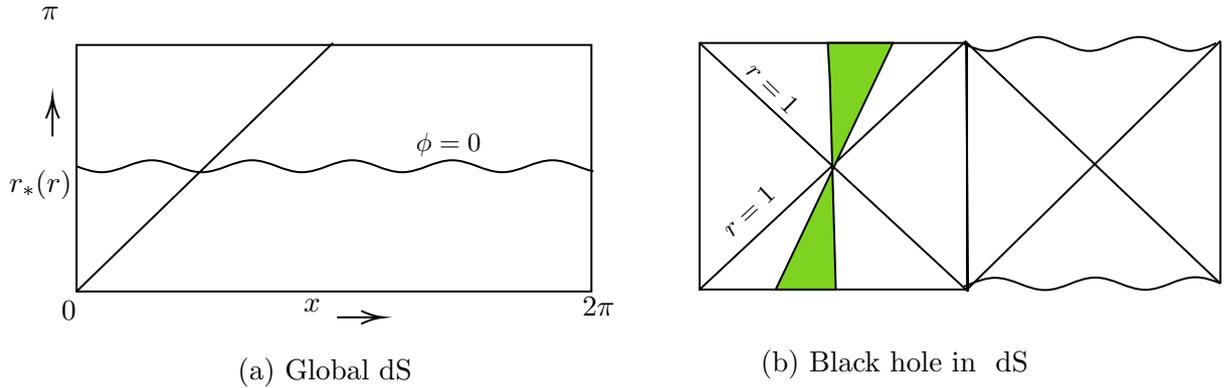
\begin{figure}

	\tikzset{every picture/.style={line width=0.75pt}} 
	
	\begin{tikzpicture}[x=0.75pt,y=0.75pt,yscale=-1,xscale=1]
		
		\draw   (37,81) -- (294,81) -- (294,205) -- (37,205) -- cycle ;
		\draw    (481,79) -- (482,205) ;
		\draw    (348,80) -- (482,205) ;
		\draw    (481,79) -- (348,204) ;
		\draw    (165,80) -- (37,205) ;
		\draw   (37,142) .. controls (41.08,143.54) and (44.98,145) .. (49.5,145) .. controls (54.02,145) and (57.92,143.54) .. (62,142) .. controls (66.08,140.46) and (69.98,139) .. (74.5,139) .. controls (79.02,139) and (82.92,140.46) .. (87,142) .. controls (91.08,143.54) and (94.98,145) .. (99.5,145) .. controls (104.02,145) and (107.92,143.54) .. (112,142) .. controls (116.08,140.46) and (119.98,139) .. (124.5,139) .. controls (129.02,139) and (132.92,140.46) .. (137,142) .. controls (141.08,143.54) and (144.98,145) .. (149.5,145) .. controls (154.02,145) and (157.92,143.54) .. (162,142) .. controls (166.08,140.46) and (169.98,139) .. (174.5,139) .. controls (179.02,139) and (182.92,140.46) .. (187,142) .. controls (191.08,143.54) and (194.98,145) .. (199.5,145) .. controls (204.02,145) and (207.92,143.54) .. (212,142) .. controls (216.08,140.46) and (219.98,139) .. (224.5,139) .. controls (229.02,139) and (232.92,140.46) .. (237,142) .. controls (241.08,143.54) and (244.98,145) .. (249.5,145) .. controls (254.02,145) and (257.92,143.54) .. (262,142) .. controls (266.08,140.46) and (269.98,139) .. (274.5,139) .. controls (279.02,139) and (282.92,140.46) .. (287,142) .. controls (289.67,143.01) and (292.26,143.98) .. (295,144.53) ;
		\draw   (348,80) -- (481.5,80) -- (481.5,204) -- (348,204) -- cycle ;
		\draw   (480,80.5) .. controls (484.08,82.29) and (487.98,84) .. (492.5,84) .. controls (497.02,84) and (500.92,82.29) .. (505,80.5) .. controls (509.08,78.71) and (512.98,77) .. (517.5,77) .. controls (522.02,77) and (525.92,78.71) .. (530,80.5) .. controls (534.08,82.29) and (537.98,84) .. (542.5,84) .. controls (547.02,84) and (550.92,82.29) .. (555,80.5) .. controls (559.08,78.71) and (562.98,77) .. (567.5,77) .. controls (572.02,77) and (575.92,78.71) .. (580,80.5) .. controls (584.08,82.29) and (587.98,84) .. (592.5,84) .. controls (597.02,84) and (600.92,82.29) .. (605,80.5) .. controls (605.33,80.35) and (605.67,80.21) .. (606,80.06) ;
		\draw   (608,200.91) .. controls (603.92,199.33) and (600.01,197.83) .. (595.49,197.84) .. controls (590.97,197.86) and (587.07,199.38) .. (583,200.99) .. controls (578.93,202.59) and (575.03,204.11) .. (570.51,204.13) .. controls (565.99,204.14) and (562.08,202.64) .. (558,201.06) .. controls (553.92,199.48) and (550.01,197.98) .. (545.49,197.99) .. controls (540.97,198.01) and (537.07,199.53) .. (533,201.13) .. controls (528.93,202.74) and (525.03,204.26) .. (520.51,204.27) .. controls (515.99,204.29) and (512.08,202.79) .. (508,201.21) .. controls (503.92,199.63) and (500.01,198.13) .. (495.49,198.14) .. controls (490.97,198.15) and (487.07,199.68) .. (483,201.28) .. controls (482,201.68) and (481.01,202.07) .. (480.02,202.43) ;
		\draw    (608,79) -- (608,201) ;
		\draw  [fill={rgb, 255:red, 126; green, 211; blue, 33 }  ,fill opacity=1 ] (412,80) -- (444.5,80) -- (386,204) -- (416,204) -- (413,98) -- cycle ;
		\draw    (608,79) -- (481.5,204) ;
		\draw    (481.5,80) -- (608,201) ;
		\draw    (25,127) -- (25,109) ;
		\draw [shift={(25,107)}, rotate = 90] [color={rgb, 255:red, 0; green, 0; blue, 0 }  ][line width=0.75]    (10.93,-3.29) .. controls (6.95,-1.4) and (3.31,-0.3) .. (0,0) .. controls (3.31,0.3) and (6.95,1.4) .. (10.93,3.29)   ;
		\draw    (167,218) -- (188,218) ;
		\draw [shift={(190,218)}, rotate = 180] [color={rgb, 255:red, 0; green, 0; blue, 0 }  ][line width=0.75]    (10.93,-3.29) .. controls (6.95,-1.4) and (3.31,-0.3) .. (0,0) .. controls (3.31,0.3) and (6.95,1.4) .. (10.93,3.29)   ;
		
		\draw (116,237) node [anchor=north west][inner sep=0.75pt]   [align=left] {(a) Global dS};
		\draw (377,233) node [anchor=north west][inner sep=0.75pt]   [align=left] {(b) Black hole in \ dS};
		\draw (28,208.4) node [anchor=north west][inner sep=0.75pt]    {$0$};
		\draw (288,206.4) node [anchor=north west][inner sep=0.75pt]    {$2\pi \ $};
		\draw (18,60.4) node [anchor=north west][inner sep=0.75pt]    {$\pi \ $};
		\draw (149,207.4) node [anchor=north west][inner sep=0.75pt]    {$x\ $};
		\draw (2,142.4) node [anchor=north west][inner sep=0.75pt]    {$r_{*}( r)$};
		\draw (376.53,86.37) node [anchor=north west][inner sep=0.75pt]  [font=\small,rotate=-44.59]  {$r=1$};
		\draw (356.09,171.83) node [anchor=north west][inner sep=0.75pt]  [font=\small,rotate=-316.78]  {$r=1$};
		\draw (205,123.4) node [anchor=north west][inner sep=0.75pt]  [font=\small]  {$\phi =0$};

	\end{tikzpicture}
\label{dsfig}
\caption{Penrose diagrams for (a) Global dS (b) Black hole in dS}
\end{figure}

One solution is global de Sitter space with metric, 
\be
\label{globalds}
ds^2=-{dr^2\over r^2+1}+(r^2+1) dx^2
\ee
where  $x$ parametrises a circle with, $x\simeq x+ 2\pi$.
The  dilaton takes the form,
\be
\label{dilpro}
\phi=A r
\ee 
The penrose diagram for global dS is shown in fig.\ref{dsfig}(a).
More general solutions are given by 
\begin{align}
	ds^2&=-{dr^2\over r^2-m}+ (r^2-m) \,dx^2,\quad x\sim x+{1}\,\nonumber\\
	\phi&=Ar	\label{dsmilne}
\end{align}
where $m$ can be both positive or negative. 
These solutions are characterized by two parameters, $m$ and $A$.

The solutions with $m>0$ are black holes with a cosmological horizon at $r=\sqrt{m}$ and a black hole horizon at $r=-\sqrt{m}$.
Here we will consider dS space with spatially compact slices. By a rescaling of $m, A,$ we can always adjust $x$ in eq.(\ref{dsmilne}) to lie in the range
\be
\label{rangex}
0< x\le 1
\ee
The resulting spacetime consists of a cone in the upper and lower Milne patches of the black hole geometry, see fig.\ref{dsfig}(b), with an orbifold singularity at $r=0$.
When $m<0$, eq.(\ref{dsmilne}), we can  again after a rescaling, take $x$ to lie in the range eq.(\ref{rangex}). 
It is easy to see that  this solution is related to  global dS solution eq.(\ref{globalds}) with the range of $x$  being different in general. 
For $m=-4\pi^2$ we get global dS described, after a suitable rescaling, by   eq.(\ref{globalds}). In this case the Penrose diagram is a rectangle and a signal starting from a point on the circle in the asymptotic far past, only makes it half way across the circle till asymptotic future infinity. A smaller or larger  magnitude for $m$ means that the signal reaches more or less  than half way across, respectively. In particular if $m<- \pi^2$ the signal can come back to the starting point before it reaches future infinity.

It is conventional in the study of JT gravity to  bound the  dilaton and allow it to take only values bigger than  some negative constant 
$\phi>-\phi_1$. This is a reasonable requirement when one views the JT theory arising from dimensional reduction, starting from say $4$ dimensions.
As mentioned above the dilaton is then related to the volume of the transverse sphere $S^2$ which cannot become negative. 
  Here we will for simplicity take $\phi_1=0$ and require the dilaton to satisfy the condition 
  \be
  \label{conddila}
  \phi\ge0, 
  \ee
  in effect requiring that the two dimensional Newton constant $G$ in eq.(\ref{jtact}) is positive. 
  Much of the resulting analysis will not be sensitively dependent on this precise choice. 
  Once eq.(\ref{conddila}) is imposed a boundary appears in spacetime at $\phi=0$ and  regions of spacetime where it is not met are  removed. 
  
  In the subsequent discussion we take the constant $A$ appearing in the dilaton solution, eq.(\ref{dilpro}) and eq.(\ref{dsmilne}) to be positive. 
  To meet eq.(\ref{conddila}) then,  for global de Sitter, only the region $r\ge 0$ is included ({the boundary $r=0$ is indicated by a wave line in fig.\ref{dsfig}(a)}).
  In the more general case eq.(\ref{dsmilne}), for $m>0$, both the upper and lower Milne patches, after identification, are included (these are shown as the shaded regions in fig.\ref{dsfig}(b)),  while for $m<0$ only the region $r\ge 0$ is included. 
  
  Two more comments are  in order. 
  First, in terms of the value of the dilaton, the length of the spatial circle along $x$ for eq.(\ref{dsmilne}) can be written as 

\begin{align}
 l= \sqrt{\frac{\phi^2}{A^2}-m}\label{leninmil}
\end{align}
  
  Second, the general solutions eq.(\ref{dsmilne}) asymptotically, as  $r\rightarrow \infty$, all  have a space-like boundary, ${\cal I}^+$.
  Conventionally this boundary is called future null infinity, since all null geodesics terminate here \cite{Bousso:2001mw}\footnote{The spacetime eq.(\ref{dsmilne}) also has a past boundary ${\cal I}^-$, as $r\rightarrow -\infty$. {When $m>0$ the spacetime has a past boundary, ${\cal I}^-$. However for $m<0$, this boundary is not present in the physical spacetime as $\phi$ becomes negative at this boundary and is removed due to the condition eq.\eqref{conddila}..}}.
  In AdS space one has, in contrast, a time-like boundary and a Brown-York stress tensor can be defined which is conserved with respect to time translations. 
  Here, in $dS_2$ space,  one can similarly define a  charge which is conserved with respect to spatial translations along the boundary. Since the boundary is space-like,  the conserved charge gives the momentum, instead of  energy, and generates translations along the spatial boundary at ${\cal I}^+$. 
  As shown in appendix \ref{symcur} this momentum is given by 
  \be
  \label{momadm}
  P_{\text{ADM}}=\half mA
  \ee
  (here we have set $8 \pi G=1 $).
   For $m>0$ cases, corresponding to  black hole solutions, the momentum is positive, while for  $m<0$, which corresponds to global dS with differing sizes of the spatial circle, it is negative. 
  
 \subsubsection{Classical Phase space and A Symplectic Form}
 \label{cpssym}
 The most general solution, as mentioned above, has two constants, $m,A$. This shows that the classical phase space of this system is two dimensional \footnote{We are grateful to Alok Laddha, Shiraz Minwalla and  Onkar Parrikar for patiently explaining various points pertaining to  this subsection to us.}. 
 A symplectic form on this phase space can be constructed, using covariant methods,  following \cite{Crnkovic:1986ex}. As discussed in appendix \ref{symcur} the symplectic 
 current,  $J^{\mu}$,  in JT gravity takes the form 
 \begin{align}
 	\sqrt{-g}J^\mu=-\delta \Theta^\mu, \quad\Theta^\mu=\half \sqrt{g} (g^{\nu\alpha} g^{\mu\beta} - g^{\mu \nu} g^{\alpha \beta} )\left(\phi \, \nabla_\nu \delta g_{\alpha \beta}-\nabla_\nu\phi\, \delta g_{\alpha \beta}\right).
 	\label{presymp1}
 \end{align}
 Working in Fefferman-Graham gauge with the general solutions eq.(\ref{dsmilne}) one then finds, see appendix \ref{fgforcpq},  that the corresponding  symplectic form is given by  eq.\eqref{current}
 \be
 \label{symf}
 \omega=\half \delta A \wedge \delta m.
 \ee
 In terms of the conserved momentum discussed above, the symplectic form is given by, 
 \begin{align}
 	\omega=\frac{1}{A} \delta A\wedge \delta P_{\text{ADM}}\label{sympinadm}.
 \end{align}
 
 In the next section we will canonically quantise the theory, and towards the end of the section show that the resulting behaviour in the classical regime where $\phi\rightarrow \infty$, $l \rightarrow \infty,$ is in accord with this understanding of the  classical phase space.

\subsection{Canonical Quantization}
\label{cllimiwf}
Next we turn to quantise the theory. 
		We shall  canonically quantise the theory by working in ADM-like gauge. 
	
		We follow  the approach  of \cite{Hennauxjt}, see also \cite{Verlinde:2020zld},  in the 	quantization procedure. In  ADM gauge the metric is given by 
		\begin{align}
			ds^2=g_{\mu\nu}dx^\mu dx^\nu=-N^2(x,t) dt^2+g_{1}(x,t)(dx+N_{\perp}(x,t)dt)^2\label{adm}
		\end{align}
		We mostly work in the units  of $8\pi G=1$ and reinstate this factor  wherever necessary. The action eq.\eqref{jtactb}, in this gauge becomes 
		\begin{align}
			I_{\text{JT}}=&\int d^2x \left[\frac{\dot{\phi}}{N}\left(\frac{N_{\perp}}{2\sqrt{g_1}}{g_1'}+\sqrt{g_1} N_{\perp}'-\frac{\dot{g}_1}{2\sqrt{g_1}}\right)\right]\nonumber\\
			&+\int d^2x\left[ \frac{\phi'}{N}\left(\frac{NN'}{\sqrt{g_1}}-\sqrt{g_1}N_{\perp}N_{\perp}'+\frac{N_{\perp}}{2\sqrt{g_1}}\dot{g}_1-\frac{N_{\perp}^2}{2\sqrt{g_1}}g_1'\right)-\frac{1}{2}NU(\phi)\sqrt{g_1}\right]\label{jtadmact}
		\end{align}
	 It is easy to see from eq.\eqref{jtadmact} that the variables $N_{\perp},N$ are non-dynamical since they do not have time derivatives, whereas $\phi, g_1$ are dynamical variables. The canonically conjugate  momenta  can be evaluated from the above action to be,
		\begin{align}
			\pi_{N}&=0=\pi_{N_{\perp}},\nonumber\\
			\pi_\phi&=\frac{N_{\perp}}{2N\sqrt{g_1}}g_1'+\frac{N_{\perp}'\sqrt{g_1}}{N}-\frac{\dot{g}_1}{2N\sqrt{g_1}}\nonumber\\
			\pi_{g_1}&=-\frac{\dot{\phi}}{2N\sqrt{g_1}}+\frac{N_{\perp}\phi'}{2N\sqrt{g_1}}.\label{canmomall}
		\end{align}
	In the gauge $N_{\perp}=0$, we find
	\begin{align}
		\pi_{\phi}=-\frac{\dot{g}_1}{2N\sqrt{g_1}},\quad \pi_{g_1}=-\frac{\dot{\phi}}{2N\sqrt{g_1}}\label{cmominnpz}
	\end{align}
		
		Also, the constraint equations, the equations of motion obtained by varying with respect to $N,N_{\perp}$, are, 
		\begin{align}
			0=\mathcal{H}\equiv{\frac{\delta 	I_{\text{JT}}}{\delta N}}&=-\frac{\pi_\phi\dot{\phi}}{N}-\left(\frac{\phi'}{\sqrt{g_1}}\right)'-\frac{\sqrt{g_1}U(\phi)}{2}\nonumber\\
			&=2\pi_\phi \pi_{g_1}\sqrt{g_1}-\left(\frac{\phi'}{\sqrt{g_1}}\right)'-\frac{\sqrt{g_1}U(\phi)}{2}\nonumber\\
			0=\mathcal{P}\equiv\frac{\delta 	I_{\text{JT}}}{\delta N_{\perp}}&=\frac{\dot{\phi}g_1'}{2N\sqrt{g_1}}-\left(\frac{\dot{\phi}\sqrt{g_1}}{N} \right)'+\frac{\phi'\dot{g}_1}{2N\sqrt{g_1}}\nonumber\\
			&=2g_1\pi_{g_1}'+\pi_{g_1}g_1'-\pi_\phi \phi'\label{hammomcon}
		\end{align}
		The physical wavefunction has to satisfy the constraints above, $\mathcal{H}=0$ and $\mathcal{{P}}=0$, the Hamiltonian and Momentum constraints respectively, i.e
		\begin{align}
			\mathcal{H}\Psi=0\nonumber\\
			\mathcal{P}\Psi=0\label{hmwfcon}
		\end{align}
		The Hamiltonian constraint  is also referred to as the Wheeler De-Witt (WDW) equation.
		 
	By appropriate coordinate transformation we can make a gauge choice such that
	\begin{align}
		N=1,N_{\perp}=0\label{nnperg}.
	\end{align}
	  Even after this gauge fixing, we still have the freedom to make coordinate transformation that preserve this choice of gauge. These are the residual gauge symmetries and	 are of the form
	\begin{align}
		t\rightarrow t+\epsilon_1(x),\,x\rightarrow x+\epsilon_2(t,x).\label{txrestran}
	\end{align}
	The constraints eq.\eqref{hmwfcon} imply that the physical wavefunction is invariant under these residual gauge transformations.   We will have a more detailed discussion  of the constraints,  in an upcoming work\cite{nanda:2023yta}. 
	
	As noted in \cite{Hennauxjt}  there are two issues one needs to deal with in  quantizing this system. First, the Hamiltonian and Momentum constraints suffer from operator ordering ambiguities which need to be resolved. Second, the Hamiltonian constraint involves a double functional derivative, $\pi_\phi \pi_{g_1}$,  and  this gives rise to $\delta(0)$ type of singularities which must be  regularised in some fashion. 
	
	To deal with these issues \cite{Hennauxjt}  adopted a particular strategy. The constraints were first  manipulated classically  by taking linear combinations  to express $\pi_{g_1}$ directly in terms of the 
	position variables, $g_1,\phi$. The resulting relation was then taken to hold for physical wave functions - this gets rid of the $\delta(0)$ singularities. It was then  found that these wave functions also  automatically satisfy the momentum constraint, for  one particular choice of operator ordering. By taking this choice of operator ordering to be the  correct one, 
	solutions to both constraints could in-effect be  found. These solutions were taken to be the gauge invariant wave functions. We will follow the same strategy here.

	In the classical theory,  taking an appropriate linear combination of the Hamiltonian and Momentum constraints, we can eliminate $\pi_\phi$, as shown below
	\begin{align}
	2\pi_{g_1}\sqrt{g_1}\mathcal{P}+\phi' \mathcal{H}&=2\pi_{g_1}\sqrt{g_1}(2g_1\pi_{g_1}'+\pi_{g_1}g_1')-\phi'\pqty{\left(\frac{\phi'}{\sqrt{g_1}}\right)'+\frac{\sqrt{g_1}U(\phi)}{2}}\nonumber\\
	&=2\sqrt{g_1}\pqty{g_1\pi_{g_1}^2-\frac{1}{4}\left(\frac{\phi'}{\sqrt{g_1}}\right)^2-\frac{1}{4}{\phi}^2}'\label{lcomcon}
	\end{align}
	where we used the fact that $U(\phi)=2\phi$ for the case of dS spacetime under consideration\footnote{The canonical quantization for more general potentials is very similar to the case of dS studied here and  will be treated in \cite{nanda:2023yta}.}. 
	Since the LHS vanishes by the equations of motion,  we learn that the quantity in the brackets in the final line above should be independent of $x$, the spatial coordinate. Defining $M(t)$ as,
	\begin{align}
		M(t)=-4\left({g_1\pi_{g_1}^2-\frac{1}{4}\left(\frac{\phi'}{\sqrt{g_1}}\right)^2-\frac{1}{4}{\phi}^2}\right)\label{lconint}.
		\end{align}
	 It can further be shown in fact that the above quantity is a constant for on-shell configurations. Using eq.\eqref{cmominnpz} in eq.\eqref{lconint}, we have
	\begin{align}
		M(t)=&{-\frac{\dot{\phi}^2}{N^2}+\left(\frac{\phi'}{\sqrt{g_1}}\right)^2+\phi^2}\nonumber\\
		=&\nabla_\mu\phi\nabla^\mu\phi+\phi^2\label{mas}
	\end{align}
This quantity is independent of $t$ for configurations where  $\phi$  satisfies the operator equations of motion. Consider the quantity $\nabla_\alpha M$ given by 
\begin{align}
	\nabla_\alpha M=2{\nabla^\mu\phi}(\nabla_\alpha \nabla_\mu\phi+g_{\mu\alpha}\phi)\label{derm0}
\end{align}
The EOM for $\phi$, obtained by varying with respect to the metric $g_{\mu\nu}$, gives, 
\begin{align}
	\nabla_\mu\nabla_\nu\phi-g_{\mu\nu}\nabla^2\phi-g_{\mu\nu}\phi=0.\label{eomgmu}
\end{align} Inserting  this in eq.(\ref{derm0}) and using the condition following from the trace of eq.(\ref{eomgmu}), $\nabla^2\phi=-2\phi$,  then leads to 
\be
\label{condderm0}
\nabla_\alpha M=0.
\ee

From now on, we drop the $t$ dependence of $M$.
Let us also note that for the general classical solution, eq.(\ref{dsmilne}), we get 
\be
\label{valMa}
M=mA^2
\ee
as discussed in appendix \ref{fgforcpq}, eq.(\ref{Mcos}). 
Also, we note that $M$ is related to the conserved momentum which can be defined at space-like infinity ${\cal I}^+$, as is discussed in appendix \ref{dsadmmass} and eq.(\ref{momadm}).

 From eq.\eqref{lconint}, it follows that 
\begin{align}
	4 g_1\pi_{g_1}^2={\left(\frac{\phi'}{\sqrt{g_1}}\right)^2+{\phi}^2-M}\label{pigval}
\end{align}

Solving constraints eq.\eqref{hmwfcon} is equivalent, classically,  to imposing the conditions eq.\eqref{pigval} and the momentum constraint in eq.\eqref{hammomcon}.

In going to the quantum theory, we now impose the relations eq.(\ref{pigval}) and the momentum constraint with the choice of operator ordering such that
\begin{align}
	\mathcal{{P}}\Psi=(2g_1\pi_{g_1}'+g_1'\pi_{g_1}-\phi'\pi_\phi)\Psi=0\label{moope}
\end{align}  as conditions that physical states must satisfy in the quantum theory. 

Imposing  eq.\eqref{pigval} leads  to 
\begin{align}
	-i\frac{\del\Psi}{\del g_1}=\frac{1}{2\sqrt{g_1}}\left(\sqrt{\left(\frac{\phi'}{\sqrt{g_1}}\right)^2+{\phi}^2-M}\right)\,\,\Psi\label{henhmcom}
\end{align}
with the solution, 
\begin{align}
	\Psi[g_1(x),\phi(x)]=e^{i\mathcal{S}[g_1(x),\phi(x)]},\quad
	\mathcal{S}[g_1(x),\phi(x)]&=\int dx \pqty{\mathcal{Q}- {\phi'(x)}\tanh^{-1}\pqty{\frac{\mathcal{Q}}{\phi'(x)}}}\nonumber\\
	\mathcal{Q}&=\sqrt{(\phi(x)^2-M)g_1(x)+\phi'(x)^2}\label{wfsinw}
\end{align}
Now, we can obtain the action of $\pi_\phi$ on the above solution by using eq.\eqref{moope} which gives, 
\begin{align}
	-i\frac{\del\Psi}{\del \phi}=\frac{\sqrt{g_1}}{\mathcal{Q}}\pqty{\left(\frac{\phi'}{\sqrt{g_1}}\right)'+{\sqrt{g_1}\phi}}\Psi.\label{momconwf}
\end{align}
In appendix \ref{momcowf}, we verify the above result by explicitly computing the variation of the solution eq.\eqref{wfsinw} with respect to $\phi$. .

$M$, a constant,  which appears in the wave function is to be interpreted as the eigenvalue of a quantum operator ${\hat M}$ which is a suitably regularised version of the classical variable\footnote{In the AdS case $M$ is related to the ADM mass, \cite{Hennauxjt,Verlinde:2020zld}} eq.(\ref{lconint}). 
We note that  we have  not been    precise above about the definition of ${\hat M}$, in terms of ${\hat \pi}_\phi, {\hat \pi}_{g_1}, \phi, g_1$, except to imply  that its  action on a physical state is obtained by taking the action of the quantum operator $-4( g_1 {\hat \pi}_{g_1}^2 -{1\over 4}{( \phi')^2\over g_1}-{1\over 4} \phi^2)$ and discarding all the $\delta(0)$ terms. It would certainly be worthwhile to try and give a more precise definition of this operator.

The functions  $\phi(x),  \,g_1(x)$ which appear in $\Psi$ can be thought of as eigenfunctions of the quantum operators ${\hat \phi}(x)$, ${\hat g}_1(x)$ and the wave function is therefore the  overlap of the states with an eigenbasis of the $\hat{\phi},\hat{g}_1$ operators.
Also we observe that the eq.\eqref{pigval} is quadratic in $\pi_{g_1}$ and hence has a square root ambiguity. The result in eq.\eqref{wfsinw} is for a specific choice corresponding to the positive square root solution for $\pi_{g_1}$ in eq.\eqref{pigval}. The alternate choice of negative square root would give the corresponding solution in eq.\eqref{wfsinw} with a negative sign in the exponent leading to,
\be
\label{sol22}
\Psi[g_1(x),\phi(x)]=e^{-i\mathcal{S}[g_1(x),\phi(x)]}
\ee
General solutions would be a linear combination of eq.(\ref{wfsinw}) and eq.(\ref{sol22}).

More generally, solutions to the gauge constraints   can be written as a sum over various eigenvalues, $M$, for each of which it takes the form given above. Motivated by the fact that in the classical theory  $M$ takes real values which can  be both positive and negative, we also take the range of $M\in[-\infty,\infty]$ in the quantum theory.

Putting all this together we get that a general solution is an arbitrary linear combination of these two branches and different values of $M$  and can be written as,
\begin{align}
	\Psi=\int dM{\tilde \rho}(M)e^{i\mathcal{S}}+\int dM \rho(M)e^{-i\mathcal{S}}\label{gensol}
\end{align}
where the integral goes over all values of $M\in [-\infty,\infty]$ and $\rho(M), {\tilde \rho}(M) $ are arbitrary complex functions of $M$. 

Eq.(\ref{gensol}) is of central importance for this paper, and we will explore the consequences of any physical state being described by a wave function of this form in the subsequent sections below.

To proceed, we now make the choice of a  spacelike slice such that  such that $\phi'(x)=0$ on it, i.e.  $\phi(x)$ is independent of $x$. Then the wavefunction, eq.(\ref{wfsinw})  simplifies to
\begin{align}
	\Psi=e^{i \sqrt{\phi^2-M}\int dx\sqrt{g}_1}.\label{wfinpm}
\end{align}
Note that starting from such a  spacelike slice one can obtain the wave function on any other slice related to it by the coordinate transformations,
eq.(\ref{txrestran})  which preserve ADM gauge, by the action of the Hamiltonian and momentum constraints. Physical states are annihilated by these constraints and therefore invariant under such coordinate transformations. Thus,  the wave function on any other slice obtained through such a  coordinate  transformations remain the same.

We will choose the spatial slice to be compact as mentioned before. Denoting by $l$  the length of the spatial extent of this spacelike slice, $l=\int dx \sqrt{g_1}$, eq.(\ref{wfinpm}) becomes
\begin{align}
	\Psi=e^{il\sqrt{\phi^2-M}}.\label{psigencon}
\end{align}
The most general wavefunction eq.\eqref{gensol} will then be of the form
\begin{align}
	\Psi=\int dM {\tilde \rho}(M)e^{il\sqrt{\phi^2-M}}+\int dM \rho(M) e^{-il\sqrt{\phi^2-M}}.\label{genwfinl}
\end{align}
with the coefficients $\rho(M), {\tilde\rho}(M)$  being general complex functions of $M$.

Let us also note that the Hartle-Hawking wave function, which can be obtained by path integral methods, agrees with the general form above, 
as we will discuss further in section \ref{pbtime}. 

It is also worth noting, as is discussed in appendix \ref{dsadmmass} in more detail, that the above formula for $\Psi$, asymptotically in the limit when $l,\phi$ are large (or more precisely when the WKB approximation is valid), is  suggestive of being  a trace of the operator $e^{\pm i{\hat P}}$,  where ${\hat P}$ is the translation operator along the future boundary ${\cal I}^+$, whose eigenvalue is related to  the ADM momentum, $P_{\text{ADM}}$ discussed in appendix \ref{dsadmmass}. For the expanding branch $\rho(M)$ would  then be the density of  eigenstates of ${\hat P}$ 
whose eigenvalues lie in the range $[M, M+dM]$, with $M$ being related to $P_{\text{ADM}}$, for any given value of $A$,  by eq.(\ref{admmvsm}). Similarly for the contracting branch 
${\tilde \rho}$ would be related to the density of eigenvalues of ${\hat P}$.
Such an interpretation is also suggested by the HH wave function which we study further below. This wave function  has $\rho={\tilde \rho}$ and is related to the partition function in AdS space after an analytic continuation, which also relates ${\hat P}$ to $H$ - the AdS Hamiltonian that generates boundary time translations. However, things are more non-trivial in the dS case for general states. First,  
$\rho$ and
 ${\tilde \rho}$ are in general independent, and if such an interpretation is true one would need two separate density of states to account for both of them.  Second, as we have noted above, $\rho$ and ${\tilde \rho}$ are complex, and as we will see below, even when   real,  they are often   not positive. In that case  $\rho,{\tilde \rho}$ cannot be interpreted,  at least as   conventional density of states. We leave  a more detailed discussion of some of these issues for the future.

\subsection{Minisuperspace Quantization}
\label{minquant}
In the previous subsection we have obtained the general wavefunction satisfying the gauge  constraints. 
It is also illuminating to consider what one obtains for the wave function in the mini-superspace approximation.  This approximation amounts to considering  configurations for which $\phi, g_1$ are independent of $x$,
i.e.,  
$\phi'(x)=g_1'(x)=0$. 

As above, we will  consider compact spatial slices and without loss of generality  we take the   $x$ coordinate to satisfy the condition,  $x\sim x+1$. The proper length of  such  a slice is given by 

	\begin{align}
		l=\sqrt{g_1}\label{lg1}.
	\end{align}
Since $l$ is the physical length of the boundary, it is a non-negative quantity. We also take the dilaton, $\phi$, to be non-negative as mentioned earlier, see discussion around eq.\eqref{conddila}.

In the mini-superspace approximation the JT action, eq.\eqref{jtadmact} becomes, 
\be
\label{miniact}
I= - \int d^2x \left[\frac{1}{N} {\dot l} {\dot \phi} + N l \phi\right] 
\ee
The corresponding Hamiltonian constraint is, 
\be
\label{miniham}
H=-\pi_l \pi_\phi + l \phi
\ee

In the quantum theory, we impose the canonical commutation relations on the variables $\phi,\pi_\phi,g_1,\pi_{g_1}$:
	 \begin{align}
	 	[\phi,\pi_\phi]=i,\nonumber\\
	 	[g_1,\pi_{g_1}]=i,\label{canoncom}
	 \end{align}		
  and   $\pi_{g_1}$ and $\pi_{\phi}$ become the operators $\pi_{g_1}=-i\del_{g_1},\pi_{\phi}=-i\del_\phi$. Using eq.\eqref{lg1}, we have
  \begin{align}
  	\pi_{l}=-i\del_l=2\sqrt{g_1}\pi_{g_1}\label{pil}
  \end{align}
 	
	Setting the Hamiltonian constraint eq.(\ref{miniham}) to vanish on physical states would give

	\begin{align}
		(\del_l \del_\phi+\phi l)\hat{\Psi}[\phi,l]=0\label{dshamcon}.\end{align}
	This has  a solution
	\begin{align}
		\hat{\Psi}[\phi, l]=\frac{1}{l}e^{il\sqrt{\phi^2-M}},\label{mpsisol}
		\end{align}
	 where $M$ is an arbitrary constant. 
	 
	 We see that this solution does not agree with the solution obtained in the previous subsection in eq.\eqref{psigencon}. Denoting the solution in  eq.\eqref{psigencon} by $\Psi$ we have  that  
	 \be
	 \label{formsa}
	 \hat {\Psi}={\Psi\over l}. 
	 \ee
	 
	 The extra factor of $l$ is  tied to operator ordering ambiguities we had mentioned in the previous subsection and the method of quantization we adopted leading to eq.(\ref{psigencon}). To get agreement with our  previous result we need to change the WDW equation in  mini-superspace to instead become 
	 \begin{align}
		\left(\del_l \del_\phi-{1\over l } \del_\phi + \phi l\right)\Psi[\phi,l]=0,\label{newwdw}
	\end{align}	
	which  is then  satisfied by eq.(\ref{psigencon}). 
	
	 The reader might wonder why we are keen to retain the procedure   we followed in quantizing the theory in the previous section, along with the implicit choice for operator ordering etc that  it contains. An important reason, as was mentioned at the end of the previous subsection, and as we will comment on in more detail below,  is that the Hartle- Hawking wave function agrees with the general form of the solution which we have obtained following this procedure.

	In our subsequent discussion it will be sometime convenient to work with the variable ${\hat \Psi}$, related to the actual wave function $\Psi$ by eq.(\ref{formsa}),   since it satisfies a standard Klein-Gordon type   equation, eq.(\ref{dshamcon}). For ease of nomenclature we will sometimes refer to eq.(\ref{dshamcon}), as the WDW equation.

Also note that in obtaining eq.\eqref{dshamcon}, we have set $8\pi G=1$. Reinstating this factor, we would get eq.(\ref{dshamcon}) to be
\begin{align}
			((8\pi G)^2\del_l \del_\phi+\phi l)\hat{\Psi}[\phi,l]=0\label{dshamconwg}
\end{align}

	\subsection{General Solutions}
	\label{gensols}
	
	Let us start with a solution of the form
	\be
	\label{sola}
	\Psi=e^{-il \sqrt{\phi^2-M}}
	\ee
	When $l, \phi\gg 1$ the phase factor oscillates rapidly and the WKB approximation for eq.(\ref{newwdw})  (or eq.(\ref{dshamcon}))  becomes good, see also appendix \ref{allnfwf}. 
	The solution then corresponds to a universe which is expanding. This follows from noting that 	
	\be
	\label{actpp}
	\pi_\phi \Psi=-{l\phi \over \sqrt{\phi^2-M} }\Psi
	\ee
	and  also from the fact that  eq.(\ref{pil}) and eq.(\ref{cmominnpz}) give 
	\be
	\label{pxx}
	\pi_\phi=-{\dot l}
	\ee
	A negative value for $\pi_\phi$ therefore corresponds to ${\dot l}>0$. 
	Keeping this in mind we will refer to  the branch, 
	\be
	\label{posb}
	\Psi=\int \rho(M) e^{-il \sqrt{\phi^2-M}}
	\ee
	as the expanding  branch. 	
	
	In contrast, the branch
	\be
	\label{posc}
	\Psi=\int {\tilde \rho}(M) e^{il \sqrt{\phi^2-M}}
	\ee
	leads to a contracting universe,  i.e.  to a universe which shrinks as coordinate time increases, and we will refer to it as the contracting branch. 
	
	A general solution is a linear combination of both branches. Actually we need to be more precise. If $\phi^2<M$, we need to need to specify the meaning of the square root  which appears in the wave function. Keeping this in mind we write the  general wave function  as,
	\be
	\label{genexwf}
	\Psi=\int_{-\infty}^{\phi^2} dM \left(\rho(M) e^{-il\sqrt{\phi^2-M}} + {\tilde \rho}(M) e^{il\sqrt{\phi^2-M}}\right) + \int_{\phi^2}^\infty dM \left(\rho_1(M) e^{-l \sqrt{M-\phi^2}}+\rho_2(M) e^{l\sqrt{M-\phi^2}}\right)
	\ee
	The first two terms, which are oscillating, are the analogue of a wave function in the classically allowed region, and the last two in the classically forbidden region. In the subsequent discussion we will refer to the first two terms as belonging to the oscillatory branch and the last two as belonging to   the tunneling branch.
	
	Writing $\Psi$ in the form
	\begin{eqnarray}
			\Psi&=&\int_{-\infty}^{\phi^2} dM \,\left(\rho(M) e^{-il\sqrt{\phi^2-M}} + {\tilde \rho}(M) e^{il\sqrt{\phi^2-M}}\right) \nonumber\\
			&&+ \int_{\phi^2}^\infty dM\, \left(\rho_1(M) e^{-l \sqrt{M-\phi^2}}+\rho_2(M) e^{l\sqrt{M-\phi^2}}\right)\label{gnthefn}
	\end{eqnarray}
it is easy to see that 
	the WDW equation (\ref{newwdw}) leads to the condition 
	\be
	\label{condrho}
	\rho(\phi^2) + {\tilde \rho}(\phi^2)=\rho_1(\phi^2) + \rho_2(\phi^2)
	\ee
	which needs to be met for all $\phi^2$. 
	This means that for all positive values of $M$ they meet the condition, 
	\be
	\label{conr2}
	\rho(M) + {\tilde \rho}(M)=\rho_1(M) + \rho_2(M)
	\ee
	We see that this is only one condition among the $4$ complex coefficient functions. 
	Note also that $\rho(M), {\tilde \rho}(M)$ are unconstrained for $M<0$.

{\sl \underline {Another Representation}:}

\begin{figure}[h!]
\centering

\tikzset{every picture/.style={line width=0.75pt}} 

\begin{tikzpicture}[x=0.75pt,y=0.75pt,yscale=-1,xscale=1]
	
	\draw   (232.48,81.26) -- (363.49,79.77) -- (364.47,210.76) -- (233.46,212.25) -- cycle ;
	\draw    (232.48,81.26) -- (364.47,210.76) ;
	\draw    (233.47,212.25) -- (363.48,79.77) ;
	\draw  (456.64,90.25) -- (472.91,73.08)(441.13,72.28) -- (460.17,90.33) (464.46,74.72) -- (472.91,73.08) -- (471.72,81.6) (442.77,80.73) -- (441.13,72.28) -- (449.65,73.47)  ;
	
	\draw (307.74,110.16) node [anchor=north west][inner sep=0.75pt]  [rotate=-314.81]  {$\mathcal{H}_{1}$};
	\draw (269.82,88.74) node [anchor=north west][inner sep=0.75pt]  [rotate=-44.73]  {$\mathcal{H}_{2}$};
	\draw (423,63.4) node [anchor=north west][inner sep=0.75pt]    {$u$};
	\draw (477,61.4) node [anchor=north west][inner sep=0.75pt]    {$v$};

\end{tikzpicture}
\caption{Milne wedge and the horizons}
\label{h2h1hor}
\end{figure}
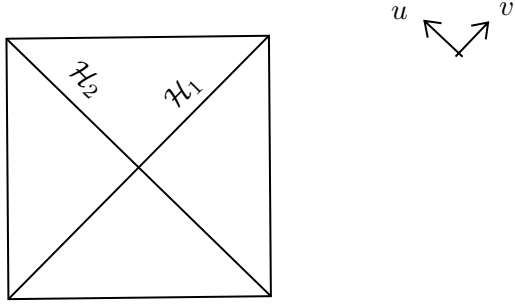

Let us end this section by discussing other ways, besides   eq.\eqref{genexwf}, of representing solutions to the constraint equations. One other     representation
which is helpful   is as follows, \cite{Maldacena:2019cbz}. The WDW equation eq.(\ref{dshamcon}) can be also written as 
\be
\label{formwdwa}
\partial_u\partial_v{\hat \Psi} +{1\over 4} {\hat \Psi}=0
\ee
where 
\be
\label{defuv}
u=l^2, \ \, v=\phi^2
\ee
This is the wave equation for a free scalar of mass $m^2=1$ in the flat metric 
\be
\label{fmet}
ds^2=-du\, dv
\ee
We see that $u,v$ are null coordinates\footnote{These observations might lead the reader to ask why we did not choose as our notion of physical time 
the variable $u+v =l^2+\phi^2$, instead of $\phi$. The reason is  that an acceptable   notion of time must be defined locally in spactime, while the length of a spacelike slice cannot be defined locally and needs information about the full slice. } and the condition, $u,v>0$ means we are working in the upper Milne patch, see Fig \ref{h2h1hor}. 

Defining coordinate 
\be
\label{defcona}
u=\xi\, e^{-\theta}, \ v= \xi \,e^{\theta},
\ee
eq.(\ref{formwdwa}) becomes, for  a mode ${\hat \Psi}_k\sim e^{i k \theta}$, 
\be
\label{genwdw}
\xi^2 {\hat \Psi_k}''+\xi {\hat \Psi_k}'+(\xi^2  + k^2 ) {\hat \Psi_k}=0
\ee
The general solutions to the above are the Bessel function $J_{i k}( \xi), J_{-ik}( \xi)$.
The two horizons, ${\cal H}_1, {\cal H}_2$ are given by  
\begin{eqnarray}
\label{twohor}
{\cal H}_1 & :   & \theta\rightarrow \infty, \xi \rightarrow 0,  \ {\rm keeping \  v \ fixed} \\
{\cal H}_2 & : & \theta\rightarrow -\infty, \xi \rightarrow 0, \ {\rm keeping\  u \ fixed}.
\end{eqnarray}

As discussed in appendix \ref{rindlerapp} a general solution of the KG equation can then be expanded as 
\be
\label{exppsi}
{\hat \Psi}=\int_{-\infty}^\infty dk \bigl[a(k) e^{i k \theta}J_{-i |k|}( \xi)+ b(k) e^{-i k \theta}J_{i |k|}( \xi)\bigr ]
\ee
where $a(k), b(k)$ are general complex coefficients. For $\xi\rightarrow 0$, $J_{\pm i |k|}\sim e^{\pm i |k| \ln(\xi)}$ , and it is easy to see therefore 
that near ${\cal H}_2$ the modes with $k>0$ behave like $u^{\pm i |k|}$, and similarly near ${\cal H}_1$ the modes with $k<0$ behave like 
$v^{\pm i |k| }$. Thus by examining the behaviour of ${\hat \Psi}$ near the two horizons   we can obtain the coefficients $a(k), b(k)$,   appendix \ref{rindlerapp}.

It also follows from this discussion   that  the  modes in eq.(\ref{exppsi}) provide  a complete set of  solutions to  eq.(\ref{dshamcon}). In particular the solutions discussed earlier, eq.(\ref{genexwf}), for both positive and negative values of $M$, can be expanded in terms of these modes, as is discussed in appendix \ref{rindlerapp}. 

Let us also note the following. Instead of considering modes going like $e^{ik \theta}$ if we took a mode going like $e^{m \theta}$ one finds the equation 

\begin{align}
	\xi^2{\hat \Psi_m}''+\xi {\hat \Psi_m}' +(\xi^2-m^2)\hat{\Psi}_m	=0\label{tsigmpsdf}
\end{align}
with a solution in terms of the Hankel functions, 
\begin{align}
	\hat{\Psi}_m=A_m H_m^{(1)}(\xi ) e^{m \theta}+B_m H_m^{(2)}(\xi) e^{m \theta}\label{psihanks}
\end{align}
where $A_m,B_m$ are complex coefficients. More generally one can consider solutions which are linear combinations
\begin{align}
	\hat{\Psi}(\phi, l)=\int dm\, e^{m\theta} (A_m H_m^{(1)}(\xi)+B_m H_m^{(2)}(\xi))\label{mgsh1h2}
\end{align}

To  generalise further,  we can consider shifting the $u,v$ variables and define new   variables
\be
	u=l^2+c_2,v=\phi^2+c_1\label{uvshifts}
	\ee
	with $\xi,\theta$ being related to these shifted $u,v$ variables by
	\be
	\xi =\sqrt{uv},\,e^\theta=\sqrt{\frac{v}{u}}\label{uvshiftsb}
	\ee

Then it is easy to see that a function  of the form eq.(\ref{psihanks}), in terms of the shifted variables, and more generally, eq.(\ref{mgsh1h2}),  also solves the WDW equation.

It will turn out that the Hartle-Hawking wavefunction obtained from the path integral calculation can be expressed as a term for a single value of $m$ in eq.\eqref{mgsh1h2} for the choice of $u,v$ in eq.\eqref{uvshifts} with $c_1=0, c_2=-(2\pi)^2$.

\section{Physical Properties and The Problem of Time}
\label{pbtime}

In the previous section we canonically quantised the theory and constructed the most general solution to the constraint equations. 
Here we will study these solutions in further detail and understand their properties. For doing so we will need to address the  important issue of the problem of time.  As discussed in the introduction, we will explore the idea   that the dilaton can provide a physical notion of time, i.e. a good clock, with which to view the evolution of the universe.   
The different solutions found in the previous section  lead to   different states, at a given instance of the dilaton. We  will show how  a norm for these states can be defined  and study  its properties. 
We  find that some but not all states have a  finite  norm which is conserved in time, and among them some have  an associated probability density which  is positive. 
Similarly, when we consider expectation values,   like for  moments of the length operator $l^n$ or its canonical conjugate $\pi_l^m$, these will be well defined for some but not all the states. Requiring a good classical limit when $\phi\rightarrow \infty$ will impose additional conditions on the states. 

Let us add one more comment. Other choices of physical time, besides the dilaton,  are also possible. For example one could consider the extrinsic curvature, $K$ as providing such a choice\footnote{We thank Onkar Parrikar for drawing attention to this possibility.}.
{For a given classical solution a fixed value of $\phi$ also corresponds to a fixed value of $K$ so the two sets of slices would be the same, however away from the classical limit the  behaviour of states constructed with these two choices  of time can be different. We leave a more detailed study of this question for the future.}

\subsection{Norm and Probability density}
\label{normproden}
 To proceed, we note that the wavefunction satisfies a Klein-Gordon like equation, eq.(\ref{dshamcon}) and (\ref{newwdw}),  in terms of the  coordinates $l,\phi$, (which are analogous to light-like coordinates in the  Klein-Gordon equation). Since this equation is linear in $\Psi$  the solutions lie in a  vector space  which  is  the space of solutions to this equation at any value of $\phi$. And the natural norm to consider  is a  Klein-Gordon (KG) type  norm 
 
   given by 
\begin{align}
	\langle \hat\Psi, \hat\Psi\rangle =\int_{0}^{\infty} dl \,\,i({\hat\Psi }^*\del_l \hat{\Psi} -\hat{\Psi}\del_l {\hat{\Psi}}^*),\label{kgnorm}
\end{align}
where the integral is  evaluated at  a fixed value of $\phi$. Actually one can consider the RHS with either a positive or negative sign which can be chosen to ensure that the norm is positive, as long as it is non-vanishing and finite. 

The condition that the norm is  conserved is,   
\begin{align}
\del_\phi \langle{\hat  \Psi},{\hat \Psi}\rangle =0.	\label{normcons}
\end{align}
This norm would lead to the probability density for the universe having  size $l$,  when dilaton takes value $\phi$, $p(l,\phi)$, being given by,  
\begin{align}
	p(l,\phi)=\pm  {\cal N}^{2} \   i({\hat\Psi }^*\del_l \hat{\Psi} -\hat{\Psi}\del_l {\hat\Psi}^*)\label{pbden}
\end{align}
where 
\be
\label{defn}
{\cal N}=\left( \sqrt{| \langle {\hat \Psi}|{\hat \Psi} \rangle|}\right)^{-1}
\ee
Eq.(\ref{pbden}) can also be written as
\be
\label{repbd}
p(l,\phi)=\pm {  i({\hat\Psi }^*\del_l \hat{\Psi} -\hat{\Psi}\del_l {\hat\Psi}^*)\over \langle {\hat \Psi},{\hat \Psi}\rangle}
\ee
For  this probability density to be   non-negative,  we get the condition, 
\be
	p(l, \phi)\geq 0\qquad  \forall  \,\,l>0, \phi>0\label{pbdenl}.\ee
	
	Finiteness of the norm requires that $ \langle \hat\Psi, \hat\Psi\rangle$ is finite and non-vanishing. 
	Given  a finite norm one can adjust the sign in eq.(\ref{pbden}) to make it  positive and then  examine  with this choice if $p(l,\phi)$ is positive for all $ l, \phi$.  

Let us now turn to studying these conditions in more detail. 

\subsection{Conservation of Norm}
\label{consnorm}

 Using the fact that $\hat{\Psi}$ satisfies WDW equation, we find
 \begin{align}
 	\del_{\phi}\langle \hat{\Psi},\hat{\Psi}\rangle &=i\int dl\,( \del_{\phi}\hat{\Psi}^*{\del_{l}}\hat{\Psi}+\hat{\Psi}^*{\del_{\phi}\del_{l}}\hat{\Psi}- \del_{\phi}\del_{l}\hat{\Psi}^*\hat{\Psi}-\del_{l}\hat{\Psi}^*\,{\del_{\phi}}\hat{\Psi} )\nonumber\\
 	&=i\int dl ( \del_{\phi}\hat{\Psi}^*{\del_{l}}\hat{\Psi}+{{\phi}{l}}\hat{\Psi}^*\hat{\Psi}- {\phi}{l}\hat{\Psi}^*\hat{\Psi}-\del_{l}\hat{\Psi}^*{\del_{\phi}}\hat{\Psi} )\nonumber\\
 	&=i ( \hat{\Psi}\del_{\phi}\hat{\Psi}^*{}\,-\hat{\Psi}^*{\del_{\phi}}\hat{\Psi} )\bigg\vert_{l=0}^{\infty}\label{glucond}
 \end{align}
 where in obtaining the first line from the second we used the fact that $\hat{\Psi}$ satisfies the WDW equation and the third line is obtained from the second by an integration by parts and using WDW equation again. Thus the requirement that the norm be conserved translates to the condition, denoted $\mathcal{C}_N$, that 
 \begin{align}
 \mathcal{C}_N\equiv	i ( \hat{\Psi}\del_{\phi}\hat{\Psi}^*{}\,-\hat{\Psi}^*{\del_{\phi}}\hat{\Psi} )=0 \quad \text{at}\,\,{ l=0,\infty}\label{normccond}
 \end{align}

More generally it is enough to ensure that 
\be
\label{mgennc}
i ( \hat{\Psi}\del_{\phi}\hat{\Psi}^*{}\,-\hat{\Psi}^*{\del_{\phi}}\hat{\Psi} )|_{l\rightarrow\infty} - i ( \hat{\Psi}\del_{\phi}\hat{\Psi}^*{}\,-\hat{\Psi}^*{\del_{\phi}}\hat{\Psi} )|_{l\rightarrow 0}=0
\ee

In the discussion below we study the case where eq.(\ref{normccond}) vanishes separately at $l=0,\infty$. 
Note that we need  these conditions to be met at all $\phi>0$. The condition in eq.\eqref{mgennc} is the statement that the probability flux does not leak out from the two ends, $l=0,\infty$. 

Starting from the general form eq.(\ref{genexwf}), with ${\hat \Psi}$ related to $\Psi$ by eq.(\ref{formsa}), and with  the coefficients  satisfying eq.(\ref{conr2}), one finds that eq.(\ref{normccond})   imposes a very non-trivial condition on the wave function. To make the condition more tractable let us consider the case where the coefficients $\rho, {\tilde \rho}, \rho_{1},\rho_2$ have 
compact support in the range $|M|<M_0$. Then  we get, for $l\rightarrow 0$,
\be
\label{nearl}
{\hat \Psi}  \simeq  {1\over l} \int _{-M_0}^{\phi^2}(\rho+{\tilde \rho})  +\frac{1}{l}\int_{\phi^2}^{M_0} (\rho_1+\rho_2)
 -i \int_{-M_0}^{\phi^2}( \rho-{\tilde \rho}) \sqrt{\phi^2-M} -\int_{\phi^2}^{M_0} (\rho_1- \rho_2)\sqrt{M-\phi^2} 
\ee
More generally, when the support is not compact,  such an expansion near $l=0$ {can also be done  as long as the integrals over $M$ are sufficiently convergent.}

In the case with compact support eq.(\ref{nearl}) is actually valid for $\phi^2<M_0$. Beyond that, when $\phi^2>M_0$ we get, 
\be
\label{nearl2}
{\hat \Psi}  \simeq  {1\over l} \int _{-M_0}^{M_0}(\rho+{\tilde \rho})  dM
 -i \int_{-M_0}^{M_0}( \rho-{\tilde \rho}) \sqrt{\phi^2-M} dM
\ee

Requiring $\mathcal{C}_N$ to vanish, as $l \rightarrow 0$, gives from eq.(\ref{nearl})  the condition at leading order in $l$, 
\be
\label{condlo}
{\phi\over l} \left[\left( \int_{-M_0}^{M_0}( \rho^*+{\tilde \rho}^*) \right)  \left(-i \int_{-M_0}^{\phi^2} {\rho-{\tilde \rho}\over\sqrt{\phi^2-M}}+\int_{\phi^2}^{M_0}{\rho_1-\rho_2\over \sqrt{M-\phi^2}} \right)- c.c \right]=0.
\ee
and for $\phi^2>M_0$,
\be
\label{condloo}
{\phi\over l} \left[\left( \int_{-M_0}^{M_0}( \rho^*+{\tilde \rho}^*) \right)  \left(-i \int_{-M_0}^{M_0} {\rho-{\tilde \rho}\over\sqrt{\phi^2-M}} \right)- c.c \right]=0.
\ee
One way to meet this restrictive condition, and avoid the $1/l$  divergence as $l\rightarrow 0 $, is to take $\rho, {\tilde \rho}$ to meet the condition,
\be
\label{condrr}
\int_{-M_0}^{M_0} (\rho+{\tilde \rho}) dM=0.
\ee

Once this condition is imposed the wave function takes the form, to leading order as $l\rightarrow 0$, for $\phi^2<M_0$,
\be
\label{lof}
{\hat \Psi}\simeq -i \int_{-M_0}^{\phi^2} dM\,(\rho-{\tilde \rho} ) \sqrt{\phi^2-M}  -\int_{\phi^2}^{M_0}dM\,(\rho_1-\rho_2)\sqrt{M-\phi^2},
\ee
and the condition that $\mathcal{C}_N$ vanishes becomes, 
\begin{align}
\label{condva}
&\bigg(i\int_{-M_0}^{\phi^2}(\rho-{\tilde \rho})^*\sqrt{\phi^2-M}\int_{\phi^2}^{M_0}{\rho_1-\rho_2\over\sqrt{M-\phi^2}}
+i\int_{\phi^2}^{M_0}(\rho_1-\rho_2)^*\sqrt{M-\phi^2} 
\int_{-M_0}^{\phi^2}{\rho-{\tilde \rho}\over \sqrt{\phi^2-M}}\nonumber\\
&+\int_{-M_0}^{\phi^2}(\rho-{\tilde \rho})^{*}\sqrt{\phi^2-M} \int_{-M_0}^{\phi^2}{\rho-{\tilde \rho}\over \sqrt{\phi^2-M}} 
- \int_{\phi^2}^{M_0}(\rho_1-\rho_2)^*\sqrt{M-\phi^2}\int_{\phi^2}^{M_0}{\rho_1-\rho_2\over \sqrt{M-\phi^2}}\bigg)-c.c= 0
\end{align}
where the complex conjugation refers to taking the complex conjugate of all the terms which have been written down. 
For $\phi^2>M_0$ this  becomes, 
\begin{align}
\label{condvaa}
&\bigg(\int_{-M_0}^{M_0}(\rho-{\tilde \rho})^{*}\sqrt{\phi^2-M} \int_{-M_0}^{M_0}{\rho-{\tilde \rho}\over \sqrt{\phi^2-M}} \bigg)
-c.c= 0
\end{align}
The reader will note that these too are   very non-trivial condition to satisfy for all $\phi$. There are some straightforward ways to meet them, though. 
In this subsection we will discuss three of them. 

\begin{enumerate}
\item
	We  take $\rho$ to have compact support only for negative argument and all other coefficient functions to vanish, i.e., 
	\begin{align}
		&\rho\neq 0\qquad  \text{for \,\,\, }M<0, \nonumber\\
		&\tilde{\rho}=\rho_1=\rho_2=0\label{rhogaus}
	\end{align}
	The solution then only has an expanding branch.
	In addition we take $\rho$ to be real and take its integral to { satisfy eq.(\ref{condrr})}. 
	Firstly, we see that  eq.\eqref{condrho} is automatically satisfied since $\rho$ vanishes for positive argument.  Furthermore, we see from eq.\eqref{condva} that the only non-zero term in this case is the first term in the second line and its complex conjugate. Since $\rho$ is real eq.(\ref{condva}) and also eq.(\ref{condvaa})  are  met.
	Thus the norm is conserved. 
	
	Similarly we can exchange ${\rho}$ with ${\tilde \rho}$, now only taking ${\tilde \rho}$ to be non vanishing and construct an example where only the contracting branch is present, and the norm is conserved.

	\item
	It is easy to see from the condition eq.\eqref{normccond} that the conservation of the norm holds when $\Psi$ is real upto a complex phase independent of $\phi$, i.e,
		\begin{equation}
			\Psi = e^{i \alpha} \Psi^*
		\end{equation}
		where $\alpha$ can be a function of $l$ but should be independent of $\phi$ for the conservation of norm to hold.
		
		For a wavefunction of the form given in eq.\eqref{genexwf}, this can be achieved by choosing the coefficient functions as follows
\begin{align}
	\tilde{\rho}&=e^{i\alpha}\rho^*\nonumber\\
	\rho_1&=\rho+\tilde{\rho}=\rho+e^{i\alpha}\rho^*\nonumber\\
	\rho_2&=0\label{rhospsirel}
\end{align}	
With the above choice of $\rho,\tilde{\rho},\rho_1,\rho_2$, the wavefunction can be explicitly written as,
\begin{align}
	\hat{\Psi}=\frac{2e^{\frac{i\alpha}{2}}}{l}\left( \int_{-\infty}^{\phi^2} dM\Re(\rho e^{-\frac{i\alpha}{2}-il\sqrt{\phi^2-M}})+\int_{\phi^2}^{\infty} dM\Re(\rho e^{-\frac{i\alpha}{2}})e^{-l\sqrt{M-\phi^2}} \right)\label{wfuptophase}
\end{align}
As mentioned, the wavefunction is real upto the phase $e^{i\alpha\over 2}$.

A few important special cases of the above are worth mentioning. 
\begin{enumerate}
	\item The wavefunction is real, i.e., $\alpha=0$. In this case we have
		\begin{align}
		\rho &= \tilde{\rho}^* \, ,\forall M\nonumber\\
		\rho_1 (M) &= 2 \Re(\rho(M)) \, , \,M>0
	\end{align}
	Actually in this case the norm just vanishes identically since $\Psi$ is real, 
	\be
	\label{condrepsi}
	\hat{\Psi}=\hat{\Psi}^*
	\ee
	and  time reversal invariant. 
	
	\item The wavefunction is purely imaginary, i.e., $\alpha=\pi$. This would correspond to the case
	\begin{align}
	\label{condccr}
	{\tilde \rho}&=-\rho^*\nonumber\\
\rho_1&=\rho-\rho^*
	\end{align}
	Then we see that again all the required  conditions eq.(\ref{condrho}), eq.(\ref{condva}), eq.(\ref{condvaa}) are  met. 
	In this case the wave function is imaginary and satisfies the condition
	\be
	\label{condpsr}
		\hat{\Psi}=-\hat{\Psi}^*
	\ee
	and again, like in the previous example the norm vanishes identically
\end{enumerate}

It has to emphasized that when  $\alpha$ is a function of $l$, the norm will not vanish.  At large values of $\phi$, $\phi^2>M_0$ (for cases where $\rho$ has compact support in range $[-M_0, M_0]$) only the oscillatory terms remain in the wave function, and  one can show that the 
interference term between the expanding and the contracting branches vanishes. This happens for the probability density $p(l,\phi)$ which is integrated to get the norm. 
Since the interference terms vanish one can consider the expanding and oscillatory branches independently, and define a norm independently for both branches. 
This situation is analogous to the one studied in section \ref{positive} below.

 \item  Yet another way to satisfy the condition of conservation of norm is to require that $\hat{\Psi}$ vanishes as $l\rightarrow 0$. Assuming that $\hat{\Psi}$ also decays sufficiently fastly as $l\rightarrow\infty$, it can easily be seen that if $\hat{\Psi}\rightarrow 0$ as $l\rightarrow 0$, eq.\eqref{consnorm} can be satisfied. To ensure the decay of $\hat{\Psi}$ at $l=\infty$, we set $\rho_2=0$. From the expression for $\hat{\Psi}$ near $l\rightarrow 0$ in eq.\eqref{lof}, we find the requirement that it vanishes leads to
	\begin{equation}
		\rho = \tilde{\rho}, \,\,\rho_1=0=\rho_2\label{psilzer}
	\end{equation}
	This essentially means that the coefficient functions have support only over negative values of $M$. The wavefunction, with eq.\eqref{psilzer} reads
	\begin{equation}
		\hat{\Psi} = \frac{2}{l} \int_{-M_1} ^{-M_2} \rho(M) \cos{( l \sqrt{\phi^2-M)}}\label{psihcom}
	\end{equation}
	for some $M_1,M_2\geq 0$. The function $\rho(M)$ is an arbitrary complex function of $M$.  Restricting $\rho(M)$ to be a real function of $M$, upto an overall phase independent of $M$,  leads to a wavefunction which has a vanishing norm.\footnote{We thank Shiraz Minwalla for pointing out this possibility to us.}
	
	\end{enumerate}

 To complete our discussion of norm conservation we are still left with the analysis of the condition in eq.\eqref{normccond} for $l\rightarrow\infty$. 
  
  In cases where $\rho_2$ is absent and where the coefficient functions $\rho,{\tilde \rho}, \rho_1$ all have compact support, one can argue that ${\cal C}_N$ vanishes as $l\rightarrow \infty$. At   large $l$ we 
 see, after noting  ${\hat \Psi}={\Psi\over l}$, that ${\hat \Psi}$  will decay at least as fast as $1/l$, as a result  the contribution to  $\mathcal{C}_N$ from $l\rightarrow \infty$ vanishes.

 In passing we note that if $\rho_2$ were present it would dominate the behaviour as $l \rightarrow \infty$ and the dominant contribution to $\mathcal{C}_N$ comes from the exponentially growing branch, leading to the condition,
 \be
 \label{condlargel}
 \int_{\phi^2}^{M_0} \rho_2^*\, e^{l \sqrt{M-\phi^2}} \int_{\phi^2}^{M_0}\rho_2\, {e^{l \sqrt{M-\phi^2}}\over \sqrt{M-\phi^2}}-c.c.=0
 \ee
 This condition can be met by taking $\rho_2$ to be real. One would have to also worry about subleading terms though in such cases. 
 
 In summary, although the conservation of the norm imposes very stringent conditions on the allowed wave functions these conditions can be met as we have discussed above.

 A few comments are now in order. First note that we have not analysed the general case where the coefficient functions do not have compact support. 
 One such example is the Hartle-Hawking wave function which in fact does not meet the condition eq.(\ref{normccond}); in fact it does not have a finite norm at all, as we will see below.

Secondly,  instead of requiring that one has a conserved norm for all values of $\phi$, one can ask whether there are states in which this property holds true only for sufficiently large  $\phi$. It is easy to construct examples of this type. For example take a wave function where ${\tilde \rho}=0, \rho_2=0$,
and $\rho$ has compact support for  $M$ in the range  $|M|\le M_0$. 
We also take $\rho$ to be real. From eq.(\ref{conr2})we then have 
\be
\label{condnna}
\rho_1=\rho, \forall 0<M<M_0 . 
\ee
Finally we also assume that eq.(\ref{condrr})  is met. 
The condition that $\mathcal{C}_N$ vanishes at $l=0$, eq.(\ref{normccond})  then leads to 
\be
\label{condlate}
\int_{-M_0}^{\phi^2}\rho\sqrt{\phi^2-M}\int_{\phi^2}^{M_0}{\rho\over\sqrt{M-\phi^2}}
+\int_{\phi^2}^{M_0}\rho\sqrt{M-\phi^2} 
\int_{-M_0}^{\phi^2}{\rho\over \sqrt{\phi^2-M}}=0
\ee
Generically this condition will not be met.
However, once $\phi^2$ becomes larger than $M_0$ the LHS trivially vanishes and the condition is met. 
Note in this case the  constraint coming from $l\rightarrow \infty$ for norm conservation is also  met as discussed above. Thus we see in this example that a conserved norm will arise, but only after $\phi^2>M_0$. 
For earlier times the norm is not conserved and a key property for a satisfactory notion of physical time is therefore  not being met. A related discussion along these lines is presented in subsection \ref{gausswf1}, see also  the appendix section \ref{normconss}, for a  $\rho$ with noncompact support;  there we will see that the norm is conserved only approximately at large $\phi$.

Finally,  it is also worth examining the restrictions imposed by the conservation of the norm in the representation, eq.(\ref{exppsi}). $l\rightarrow 0$ corresponds to $u\rightarrow 0$, i.e. to the horizon ${\cal H}_1$. One way to meet ${\cal C}_N\rightarrow 0$ at $l\rightarrow 0$ is then to simply set all coefficients $a(k), b(k)$,  with $k<0$, to vanish.
Imposing the condition that ${\cal C}_N\rightarrow 0$ as $l\rightarrow \infty$ on the remaining coefficients $a(k), b(k), k>0$ would then lead to a conserved norm.

  \subsection{Finiteness of Norm}
\label{finohim}

The KG norm introduced in eq.\eqref{kgnorm} need not be  finite.

Two potential regions from where divergences can arise in the integral in eq.(\ref{kgnorm}) are $l\rightarrow 0, \infty$. In some cases these can be ruled out in a straightforward manner. An example is the case we discussed in subsection \ref{gensols} where the coefficient functions in eq.(\ref{genexwf}) have compact support and  the norm is conserved. We satisfy this  latter condition by taking ${\cal C}_N$ to vanish separately as $l\rightarrow 0$ and $l \rightarrow \infty$. For  $l\rightarrow 0$ we  impose the condition, eq.(\ref{condrr}); the behaviour of ${\hat \Psi}$ near $l\rightarrow 0$ is then given by eq.(\ref{lof}), showing that it is finite. For $l\rightarrow \infty$, we ensure that ${\cal C}_N$ vanishes by ensuring $\hat{\Psi}$ decays sufficiently fast. 
Once we have made sure that the norm is conserved we can go to large values of $\phi$ to  evaluate it. 
 At sufficiently large $\phi$ only the oscillatory branches are present in eq.(\ref{genexwf}) and  as long as $\rho, {\tilde \rho}$ are bounded functions one can argue that the norm is finite. 
If it is non-vanishing, by choosing the overall sign in front of eq.(\ref{kgnorm})  we can then make the norm   positive.

When the coefficient functions do not have compact support, or are  unbounded,  one needs to be more careful. For instance, a divergence can occur at finite $l$ due to $\rho(M)$ diverging when $M\rightarrow \infty$. An  example of this is provided by the Hartle-Hawking wave function which we discuss in more detail below.

 It is also worth discussing the  wave function in the representation eq.(\ref{exppsi}). As was mentioned towards the end of subsection (\ref{consnorm}) one way to ensure norm conservation is to take all the $k<0$ modes in eq.(\ref{exppsi}) to vanish, and impose the required conditions at $l\rightarrow \infty$ on the $k>0$ modes so that ${\cal C}_N$ vanishes at $l \rightarrow \infty$. Having done so, we can evaluate the norm as $\phi\rightarrow 0$ by examining the behaviour of ${\hat \Psi}$ at the horizon ${\cal H}_2$. It is easy to see that near ${\cal H}_2$ only the right movers (functions of $u$ appear)
 \be
 \label{valn}
 {\hat \Psi}=\int_{k>0} dk [a(k) C_k e^{-i k \ln(u)}+ b(k) C_k^{*} e^{i k \ln(u)}]
 \ee
 { where $C_k$ are some complex constants.}
 As a result the norm is given by 
 \be
 \label{valnab}
 \langle {\hat \Psi}|{\hat \Psi}\rangle =4 \int_{k>0} dk  (|a(k)|^2 -|b(k)|^2)k 
 \ee
 As long as this integral converges the norm is finite and can be made positive by adjusting the overall sign as was mentioned above.

\subsection{Positivity of Probability and Expectation Values of Operators}
  \label{positive}
  It is also worth examining when the probability density $p(l, \phi)$, defined in eq.(\ref{pbden}), for an appropriate choice of sign in front  of the RHS, is positive.  This is a more restrictive condition, since if $p(l,\phi)$ is positive the norm will also be positive, as long as it is finite. 
In general  it is quite non-trivial to decide whether $p(l, \phi)$ can be made  positive everywhere, even on one fixed $\phi$ slice.
 As an example consider the case 1) above where ${\hat \Psi}$ is given by 
 \be
 \label{negswfa}
 {\hat \Psi}={1\over l}\int dM \rho(M) e^{-il\sqrt{\phi^2-M}}
 \ee
  (with $\rho$ being real, having compact support in the region $M<0,\, -M_2<M<-M_1$, and meeting eq.(\ref{condrr})). We get, eq.(\ref{pbden}) 
 \be
 \label{valp}
 p(l,\phi)=\pm   {{\cal N}^{2}\over l^2} \left( \int dM \rho(M) e^{i l \sqrt{\phi^2-M}}  \int dM'\rho(M') \sqrt{\phi^2-M'}e^{-i l \sqrt{\phi^2-M'}} + c. c. \right)
  \ee
  with the RHS being a non-trivial integral. 
   The large $\phi$ behaviour simplifies somewhat but is still non-trivial. In particular  an easy error one can make, when $\phi^2\gg M_2$,  which is worth warning the reader about, is to expand the exponent in inverse powers of $\phi$, e.g.
   \be
   \label{exep}
   e^{-i l \sqrt{\phi^2-M}}\simeq e^{-i l \phi } e^{ {il M \over 2 \phi}}.
   \ee
   This approximation is not valid at small enough values of $l$ since the terms we have ignored   give linear corrections in $l$ which are more important than the quadratic term in $l$ arising from the second exponent  $e^{ {il M \over 2 \phi}}$, at sufficiently small $l$  when {$l<\phi^{-1}$}. 
   An example where the norm  is positive but $p(l,\phi)$ takes both positive and negative values as a function of $l$ for large $\phi$, is given in \ref{condest}. 
   
   Next we discuss     how to compute  expectation values of various operators. 
 Consistent with our interpretation of eq.(\ref{pbden}) giving the probability density $p(l,\phi)$, we take the expectation value of various moments $\langle l^n\rangle $ of the length to be given by 
  \be
 \label{lenmom}
 \langle l^n \rangle=\int dl \,l^n\, p(l,\phi)
 \ee
 where the integral is at a constant value of $\phi$. 
  Using the definition of $p(l,\phi)$, eq.\eqref{pbden}, it is easy to see that $\langle l^n\rangle$ is real. 
  Here we have set the normalisation ${\cal N}$, eq.(\ref{defn}),  to be unity. 
  
  For the conjugate momentum we take 
  \be
  \label{conjmom}
  \langle \pi_l\rangle = -{1\over 2}\int dl {\hat \Psi}^* (-i  (\overrightarrow{\partial}-\overleftarrow{\partial})) (-i  (\overrightarrow{\partial}-\overleftarrow{\partial})) {\hat \Psi},
  \ee
  more generally, 
  \be
  \label{mommom}
  \langle \pi_l^n\rangle =-{1\over 2^n} \int dl {\hat \Psi}^* (-i  (\overrightarrow{\partial}-\overleftarrow{\partial}))^{n+1} {\hat \Psi}.
  \ee
  It is easy to see that these expectation values are also real. We remind the reader that the above expressions are written for normalized wavefunctions. 
  
 It is worth commenting, with the choice of dilaton as the clock, that the complete set of  operators which commute with the clock are $g_1,\pi_{g_1}$ or equivalently $g_1, \hat{M}$ since $\hat{M}$ can be expressed in terms of $\pi_{g_1}$ at any instant $\phi$, eq.\eqref{lconint}. We remind the reader that for constant $\phi$ slices $g_1$ is related to $l$ by eq.\eqref{lg1}. 
   
   Note that even though the norm of a state may be finite, the expectation values of operators, for example $\langle l^n\rangle$, might diverge, for sufficiently large values of $n$. 
   In some of the examples we study below there  will both be cases where the norm is finite  and expectation values of all powers of $l$ are finite, and also cases  where only some of the moments are finite. 
 
\subsection{Decoherence Between Branches}
   \label{decohere}

   Let us end with a few observations about when different branches in the wave function can stop interfering and decohere. 
   Some examples of this type were already commented on  above in cases 2) and 3) of subsection \ref{consnorm}. Here we will consider some other instances and elaborate on them more. 
    
   Take the case where one has only the oscillatory components in the wavefunction, eq.(\ref{genexwf}) , with $\rho_1,\rho_2=0$
   and with  $\rho, {\tilde \rho}$ having  support only over negative $M$, $\abs{M}\leq M_0$, so that the wave function is oscillatory for all $\phi$. We also take these coefficients to meet   eq.(\ref{condrr}). 
   Then it is easy to show that $p(l,\phi)$, upto the overall normalisation ($\pm {\cal N}^2$),  takes the value
   \begin{eqnarray}
   \label{plosc}
   p(l,\phi) = & {1\over l^2} \int dMdM' \{(\sqrt{\phi^2-M'}+\sqrt{\phi^2-M}) \rho(M)\rho(M')^*e^{-il\sqrt{\phi^2-M}+il\sqrt{\phi^2-M'}}\nonumber \\
   & -(\sqrt{\phi^2-M'}+\sqrt{\phi^2-M}) \tilde{\rho}(M)\tilde{\rho}(M')^*e^{il\sqrt{\phi^2-M}-il\sqrt{\phi^2-M'}}\nonumber\\
  & - (\sqrt{\phi^2-M'}-\sqrt{\phi^2-M})  {\rho}(M)\tilde{\rho}(M')^*e^{-il\sqrt{\phi^2-M}-il\sqrt{\phi^2-M'}} \nonumber \\
  & + (\sqrt{\phi^2-M'}-\sqrt{\phi^2-M}) \tilde{\rho}(M){\rho}(M')^*e^{il\sqrt{\phi^2-M}+il\sqrt{\phi^2-M'}}
  \} 
   \end{eqnarray}
where the  terms in the last two lines  on the RHS are interference terms between the expanding and contracting branches.
We see that in general these interference terms do not vanish. However if $\rho, {\tilde \rho}$ have compact support\footnote{We have not analysed the more general case with non-compact support  for $M<-M_0$ in full detail.} then for  late time, $\phi^2\gg M_0$ the interference terms approximately vanish. This follows from noting that factors of the type  $\sqrt{\phi^2-M}$, when not in exponents, can be expanded for late enough time, and approximated to be $\phi$, which is independent of $M$. 
 The remaining terms, to good approximation, at late time are thus only the first two on the RHS which do not involve any cross terms between $\rho,{\tilde \rho}$. Hence  at late enough time we see from $p(l,\phi)$ that the expanding and contracting branches decohere and evolve independently to good approximation - as one might expect if a classical limit is being approached. 
 
 If there are non-oscillatory components in the wave function the comments above do not change as long as the coefficient functions have compact support. At late enough time only the oscillatory terms in eq.(\ref{genexwf}) will survive and the argument above shows that the expanding and contracting branches will approximately decohere and stop interfering. 
 
 Once the expanding and contracting branches have stopped interfering we can assign a probability density to each of them given by the first and second term in the first line
 on the RHS of eq. (\ref{plosc}) respectively. It is also worth mentioning that  from the analysis similar to eq.(\ref{condva}) we can see that at these large values of $\phi$ the condition for ${\cal C}_N$ to vanish, at $l\rightarrow 0$,  is approximately met individually by the two branches as well. Exactly satisfying this condition  would have required for example that $\Im(\rho(M) -{\tilde \rho}(M))=0$, however even without such a relation between $\rho,{\tilde \rho}$, at late times approximately ${\cal C}_N(l=0) $ vanishes for both branches individually.  
 
 The two branches make opposite sign contributions to $p(l,\phi)$; taking this into account we can define a probability in the expanding and contracting  branches at late times to be respectively,
 \begin{eqnarray}
 \label{excop}
  p_{\text{exp}}(l,\phi) & \simeq & {\cal N}_{\text{exp}}^2 \  {\phi \over l^2}\int dM \rho^*(M) e^{il\sqrt{\phi^2-M}}\int dM'\rho(M')e^{-il\sqrt{\phi^2-M'}}\\
  p_{\text{cont}}(l,\phi) & \simeq & {\cal N}_{\text{cont}}^2  \ {\phi \over l^2} \int dM {\tilde \rho}^*(M) e^{-il\sqrt{\phi^2-M}} \int dM'{\tilde \rho}(M')e^{il\sqrt{\phi^2-M'}} 
  \end{eqnarray}
  where ${\cal N}_{\text{exp}}, {\cal N}_{\text{cont}}$ are overall normalisations which give the total probability in the expanding and contracting branches to each be unity. 
  We will make some comments about the behaviour of the system  in the presence of matter in appendix \ref{diswmat}. While we have not been able to solve for the general gauge invariant states in these cases it would be  well worth investigating in the future if such a process of decoherence between various branches of the wave function  occurs, and if it is in fact  accentuated by the presence of local degrees of freedom, especially when they are present in large numbers.
   
   Two more comments are  in order. 
   First, note that for the case discussed above, eq.(\ref{plosc}), with  $\rho_1,\rho_2=0$, and compact support for $\rho, {\tilde \rho}$, in the range,  
   $|M|<M_0$, 
   once the interference terms become unimportant the probability density for the expanding branch is given by 
   \be
   \label{proexp}
   p(l,\phi) \simeq {2\phi \over l^2} \bigg\vert\int dM \rho(M)e^{-il \sqrt{\phi^2-M}}\bigg\vert^2
   \ee
   This shows that the probability becomes manifestly positive at sufficiently large values of $\phi$. 
   It also follows that the  norm becomes well approximated by an  $L_2$ norm,
   \be
   \label{nl2}
   \int dl \,p(l,\phi)\simeq 2\phi\int {dl \over l^2} \bigg\vert\int dM \rho(M)e^{-il \sqrt{\phi^2-M}}\bigg\vert^2
   \ee
   Similar comments apply to the contracting branch - with a change in sign. 
   Thus we see, as was mentioned earlier,  that in such cases at sufficiently large values of $\phi$ the KG norm becomes well approximated by an $L_2$ norm. 
   
   Second, in section \ref{gausswf1} we will study coefficient functions which have a Gaussian profile. In some cases we will see that, depending on the parameters determining the coefficient functions, the different branches, expanding and contracting ones, or two expanding ones, decohere and a good classical limit is obtained at late times, while in other cases this does  not happen and interference terms  persist even at late times. 

    \section{Examples}
    \label{exsec}
    In this section we consider various examples which will give us a better understanding of various physical states which arise in the system, and also when the classical limit can be attained. 
    
  \subsection{The Hartle-Hawking Wave Function and A Check}
\label{exmp}
We start in  this subsection with the Hartle-Hawking (HH) wave function. 
This wave function can be obtained exactly by path integral methods,  in the asymptotic dS limit, $\phi, l\gg1$, with $ {\phi\over l}$ held fixed, \cite{Maldacena:2019cbz,Stanford:2017thb,Moitra:2022glw}. 
As discussed in \cite{Maldacena:2019cbz}, in $2$ dimensions, the HH wave function can be computed by using either a contour with a continuation from the Euclidean sphere 
with signature $(2,0)$ or from $-AdS_2$ with signature $(0,2)$. In \cite{Moitra:2021uiv}, working in the second order formalism, it was also shown that all the determinants involved can be continued from one to the other contour without any singularities so  that the two results do indeed agree. 
The path integral in $AdS_2$ was argued to be one-loop exact in the asymptotic AdS limit,\cite{Stanford:2017thb},  it then follows from  the above chain of arguments that the path integral in dS 
and therefore the HH wave function, obtained at  one loop, is in fact exact in the asymptotic dS limit.

The precise nature of this calculation provides  an important consistency  check on the procedure we followed above for canonical quantisation. 
The reader will recall that we  had to find a way around the divergence due to the presence of two functional derivatives in the WDW equation and also choose a prescription for ordering operators, so as   to obtain well defined constraint equations in the quantum theory. The procedure we adopted  to do this determines the set of gauge invariant wave functions we get. We will see that  the HH wave function, obtained using path integral method as described above, satisfy the constraints we  obtained in section \ref{cllimiwf}. This serves as an important check on the quantisation procedure followed above.  

Before proceeding further let us note that some of the main points in this subsection were already discussed in \cite{Maldacena:2019cbz,Verlinde:2020zld,Vilenkin:2021awm}. 
%
The path integral calculation as outlined above is carried out by imposing the ``no-boundary" condition, so that the Euclidean $S^2$ or disk for $-AdS_2$ smoothly shrink to zero.  In the asymptotic region where $\phi, l \rightarrow \infty$,
 \cite{Moitra:2022glw,Maldacena:2019cbz},  (see equation (4.14)  of \cite{Moitra:2022glw})   it is given by 
\be
\label{HHPI1}
\Psi(l, \phi)=
 \abs{\mathcal N} e^{S_0} \left({\phi\over l}\right)^{3/2} 
\left[ e^{i \theta} e^{-{i l \phi }+ { 2i \pi^2 \phi \over   l}}
+ e^{-i \theta} e^{{i l \phi}-{2 i \pi^2 \phi \over  l}}\right]
\ee
(We are working in units where $8 \pi G=1$.)
The first term on the RHS is the expanding branch and the second is the contracting branch.
$\abs{ \mathcal{N}}$ is an overall normalisation, and $S_0$ is defined in eq.(\ref{vals0}). $\theta$ is a phase factor which we will comment on more below. 
In particular, the prefactor 
\be
\label{pf}
\left({\phi\over l}\right)^{3/2}
\ee 
arises from a  one-loop determinant.

In fact, this wave function  agrees with a solution to the constraint equations we have obtained above. 
A general solution for the constraints can be written,  eq.(\ref{mgsh1h2}),  in the form, 
\be
\label{genchh}
\Psi=\int dm e^{m\theta} (A_m H^{(1)}_m(\xi)+B_{m} H^{(2)}(\xi))
\ee
where $\theta, \xi$ are given in eq.(\ref{uvshifts}), eq.(\ref{uvshiftsb}) in terms of two constants $c_1,c_2$. 
Using the value of  the prefactor  eq.(\ref{pf}) we find that only one value, $m=2$, is present  and 
\be
\label{valsc}
c_1=0, \\ c_2=-4 \pi^2. 
\ee
This gives, 
\be
\label{HHPI}
{\Psi}(l,\phi)=\abs{\mathcal {\hat N}} e ^{S_0}  { l \phi^2\over  (l^2-4\pi^2)} \left(e^{i \alpha} H_2^{(2)}(\phi\sqrt{l^2-4\pi^2})+e^{-i\alpha}H_2^{(1)}(\phi\sqrt{l^2-4\pi^2})\right)
\ee
where the two terms  again correspond to the expanding and contracting branches respectively. In obtaining eq.(\ref{HHPI}) have set $8 \pi G=1$ and $\abs{\mathcal{\hat N}}=\abs{\mathcal N} (8 \pi G)^{3/2}$.
Also,  $H_2^{(2)}, H_2^{(1)}$ are the Hankel functions with index $2$ of the second and first types respectively, 
and  $\alpha$ is a phase factor related to the phase $\theta$ by $e^{i \alpha}=-e^{i( \theta-{\pi\over 4})}$.
 The overall phase factor in the wave function is unimportant but the relative phase factor, $e^{2i  \alpha}$, between the expanding and contracting branches has  significance, and its importance  will become clear below. 

Note that the wave function eq.(\ref{HHPI}) is real, this is a general  feature of the HH wave function. Its reality is tied to the HH wavefunction being time reversal invariant. 
\
 
 It is also worth noting that the HH wave function can also be expressed in the form, eq.(\ref{genexwf}).  
   Using the identity\footnote{We  established this identity numerically.}, 
   \be
   \label{inti}
   \int_0^{\phi^2} \sinh(2\pi \sqrt{M})e^{-il\sqrt{\phi^2-M}}+
   \int_{\phi^2}^\infty \sinh(2\pi\sqrt{M}) e^{-l\sqrt{M-\phi^2}}= - {2i  \pi^2 l \phi^2 \over l^2-4\pi^2}H_2^{(2)}(\phi\sqrt{l^2-4 \pi^2})
   \ee
    we see that the wave function in eq.(\ref{HHPI}) can be written as 
   \begin{align}
   \label{HHIntr}
   {\hat \Psi}=\frac{\abs{ \mathcal{\hat N}}}{2\pi^2 l} e^{S_0} 
   [& \int_0^{\phi^2} dM  \left(ie^{i \alpha-i l \sqrt{\phi^2-M}} -ie^{-i\alpha+i l \sqrt{\phi^2-M}} \right)\sinh(2\pi \sqrt{M})\nonumber\\
   &- 2 \sin\alpha \int_{\phi^2}^\infty  dM \sinh(2\pi \sqrt{M}) e^{-l\sqrt{M-\phi^2}}]
   \end{align}
   This  corresponds to taking the coefficient functions in  eq.(\ref{genexwf}), upto an overall multiplicative factor of $\frac{\abs{ \mathcal{\hat N}}}{2\pi^2 } e^{S_0} $ , 
   \begin{align}
   \label{valrhoa}
   &\rho(M)=i e^{i\alpha} \ \sinh(2\pi\sqrt{M}), \quad {\tilde \rho}(M)=- i e^{-i\alpha} \ \sinh(2\pi\sqrt{M})\nonumber\\
   &\rho_1(M)=-2 \sin\alpha \  \sinh(2\pi\sqrt{M}),\quad \rho_2(M)=0,
   \end{align}
   which meets the condition,
   eq.(\ref{conr2}). 

We also note that the value  obtained in \cite{Moitra:2022glw}, following the  conventions used therein  for regulating determinants, etc, gives $e^{i \alpha}=\pm i$. This corresponds to the coefficient functions, eq.(\ref{valrhoa}),  being proportional to \cite{Moitra:2022glw}
\be
\label{phase1}
\rho={\tilde \rho} ={1\over 2} \rho_1\propto \sinh (2\pi \sqrt{M})
\ee 

The coefficient functions only have support over positive values of $M$. From section \ref{cllimiwf}, eq.\eqref{dsmilne} and \eqref{valMa},  we see that classically this corresponds to summing over black hole geometries with positive $m$. The values of the coefficient functions, eq.(\ref{phase1}) in fact suggestively equal, upto a proportionality constant,  the density of states in $AdS_2$; in the region,  $M \gg 1$, this arises due  to semi-classical $AdS_2$ black hole geometries. 

{\sl \underline {\textbf{A Divergence at the Turning Point:}}}

 The wave function in eq.(\ref{HHPI})  can be rewritten in terms of the Bessel  and Neumann functions,  $J_2, N_2$, as 
 \be
 \label{HHre}
 {\hat \Psi}(l,\phi)= 2 {\abs{\mathcal{\hat N}}} e ^{S_0}    { \phi^2\over  (l^2-4\pi^2)} \left[ \cos\alpha \,J_2(\phi\sqrt{l^2-4\pi^2}) + \sin\alpha\, N_2(\phi\sqrt{l^2-4\pi^2})\right]
 \ee
 If $\alpha \neq 0,\pi$, we see that due to the presence of the $N_2$ term the wave function blows up as we approach the turning point $l\rightarrow 2 \pi$, 
  ${\hat \Psi}\rightarrow { (l-2\pi)^{-2}}$.
  
  From the integral representation, eq.(\ref{genexwf}),  we can also easily understand the source of the divergence at the turning point. 
  It is clear that the integral $\int_{\phi^2}^\infty  dM \sinh(2\pi \sqrt{M}) e^{-l\sqrt{M-\phi^2}}$ will diverge once $l\le 2\pi$. This divergence is absent if $\alpha=0,\pi,$ since then the coefficient of this integral in eq.(\ref{HHIntr}) vanishes. 
  
  Due to this divergence the wave function cannot be continued into the region $l<2\pi$, which is the classically disallowed region, unambiguously and strictly speaking we must take eq.(\ref{HHPI}) to be the form of $\Psi$ only for\footnote{ \cite{Vilenkin:2021awm} has argued that $\Psi$, eq.(\ref{HHPI})  solves the WDW equation with a source term at the turning point.} $l>2\pi$.   One could try and relate the value of $\partial_\phi{\hat \Psi}$ at $l\rightarrow 2\pi^{+}$ and $l \rightarrow 2\pi^-$ using the  continuity conditions that 
  follow from the 
  WDW equation, eq.(\ref{dshamcon}), but
  we see that the second order pole in ${\hat \Psi}$ implies that $\int_{l=2\pi-\epsilon}^{l=2\pi+\epsilon} {\hat \Psi} \, dl $ diverges as $\epsilon \rightarrow 0$, and as a result
  this approach will not work.   
  It is also difficult to carry out the full JT gravity path integral, other than in the asymptotic region, see \cite{Verlinde:2020zld} for some related work.
%
  An analytic continuation of  ${\hat \Psi}$ to $l<2\pi$ in the complex $l$ plane is also not unique, since  the Neumann function, $N_2(x)$, has a logarithmic branch cut at $x=0$.
   

   One might wonder whether the HH wave function is unphysical due to this divergence and should be disallowed. 
   This might be a premature conclusion though. The norm for the wave function eq.(\ref{HHPI}) vanishes, since it is real. However, 
   instead of considering  both the expanding and contracting branches, let us consider a wave function which say has only the expanding branch. This is analogous to the Vilenkin wave function in the classically allowed region, for the model  studied in \cite{Vilenkin:1986cy}. In this case the wave function takes the form,
    \be
    \label{expwf}
    {\hat \Psi}\propto e^{S_0} {\phi^2\over (l^2-4\pi^2)} H_2^{(2)}(\phi\sqrt{l^2-4\pi^2})
    \ee
    This wave function still  diverges at the turning point as ${ (l-2\pi)^{-2}}$. 
    In the norm eq.(\ref{kgnorm}), for this wave function, the leading divergence which goes like ${(l-2\pi)^{-5}}$ vanishes due to cancellations, but a sub-leading divergence going like ${ (l-2\pi)^{-3}}$ is still present. 
    A divergence in the norm for this case,  is also obtained from the path integral where it arises due to the presence of conformal killing vectors, as discussed in \cite{Mahajan:2021nsd}. This suggests that once matter is added the divergence will  be removed, as happens  in the string world sheet path integral, and finite results can  be obtained for correlation functions. 
   In the presence of matter one could therefore hope to  extract physically sensible quantities from the HH wave function, or at least from the individual expanding and contracting branches. We leave a more detailed study of this issue for the future.
   
   Let us end with two   comments. 
    If the phase $\alpha=0$,  the $N_2$ term in eq.(\ref{HHre}) is absent and ${\hat \Psi}$ is proportional to $J_2$, as has been noted in \cite{Verlinde:2020zld,Vilenkin:2021awm}. In this case the behaviour near the turning point is smooth and on continuing to  $l<2\pi$ we get that 
   \be
   \label{valpsi}
   {\hat \Psi}(l,\phi)=2 \abs{\mathcal{\hat N}} e ^{S_0}    { \phi^2\over  (l^2-4\pi^2)}  J_2(i \phi\sqrt{-l^2+4\pi^2})
   \ee
   We have set $8 \pi G=1$ in the expressions  above. This factor can be restored by rescaling $\phi\rightarrow {\phi\over 8 \pi G}$. 
   On doing so in  eq.(\ref{valpsi})  one gets that near the origin, $l\rightarrow 0$, in the classical limit, $G\rightarrow 0$,   ${\hat \Psi}$ takes the   form
   \be
   \label{psino}
   {\hat \Psi}\propto e^{S_0}  \phi^{3/2}   e^{\phi \sqrt{4\pi^2-l^2}\over 8 \pi G}
   \ee
   and therefore decays rapidly as one goes away from the origin $l=0$ towards the turning point. 
   This is different from the behaviour of the HH wave function in the mini-superspace model studied in \cite{PhysRevD.28.2960,Vilenkin:1986cy}. In that case the HH wave function 
   grows as one goes from the origin towards the turning point. Here the behaviour is more akin, in fact, to  that of the Vilenkin wave function in the classically disallowed region,  in the model studied in \cite{PhysRevD.28.2960,Vilenkin:1986cy}, as was also noted in \cite{Vilenkin:1986cy}. However note that in the classically allowed region $l>2\pi$ the wave function, eq.(\ref{valpsi}) has both an expanding and contracting branch.

Finally, we have included a discussion of the divergence in  the norm of the Hartle-Hawking state  from the path integral point of view in appendix \ref{nrmhh}.

\subsection{Gaussian Wave-functions}
\label{gausswf1}

\subsubsection{Gaussian Coefficient Function}
\label{gausscf}

A Gaussian form for the coefficient functions is a natural one to consider and we analyse it here. 
More specifically, 
consider a wavefunction of the form 
\be
	\hat{\Psi}={1\over l}  \int_{-\infty}^{\phi^2} dM \,\rho(M)\,\, e^{-il\sqrt{\phi^2-M}} + \frac{1}{l}\int_{\phi^2}^\infty dM \rho_1(M) e^{-l \sqrt{M-\phi^2}} \label{pshrhsh}
\ee
with $\rho(M), \rho_1(M) $ given by 
\begin{eqnarray}
	\rho(M) & = & {2  \sin (M x_0)}\,e^{-\frac{M^2}{2\sigma} }, \ M\in [-\infty,\infty] \\
	\rho_1(M) & = & {2  \sin (M x_0)}\,e^{-\frac{M^2}{2\sigma} }, \ M \in [0,\infty]
	\label{rhomgsa}
\end{eqnarray}
The  other coefficient functions, ${\tilde \rho}, \rho_2$ vanish. 
 Note that $\rho,\rho_1$ meet the condition eq.(\ref{conr2}). 
 We take 
 \be
 \label{parax0}
 x_0\sim \order(1)>0.
 \ee
 (in fact it could have been set to $1$ by a suitable rescaling). 
 
 We will find below that when $\sigma\gg1$, at late times, where $\phi^2\gg \sigma$, the state becomes classical, and its resulting behaviour agrees with  classical solutions of JT gravity at late times, discussed in section \ref{cllimiwfa}. 
 
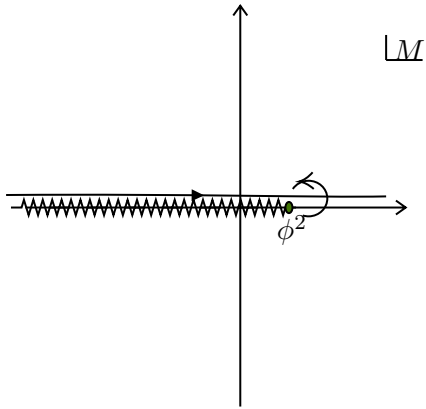
\begin{figure}[h!]
	
	\centering
	\tikzset{every picture/.style={line width=0.75pt}} 
	
	\begin{tikzpicture}[x=0.75pt,y=0.75pt,yscale=-0.7,xscale=0.7]
		
		\draw  (66.76,173) -- (338,173)(220,28) -- (220,316) (331,168) -- (338,173) -- (331,178) (215,35) -- (220,28) -- (225,35)  ;
		\draw   (164,173) -- (171.56,173) -- (173.24,167.5) -- (176.6,178.5) -- (179.96,167.5) -- (183.32,178.5) -- (186.68,167.5) -- (190.04,178.5) -- (193.4,167.5) -- (196.76,178.5) -- (198.44,173) -- (206,173) ;
		\draw   (137.12,173) -- (144.68,173) -- (146.36,167.5) -- (149.72,178.5) -- (153.08,167.5) -- (156.44,178.5) -- (159.8,167.5) -- (163.16,178.5) -- (166.52,167.5) -- (169.88,178.5) -- (171.56,173) -- (179.12,173) ;
		\draw   (110.24,173) -- (117.8,173) -- (119.48,167.5) -- (122.84,178.5) -- (126.2,167.5) -- (129.56,178.5) -- (132.92,167.5) -- (136.28,178.5) -- (139.64,167.5) -- (143,178.5) -- (144.68,173) -- (152.24,173) ;
		\draw   (83.36,173) -- (90.92,173) -- (92.6,167.5) -- (95.96,178.5) -- (99.32,167.5) -- (102.68,178.5) -- (106.04,167.5) -- (109.4,178.5) -- (112.76,167.5) -- (116.12,178.5) -- (117.8,173) -- (125.36,173) ;
		\draw   (190.88,173) -- (198.44,173) -- (200.12,167.5) -- (203.48,178.5) -- (206.84,167.5) -- (210.2,178.5) -- (213.56,167.5) -- (216.92,178.5) -- (220.28,167.5) -- (223.64,178.5) -- (225.32,173) -- (232.88,173) ;
		\draw   (217.76,173) -- (225.32,173) -- (227,167.5) -- (230.36,178.5) -- (233.72,167.5) -- (237.08,178.5) -- (240.44,167.5) -- (243.8,178.5) -- (247.16,167.5) -- (250.52,178.5) -- (252.2,173) -- (259.76,173) ;
		\draw   (56.48,173) -- (64.04,173) -- (65.72,167.5) -- (69.08,178.5) -- (72.44,167.5) -- (75.8,178.5) -- (79.16,167.5) -- (82.52,178.5) -- (85.88,167.5) -- (89.24,178.5) -- (90.92,173) -- (98.48,173) ;
		\draw  [fill={rgb, 255:red, 65; green, 117; blue, 5 }  ,fill opacity=1 ] (252.2,173) .. controls (252.2,175.21) and (253.32,177) .. (254.7,177) .. controls (256.08,177) and (257.2,175.21) .. (257.2,173) .. controls (257.2,170.79) and (256.08,169) .. (254.7,169) .. controls (253.32,169) and (252.2,170.79) .. (252.2,173) -- cycle ;
		\draw    (53,164) .. controls (185,163) and (275,166) .. (324,165) ;
		\draw [shift={(194.25,164.24)}, rotate = 180.53] [fill={rgb, 255:red, 0; green, 0; blue, 0 }  ][line width=0.08]  [draw opacity=0] (8.93,-4.29) -- (0,0) -- (8.93,4.29) -- cycle    ;
		\draw    (324,67) -- (350,67) ;
		\draw    (324,67) -- (324,49) ;
		\draw  [draw opacity=0] (257.46,159.67) .. controls (259.78,156.2) and (263.82,153.86) .. (268.45,153.77) .. controls (275.82,153.62) and (281.91,159.22) .. (282.06,166.28) .. controls (282.2,173.34) and (276.33,179.18) .. (268.96,179.33) .. controls (267.42,179.36) and (265.94,179.14) .. (264.55,178.71) -- (268.7,166.55) -- cycle ; \draw   (257.46,159.67) .. controls (259.78,156.2) and (263.82,153.86) .. (268.45,153.77) .. controls (275.82,153.62) and (281.91,159.22) .. (282.06,166.28) .. controls (282.2,173.34) and (276.33,179.18) .. (268.96,179.33) .. controls (267.42,179.36) and (265.94,179.14) .. (264.55,178.71) ;  
		\draw   (272.39,159.65) .. controls (268.85,156.55) and (265.38,154.78) .. (262.01,154.33) .. controls (265.29,153.44) and (268.49,151.22) .. (271.59,147.68) ;
		
		\draw (327,50.4) node [anchor=north west][inner sep=0.75pt]    {$M$};
		\draw (244,176.4) node [anchor=north west][inner sep=0.75pt]    {$\phi ^{2}$};

	\end{tikzpicture}
\caption{Integration contour for M}
\label{mbranch}
\end{figure}

 Note that the choice of coefficient functions above for the wave function can be written concisely in the form
 \be
 \label{psi2g}
 \hat{\Psi}={1\over l}  \int_{-\infty}^{\infty} dM \,\rho(M)\,\, e^{-il\sqrt{\phi^2-M}} 
 \ee
 with the function $\sqrt{\phi^2-M}$ being defined in the complex $M$ plane as follows. 
 We take
 \be
 \label{defa}
 \sqrt{\phi^2-M}\equiv -i \sqrt{M-\phi^2}
 \ee
 and $\sqrt{M-\phi^2}$ in turn to be defined as 
 \be
 \label{defs}
 \sqrt{M-\phi^2}=\sqrt{|M-\phi^2|} e^{i \theta\over 2}
 \ee
 where  $\theta$,  the phase of $M-\phi^2$, takes   values $\theta\in [-\pi,\pi]$, see Fig.\ref{mbranch}.
 With this definition a branch cut runs $[-\infty,\phi^2]$ along the $x$-axis in the complex M plane. 
 
 The integral in eq.(\ref{psi2g}) is taken to run  just above the $x$-axis, along $M=x+ i \epsilon$. 
 With these definitions,  the wave function is  given by a sum of two terms 
 \be
 \label{psig}
 {\hat \Psi}={1\over 2 i l} \left(\int_{-\infty}^\infty dM e^{{-M^2\over 2 \sigma}+ i M x_0 -il \sqrt{\phi^2-M}} - \int_{-\infty}^\infty dM e^{{-M^2\over 2 \sigma}- i M x_0 -il \sqrt{\phi^2-M}}\right)
 \ee
 
 We can carry out this integral by a saddle point approximation. 
 The first term
 \be
 \label{psit1g}
 {\hat \Psi}_1={1\over 2 i l} \int_{-\infty}^\infty dM e^{{-M^2\over 2 \sigma}+ i M x_0 -il \sqrt{\phi^2-M}}
 \ee
 has a saddle point at 
 \be
 \label{spt1}
 M_0=i  \sigma \left( x_0+ {l\over 2 \sqrt{\phi^2-M}}\right)
 \ee
 For now let us assume that the integral gets its dominant contribution when 
 \be
 \label{condmaa}
 M\ll \phi^2
 \ee
 The resulting saddle point  is then at 
 \be
 \label{spt1a}
 M_0\simeq i \sigma \left( x_0+ {l \over 2 \phi}\right)
 \ee
 Expanding the integrand in eq.(\ref{psit1g}) to quadratic order near $M_0$ gives, 
 \be
 \label{psivic}
 {\hat \Psi}={1\over 2 i l} e^{-il\phi} e^{M_0^2\over 2\sigma} \int_{-\infty}^\infty dM e^{-{(M-M_0)^2\over 2 \sigma}}
 \ee
 leading to 
 \be
 \label{psi1aag}
 {\hat \Psi}_1= { \sqrt{2 \pi \sigma} \over 2 i l} e^{-il \phi} e^{-{\sigma\over 2} (x_0+ {l \over 2 \phi})^2 }
 \ee
Two conditions ensure that our assumption above, that  the integral eq.(\ref{psit1g}) gets  its dominant contribution when eq.(\ref{condmaa}) is valid, is  correct.
These are:
\be
 \label{condag}
 l \sigma \ll \phi^3
 \ee
 and 
 \be
 \label{condaga}
 \phi\gg \sqrt{\sigma}, \\ \phi\gg \sigma^{1/4}
 \ee
 The  conditions in eq.(\ref{condaga})  arise from requiring that $M_0$ meets eq.(\ref{condmaa}), after noting that $x_0\sim O(1)$, and  due to   the variance of the Gaussian integral in eq.(\ref{psivic}) being
 $\sigma$. 
  (For a discussion with a bigger  range of  validity see the discussion in appendix \ref{sadapp}). 
%
 
 Similarly in the region where eq.(\ref{condag}) is true the second term in eq.(\ref{psig}), ${\hat \Psi}_2$, becomes, when the condition, 
 eq.(\ref{condaga}) is valid, 
 \be
 \label{psi2ag}
 {\hat \Psi}_2=-{ \sqrt{2 \pi \sigma} \over 2 i l} e^{-il\phi} e^{-{\sigma\over 2} (-x_0+ {l \over 2 \phi})^2 }
 \ee
 Adding eq.\eqref{psi1aag},\eqref{psi2ag} we get 
 \be
 \label{fullpsia}
 {\hat \Psi}= { \sqrt{2 \pi \sigma} \over 2 i l}e^{-il\phi} \left(e^{-{\sigma\over 2} (x_0+ {l \over 2 \phi})^2 }- e^{-{\sigma\over 2} (-x_0+ {l \over 2 \phi})^2 }\right)
 \ee
 
 Thus the wave function in this region is the sum of two Gaussians.
 For $l \rightarrow 0$ we get that 
 \be
 \label{smgapsi}
 {\hat \Psi}\simeq i{\sqrt{2\pi\sigma} \sigma x_0\over 2\phi } e^{-{x_0^2\sigma\over 2}}
 \ee
 which is well behaved and non-singular. 
 We also see that ${\cal C}_N$, eq.\eqref{normccond}, which can be obtained from eq.(\ref{smgapsi}), vanishes. As a result  there is no ``leakage" of the probability current as $l\rightarrow 0$. We will see later that this is only an approximate statement; on  going beyond the saddle point approximation, for large values of $\phi$ satisfying eq.\eqref{condag},\eqref{condaga}, the norm is not conserved  but the leakage of probability  is  exponentially small.
 
 In appendix \ref{sadapp} we show that when
 \be
 \label{condfia}
 l \gg {\phi^3\over \sigma}, 
 \ee
 and 
 \be
 \label{condfib}
 \phi\gg \sigma^{1/2}, 
 \ee
 ${\hat \Psi}$ decays rapidly, going like,
 \be
 \label{gla}
 {\hat \Psi}\sim e^{-{c_1\over \sigma} (l\sigma)^{\frac{4}{3}}}
 \ee
 where $c_1$ is a number given by $c_1=\frac{3}{\sqrt[3]{128}}e^{i\pi\over 3}$, see eq.\eqref{sadptval}. 
 (More generally, for eq.(\ref{gla}) to be valid  it is enough that $l \gg {\phi^3\over \sigma}$, and the less restrictive condition,  $l\gg \sqrt{\sigma}$
 holds, see eq.\eqref{lsgsqs} in appendix \ref{sadapp}) 
 
 From eq.(\ref{gla}) we see that ${\cal C}_N\rightarrow 0$, as $l\rightarrow \infty$ as well, and the norm is conserved.   Noting eq.(\ref{condfia}), the value in eq.(\ref{gla}) is seen to be bounded from above as
 \be
 \label{condpsik}
 {\hat \Psi}\lesssim e^{-c_1{\phi^{4}\over \sigma}}
 \ee
 which  is small for 
 \be
 \label{condpa}
 \phi\gg \sigma^{1/4}. 
 \ee
 
 Thus at sufficiently large values of $\phi$, when both eq.(\ref{condfib}) and eq.(\ref{condpa})  are met, 
 the probability density which  is given by
 \be
 \label{pds}
 p(l,\phi)=i( {\hat \Psi}^*\partial_l {\hat \Psi}- {\hat \Psi}\partial_l {\hat \Psi^*})
 \ee
 receives its dominant contribution from the region when $l$ meets the condition, eq.(\ref{condag}), with an exponentially suppressed contribution from the region, 
  $l\ge {\phi^3\over \sigma}$.
 The resulting probability density can be obtained from 
 the wave function given in eq.(\ref{fullpsia}) and   takes the form
 \be
 \label{formps}
 p(l,\phi)= 2\phi \hat       {\Psi}\hat       {\Psi}^*>0
 \ee
  It is easy to see  then that the corresponding norm 
  \be
  \label{norpsi}
  {\cal N}= \int_{0}^\infty dl \,p(l,\phi) 
  \ee
  is independent of $\phi$ and therefore conserved. Let us also note that ${\cal N}$ is  only dependent on $\sigma$, $x_0$
  and given by 
  \begin{align}
  	\mathcal{N}\simeq \frac{\pi^{\frac{3}{2}} \sqrt{\sigma}}{2 x_0^2}\label{norm}
  \end{align} 
  We can also use $p(l,\phi)$ in eq.(\ref{formps}) to calculate the moments $\langle l^n\rangle $ upto good approximation, as long as eq.(\ref{condpa}) is valid. 
  We get that the moments satisfy the relation 
  \be
  \label{opl}
  \langle l^n\rangle  ={d_n \over {\cal N}} \phi^n
  \ee
  with the proportionality  constant  $d_n$ being  only dependent on   $\sigma$ and $x_0$.   In calculating ${\cal N}$ and the coefficients $d_n$ we have to carry out integrals of the type
  \be
  \label{int1}
  I_n=\int_{0}^\infty y^{n-2} dy\left(e^{-{\sigma\over 2} (x_0+ {y \over 2 })^2 }- e^{-{\sigma\over 2} (-x_0+ {y \over 2 })^2 }\right)^2
  \ee
  (the integration variable being ${l \over \phi}$). 
  For $\sigma \gg 1$ we can approximate these by a saddle point.
  Only the second term within brackets will contribute since $l>0$. So we get 
  \be
  \label{inta}
  I_n\simeq \sqrt{\frac{\pi}{\sigma}} 2^{n-1} x_0^{n-2}\left(1+ \order{(e^{-x_0^2 \sigma})}\right)
  \ee
  This gives
  \be
  \label{rata}
  {d_n \over {\cal N}}=(2 x_0)^n \left(1+ \order{(e^{-x_0^2 \sigma})}\right)
  \ee
  The leading term on the RHS agrees with  $\langle l\rangle^n$. So we see that 
  the fluctuations about the mean value $\langle l\rangle ^n$ of the moments  are exponentially suppressed and  are small when $\sigma \gg 1$.
  
  
  For instance, for the second moment we get, 
  \be
  \label{insa}
  {\langle l^2\rangle -\langle l\rangle ^2\over \langle l\rangle ^2}\sim \order{(e^{-x_0^2 \sigma})}.
  \ee

  To summarise, at  late times, where the condition in eq.(\ref{condfib}) is met, and for  the variance $\sigma\gg 1$, the resulting wave function is classical. When $\sigma\lesssim O(1)$  we find that  the state does not become classical even at late times. We also note that when $\sigma\gg 1$, between eq.(\ref{condfib}) and eq.(\ref{condpa}), eq.(\ref{condfib}) is more restrictive and implies eq.(\ref{condpa}). 
  
 
 Let us also compare the behaviour we have found above with classical solutions in JT gravity.
 In the gauge we are working in here, eq.(\ref{nnperg}), the metric of $dS_2$ is 
  \be
  \label{metds2a}
  ds^2=-dt^2 + e^{2t} dx^2,
  \ee
  with the dilaton given by 
  \be
  \label{classdila}
  \phi=A e^t,
  \ee
  where $A$ can take any positive value. 
From eq.(\ref{metds2a})  we see that $l\propto e^t$ and therefore ${l/\phi}$ is a time independent constant. This agrees with what we found for the solutions above, eq.(\ref{fullpsia}), eq.(\ref{opl}), where ${\langle l\rangle \over  \phi}=2 x_0$, with $x_0$ taking positive values.

The behaviour at early times, when $\phi\le \sigma^{1/2}, \sigma^{1/4}$, is more complicated. On general grounds, see discussion around eq.(\ref{condlate}), one can argue that the norm of the wave function is not conserved in this example since $\rho(M)$ does not have compact support. However for large enough values of the dilaton the extent of this non-conservation is small,  with the change in the norm ${\partial {\cal N}\over \partial \phi}$, going  like $e^{-{\phi^4\over \sigma}}$ and being exponentially suppressed for $\phi\gg \sigma^{1/4}$.  More details on this can be found in appendix \ref{normconss}. At early times when $\phi\le \sigma^{1/4}, \sigma^{1/2}$ the change in the norm is  not suppressed.
In a following subsection \ref{shifgaus} we consider another choice  of Gaussian coefficient functions which has support   only for negative values of the argument and show that it gives  rise to a classical domain for a larger range  of $\phi$. The norm of the resulting wavefunction in this case is  conserved for all values of $\phi$.


 \subsubsection{Classical Limit and Decoherence}
 \label{gaudecoh}
 It is worth understanding the classical limit in the example above a little better. For  $\sigma \gg 1$ as discussed above, and at late times with $\phi$ meeting the condition, eq.(\ref{condpa}), ${\hat \Psi}$ can be approximated by eq.(\ref{fullpsia}) receiving its dominant contribution from the   region eq.(\ref{condag}). 
 It is then easy to see that the phase term  $e^{-il\phi}$ is the most rapidly varying part of ${\hat \Psi}$ in this region. As a result, to good approximation
 we will see that the expectation value for the conjugate momentum to $l$, $\pi_l$, also meets its    classical value. 
 
Taking  as our definition of $\pi_l$ to be, eq.(\ref{conjmom}),
 \be
  \label{clap}
  \langle \pi_l\rangle =  -{1\over 2\mathcal{N}} \int \,dl {\hat \Psi}^* (-i  (\overrightarrow{\partial}-\overleftarrow{\partial})) (-i  (\overrightarrow{\partial}-\overleftarrow{\partial})) \Psi,
  \ee
  where the integral is evaluated at a constant value of $\phi$,  
   gives, 
  \be
  \label{clap2}
  \langle \pi_l \rangle\simeq -\phi
  \ee
  since the derivative $\partial_l$ acts on the phase $e^{-i l \phi}$ for ${\hat \Psi}$ and similarly for ${\hat \Psi}^*$. 
From the classical solution, eq.(\ref{metds2a}), eq.(\ref{classdila}) we find that the 
    momentum $\pi_l=-{\dot \phi}$, eq.(\ref{cmominnpz}), eq.(\ref{pil}), is  given by 
  \be
  \label{valmoma}
  \pi_l=-\phi
  \ee
  which agrees with eq.(\ref{clap2}) above. 
  Higher moments, computed using, eq.(\ref{mommom}),   will also be approximately given by 
  \be
  \label{hmp}
  \langle \pi_l ^n\rangle\simeq \langle \pi_l \rangle^n
  \ee
  as is required in the classical limit. 
  
  
  Next, consider the case where the coefficient functions, eq.(\ref{rhomgsa}), are replaced by a sum of two functions, each of similar type:
  \begin{eqnarray}
  \rho^{(a)} (M) & = & {2  \sin (M x^{(a)}_{0})}e^{-\frac{1}{2\sigma^{(a)}} M^2 }, \ M\in [-\infty,\infty], a=1,2 \\
	\rho^{(a)} _{1}(M) & = & {2  \sin (M x^{(a)}_{0})}e^{-\frac{1}{2\sigma^{(a)}} M^2 }, \ M \in [0,\infty], a = 1,2
  \end{eqnarray}
  with 
   \begin{eqnarray}
  \label{finala}
    \rho(M)=\rho^{(1)}(M)+\rho^{(2)}(M)\\
    \rho_1(M)=\rho_1^ {(1)} (M)+\rho_1^{(2)} (M)
    \end{eqnarray}

  Each of the two sets of coefficient  functions, $(\rho^{(a)}, \rho_1^{(a)} )$ would have given rise to a good classical limit as long as $\sigma^{(a)}\gg 1$. 
  However for the two together to correspond to classical dS space  requires that the wave packets, each of the type in eq.\eqref{fullpsia},  do not overlap. This gives rise to the condition,   
  \be
  \label{condclass}
{ |x^{(1)}_0-x_0^{(2)}|\gg {1\over \sqrt{\sigma^{(1)}}}, {1\over \sqrt{\sigma^{(2)}}},}
  \ee
  which  ensures that the two solutions decohere at late times and no interference arises between them.  The resulting wave function can be interpreted as an incoherent sum of two classical dS spacetimes, with  probabilities determined by  the respective values of their  norm, eq.(\ref{norpsi}). 
  
  It is also interesting to consider the case where we have a linear combination of an expanding and contracting branch wave function obtained by taking the coefficient functions $\rho(M), {\tilde \rho}(M)$  of the form, 
  \begin{eqnarray}
  \rho(M) & = & {2  \sin (M x_{0})}e^{-\frac{1}{2\sigma}M^2 }, \ M\in [-\infty,\infty], \\
	{\tilde \rho}(M) & = & {2  \sin (M {\tilde x}_{0})}e^{-\frac{1}{2{\tilde \sigma}} M^2 }, \ M \in [-\infty, \infty]
  \end{eqnarray}
  with $\rho_1=\rho+{\tilde \rho}$, for $M>0$, meeting eq.(\ref{condrho}). Also $\rho_2$ is set to vanish. 
  
  This gives the   wave function to be  given by 
  \begin{align}
  \label{fullwa}
  {\hat \Psi}&= {1\over l}\int_{-\infty}^{\infty} dM \,\Theta(\phi^2-M)\left(\rho(M) e^{-il\sqrt{\phi^2-M}} + {\tilde \rho}(M) e^{il\sqrt{\phi^2-M}}\right) \nonumber\\
  &+\frac{1}{l} \int_{-\infty}^\infty dM\,\Theta(M-\phi^2)\rho_1(M) e^{-l \sqrt{M-\phi^2}}
  \end{align}
  We will also take $x_0,{\tilde x}_0\sim O(1)$ and 
  \be
  \label{condsigtwss}
  \sigma, {\tilde \sigma}\gg 1
  \ee
  
  For $\phi$ being sufficiently big, 
  \be
  \label{condpaas}
  \phi\gg \sigma^{{1\over 2}}, {\tilde \sigma}^{{1\over 2}} , 
  \ee
  we will then get that ${\hat \Psi}$ is given by a sum of two terms, an  expanding branch and a contracting branch. 
  Both have their dominant contributions in the region eq.(\ref{condag}). The expanding branch is given  by  eq.(\ref{fullpsia}), and similarly the contracting branch is given  by a sum of Gaussians,
  \be
  \label{sepsi}
  {\hat \Psi}={\sqrt{2\pi {\tilde \sigma}}\over 2 i l} e^{i l \phi} \left(e^{-{{\tilde \sigma} \over 2} ({\tilde x}_0+ {l \over 2 \phi})^2 }- e^{-{{\tilde \sigma}\over 2} (-{\tilde x}_0+ {l \over 2 \phi})^2 }\right)
  \ee
  
  We see then that  if  the condition, 
  \be
  \label{newcondx}
  (x_0-{\tilde x}_0)^2\gg {1\over \sigma}, {1\over {\tilde \sigma}}, 
  \ee
  is met, 
  the interference between the two branches will be negligible. Each of these branches will  be well described classically, at late times when eq.(\ref{condpaas}) is satisfied. The norm would be defined separately for each branch,  and  given by eq.(\ref{kgnorm}), with a positive overall sign   for the expanding branch and  a negative overall  sign for the contracting branch.   We had discussed  a similar situation of how decoherence between the expanding and contracting branches can arise for cases where the coefficient functions have compact support in section \ref{decohere}. 
  
  On the other hand if eq.(\ref{newcondx}) is not met the expanding and contracting branches will continue to interfere and a classical limit will not arise. 
  
  We end with some  more comments. Once the classical limit is attained the leading variation with $l$ of the wave function comes from the phase factor, 
  $e^{\pm i l \phi}$. This is to be expected on general grounds from the WKB approximation, see appendix \ref{allnfwf}. The KG norm,
  \be
 \label{defna}
 \langle{\hat \Psi}|{\hat \Psi}\rangle=  \pm i \int dl ({\hat \Psi}^*\partial_l {\hat \Psi}-{\hat \Psi} \partial_l {\hat \Psi}^*),
 \ee
with the positive   and negative sign respectively being chosen for the expanding and contracting branches, then  becomes the $L^2$ norm, 
  \be
  \label{l2n}
 \langle{\hat \Psi}|{\hat \Psi}\rangle= 2 \phi \int dl |{\hat \Psi}|^2,
  \ee
  as was seen in the previous subsection.

  Finally, note that the example considered at the start of this subsection  can be easily generalised. The saddle point approximation used in evaluating the wave function above is related  to the method of steepest descent which can be applied more generally. 
  Writing 
  \be
  \label{rew}
  \rho(M)=e^{\xi(M)}
  \ee
  and taking $\rho_1=\rho$ for $M>0$, with ${\tilde \rho}=\rho_2=0$  we have 
  \be
  \label{nexhp}
  {\hat \Psi}={1\over l}\int dM \rho(M) e^{-il\sqrt{\phi^2-M}}
  \ee
  with the contour of integration and the definition of $\sqrt{\phi^2-M}$ being given in Fig(\ref{mbranch}) and eq.(\ref{defa}), eq.(\ref{defs}) respectively. 
  
  A stationary point occurs at $M_0$ with 
  \be
  \label{stap}
  \xi'(M_0)=(-i) {l\over 2\sqrt{\phi^2-M_0}}
  \ee
  Assuming that $|M_0|\ll \phi^2$  we then get that 
  \be
  \label{formp}
  {\hat \Psi}\sim {e^{-il\phi}\over l\sqrt{|\xi''(M_0)+{il\over 4 \phi^3}|} } e^{\xi(M_0) + i{l\over 2\phi}M_0}
  \ee
  Thus in this region the wave function would  vary with  the phase factor $e^{-il\phi}$. As long as this phase   is varying rapidly the norm would be well approximated by 
  eq.(\ref{l2n}) and a classically limit could then arise along the lines of our discussion above for a wide range of functions $\xi(M)$. 
  

   \subsubsection{Shifted Gaussian}
   \label{shifgaus}

  Finally, let us consider the wavefunction which arises when  $\rho(M)$ is given by 
  \begin{align}
  	\rho(M)={2  \sin ((M+M_0) x_0)}e^{-\frac{1}{2\sigma}(M+M_0)^2 } \label{rhomgs}
	   \end{align}
  in the region
  \be
  \label{res}
  -(\Delta + M_0)<M<\Delta-M_0,
  \ee
  and is vanishing outside it. 
  
  We take 
   \begin{align}
   	\Delta-M_0<0\label{dlmo}
   \end{align}
   so that $\rho$ has support only for negative values of $M$, and also take all the other coefficient functions, ${\tilde \rho}, \rho_1, \rho_2$ to vanish. 
This  example falls in the class 1), eq.(\ref{rhogaus}), of subsection \ref{consnorm}. The 
     general arguments of subsection \ref{consnorm} then hold and it  follows that the norm is exactly conserved.
   
   In fact the wavefunction can be evaluated quite explicitly.  It  is given by 
 \begin{align}
 \hat{\Psi}={2\over l}  \int_{-M_0-\Delta}^{-M_0+\Delta} dM \,{  \sin ((M+M_0) x_0)}e^{-\frac{1}{2\sigma}(M+M_0)^2 }\,\, e^{-il\sqrt{\phi^2-M}} \label{pshrhshgf}
 \end{align}
Rewriting it in terms of the variable $\tilde{M},\tilde{\phi}$ 
\begin{align}
	\tilde{M}&=M+M_0,\,\nonumber\\
	\tilde{\phi}^2&=\phi^2+M_0\label{mtphitvar}
\end{align}
we can rewrite the above integral in eq.\eqref{pshrhshgf} as
\begin{align}
	 \hat{\Psi}={2\over l}  \int_{-\Delta}^{\Delta} d\tilde{M} \,{  \sin(\tilde{M} x_0)}e^{-\frac{1}{2\sigma}\tilde{M}^2 }\,\, e^{-il\sqrt{\tilde{\phi}^2-\tilde{M}}} \label{pshgf}
\end{align}

As before, we consider $x_0\sim \order{(1)}$.  In addition we take 
\begin{align}
	\Delta^2\gg \sigma\label{delsign}
\end{align}
The integral in eq.(\ref{pshgf}) for $\tilde{M}$ can then be extended from the range $[-\Delta, \Delta]$ to $[-\infty,\infty]$. The error in doing so will be given by 
\begin{align}
	\hat{\Psi}_{\text{error}}=\frac{2}{l}\int_{-\infty}^{-\Delta}d\tilde{M}{  \sin(\tilde{M} x_0)}e^{-\frac{1}{2\sigma}\tilde{M}^2 }\,\, e^{-il\sqrt{\tilde{\phi}^2-\tilde{M}}}+\frac{2}{l}\int^{\infty}_{\Delta}d\tilde{M}{  \sin(\tilde{M} x_0)}e^{-\frac{1}{2\sigma}\tilde{M}^2 }\,\, e^{-il\sqrt{\tilde{\phi}^2-\tilde{M}}}\label{errorgwf}
\end{align}
This error term is of the order of $e^{-\frac{\Delta^2}{2\sigma}}$ and is small when eq.(\ref{delsign}) is met. 
As a result,    the integral in eq.\eqref{pshgf}  can be approximated to be 
\begin{align}
	\hat{\Psi}\simeq{2\over l}  \int_{-\infty}^{\infty} d\tilde{M} \,{  \sin(\tilde{M} x_0)}e^{-\frac{1}{2\sigma}\tilde{M}^2 }\,\, e^{-il\sqrt{\tilde{\phi}^2-\tilde{M}}} \label{pshgfinf}
\end{align}

 This integral is  of the form  discussed in subsection \ref{gausscf}, in terms of the shifted variables, ${\tilde M}, {\tilde\phi}$. 
 As long as ${\tilde \phi}$ meets the conditions, 
 \be
 \label{condff}
 {\tilde \phi}\gg \sqrt{\sigma}\\, {\tilde \phi}\gg \sigma^{1/4}.
 \ee
  we can conclude  that the wave function will receive its dominant support from the region 
 \be
 \label{ds}
 l\sigma  \le  {\tilde \phi}^3
 \ee
 where it takes the form
 \begin{align}
	\hat{\Psi}= { \sqrt{2 \pi \sigma} \over 2 i l}e^{-il\tilde{\phi}} \left(e^{-{\sigma\over 2} (x_0+ {l \over 2 \tilde{\phi}})^2 }- e^{-{\sigma\over 2} (-x_0+ {l \over 2 \tilde{\phi}})^2 }\right)\label{psiinlpht}
\end{align}
Note that in this case the conditions in eq.(\ref{condff}) can be met for all $\phi$, as long as 
\be
\label{condmv}
M_0\gg\sigma, \sqrt{\sigma}. 
\ee
The wave packet will be tightly  centered around $l=2 |x_0| {\tilde \phi}$ as long as $\sigma \gg 1$, and the classical limit will have  a much wider range of validity  since 
eq.(\ref{delsign}) is also true. 

In section \ref{cllimiwf} we discussed the operator ${\hat M}$ which is a suitably regularised quantum version of the classical variable given in eq.(\ref{pigval}).
Since we are working on slices where $\phi'=0$ it is reasonable to take  the expectation value of ${\hat M}$ to be 
\be
\label{exmh}
\langle {\hat M}\rangle=\phi^2-\langle \pi_l^2\rangle
\ee
where $\langle\pi_l^2\rangle$ is given by eq.(\ref{mommom}). 
For the wavefunction eq.\eqref{psiinlpht} we find to good approximation,  in the classical limit when eq.(\ref{condff}) is  met, that 
\be
\label{bvalm0}
\langle{\hat M}\rangle=-M_0
\ee
This result is in accord with section \ref{cllimiwf} where  we had mentioned that the solutions $e^{\pm i l \sqrt{\phi^2-M}}$  should be interpreted as being eigenstates of 
 ${\hat M}$ with eigenvalue $M$. Eq.(\ref{bvalm0}) is then expected since the coefficient function eq.(\ref{rhomgs})  is a  Gaussian superposition  of various values of $M$ centered around $-M_0$. In appendix \ref{expvalm}, we give more details on the computation of this expectation value for more general coefficient functions. 

We also note that in section \ref{cllimiwfa} where we discussed classical solutions we observed that for $M<0$, corresponding to $m<0,A>0$ of the configuration eq.\eqref{dsmilne}, we have a geometry with no singularity, related to global dS by changing the range of the variable $x$. This dovetails nicely with the fact  that in the quantum theory we find above that the classical description can be valid, when $M_0<0$ , for all values of $\phi\ge 0$, as long as $\sigma \gg 1$ and $\Delta$ meets the condition eq.(\ref{delsign}). 

In fact, quite strikingly, the wavefunction can be extended smoothly to the region $\phi<0$ in this case. 
We had mentioned above that we will take the range of $\phi$ to be positive, eq.(\ref{conddila}), keeping in mind the classical solutions of the system. However we see  here that the quantum wave function of these shifted Gaussian state does not show any breakdown in its description at $\phi=0$  and can be extended to arbitrarily negative values of $\phi$. In fact the wave function eq.(\ref{pshrhshgf}) is an even function of $\phi$.
Taking the range of $\phi$  to go from $[-\infty,\infty]$, we find,  when eq.(\ref{condmv}) is true,  that eq.(\ref{psiinlpht}) is  valid for all values of $\phi$
 (with  ${\tilde \phi}=\sqrt{\phi^2+M_0}$). We also see from  eq.(\ref{psiinlpht}) that the 
wavefunction corresponds to a classical solution which has a contracting branch in the  past, where $\phi$ is negative, that  then smoothly  evolves into an expanding branch, where $\phi$ become positive. 

A few more comments are in order. 
As was also mentioned towards the end of section \ref{consnorm}  in the example we are considering here the norm is conserved for all values of $\phi$ since $\rho(M)$ has support only for negative values of its argument.
Instead  had we taken $M_0=0$ in eq.(\ref{rhomgs})  the coefficient function $\rho(M)$ would  have support for  positive values of its argument  and from the general considerations in section \ref{consnorm} (around eq.(\ref{condlate}))  it  follows that the norm would  only  be conserved after $\phi^2 > \Delta$.
Noting the condition in eq.\eqref{delsign} this leads to 
\be
\label{conddilabc}
\phi\gg \sigma^{1/4}. 
\ee
This   condition ties in with the fact that in the example we discussed in section \ref{gausscf},
where $\rho(M)$ had support extending till $M\rightarrow \infty$, the norm was exponentially suppressed once the dilaton became big enough and met  
eq.((\ref{conddilabc}).

Also note that we could have considered a Gaussian with the opposite sign for the shift, $M_0$. 
As an example, consider  a shifted Gaussian  where $\rho$ is now given by 
\be
\label{sshf}
\rho(M)=2\sin((M-M_0)x_0)e^{-{1\over 2 \sigma}(M-M_0)^2}
\ee
with $M\in[-\infty,\infty]$ and with $M_0>0$. 
We also take  $\rho_1(M)=\rho(M)$ for $M>0$ and the remaining coefficients ${\tilde \rho}$ and $\rho_2$ to vanish. 
${\hat \Psi}$ is then obtained by carrying out the integral along the contour shown in Fig(\ref{mbranch}) with the definitions eq.(\ref{defa}) and eq.(\ref{defs}) 

In contrast to the case above eq.(\ref{rhomgs}), it is easy to see that here the classical limit, which  requires  $\sigma \gg 1$, has a smaller range of validity 
and arises only at 
sufficiently late times when 
\be
\label{condahs}
\phi^2\gg \sigma+M_0.
\ee

In the classical limit we get for this case that 
\be
\label{exm0}
\langle{\hat M}\rangle=M_0>0, 
\ee
as is expected. 
From section \ref{cllimiwfa}  we see that this corresponds to a classical solution which is  an orbifold of a black hole geometry, since, eq.(\ref{valMa}),  $m={M\over A^2}>0$. 
From eq.(\ref{dsmilne}) the orbifold singularity occurs when $r=\sqrt{m}$ and $\phi=A \sqrt{m}$. Noting eq.(\ref{exm0}), this would imply that  for the classical solution corresponding to the quantum state above,   the singularity occurs at 
$\phi=\sqrt{M_0}$. Near the  singularity, the universe shrinks to zero size (along constant $\phi$ slices) and one would expect quantum effects to become important.  
This is consistent with what we see in the quantum theory, where the   classical limit arises  at much later times, for values of $\phi$ which are larger than $\sqrt{\sigma +M_0}$, eq.(\ref{condahs}).

Finally it is worth mentioning that  by varying the parameter $M_0$, eq.(\ref{rhomgs}), eq.(\ref{res}),  with $\Delta,x_0,$ held fixed, one can construct an infinite number of linearly independent solutions 
of the WDW equation which are normalisable with well defined expectation values for the moments $\langle l^n \rangle, \langle \pi_l^n \rangle$. 
The essential point is more simply illustrated in the next subsection \ref{deldens} see  the discussion eq.(\ref{valra}) onwards. In the Gaussian solutions we are considering here we take $M_0\in (\Delta,\infty]$ and  obtain the family of solutions. The  wave functions for different values of $M_0$ are linearly independent since they contain some eigenvectors of the operator ${\hat M}$, eq.(\ref{formM}), which are different. Since $M_0$ varies continuously we get an infinite number of such solutions.

\subsection{$\rho(M)$:  A Delta function }
\label{deldens}
We end our study in this section  by considering two more examples, in this subsection and the next with rather simple coefficient functions.  
Here we shall discuss a simple example which has a conserved norm and a positive probability density $p(l) \,\,\forall l$. 
The coefficient function $\rho(M)$  is given by
\begin{align}
	\rho(M)=\delta(M+M_1)-\delta(M+M_2),\quad M_1,M_2>0\label{rhomdel}
\end{align}
and ${\tilde \rho}, \rho_1,\rho_2$ are taken to vanish. 
We see that eq.(\ref{condrr}) is met, and since $\rho$ is real eq.(\ref{condva}) is also valid and therefore ${\cal C}_N$ vanishes at $l\rightarrow 0$. 
The wavefunction is given by
\begin{align}
	\hat{\Psi}=\frac{1}{l}\left(\exp(-il\sqrt{\phi^2+M_1})-\exp(-il\sqrt{\phi^2+M_2})\right)\label{wfdel}
\end{align}
We can compute the quantity $\mathcal{C}_N$ directly which turns out to be
\begin{align}
	\mathcal{C}_N=-\frac{4 \phi  \left(\sqrt{M_1+\phi ^2}+\sqrt{M_2+\phi ^2}\right)}{l \sqrt{\left(M_1+\phi ^2\right) \left(M_2+\phi ^2\right)}} \sin ^2\left(\frac{l}{2}  \left(\sqrt{M_1+\phi ^2}-\sqrt{M_2+\phi ^2}\right)\right)\label{cndelval}
\end{align}
We see that it vanishes for $l\rightarrow 0$ and also for $l\rightarrow \infty$. This shows that the norm is conserved. 

The probability density  $p(l)$, which follows from eq.(\ref{wfdel}) is 
\begin{align}
	p(l)=\frac{4}{l^2} \left(\sqrt{M_1+\phi ^2}+\sqrt{M_2+\phi ^2}\right) \sin ^2\left(\frac{1}{2} l \left(\sqrt{M_1+\phi ^2}-\sqrt{M_2+\phi ^2}\right)\right)\label{pldelwf}
\end{align}
It is manifestly   non-negative for all values of $l,\phi>0$. The norm of the wavefunction can then obtained immediately and we find
\begin{align}
	\langle \hat{\Psi},\hat{\Psi}\rangle=\pi\abs{M_1-M_2}\label{normdefwf}
\end{align}
This also shows explicitly that the norm is conserved. 

Turning to the moments, $\langle l^n\rangle$, we see from eq.\eqref{pldelwf}, that all these moments diverge for $n\geq 1$.  We can however calculate moments of $\hat{M}, \hat{\pi}_l$ and they turn out to be finite. The details are presented in \ref{expvalm}.

{\it \underline {An Infinite Family of Solutions:}}

Starting from the example above we can construct an infinite number of solutions to the WDW equation which are linearly independent, and have finite norm.
Take a one-parameter family of solutions labelled by $M_1$ where 
\be
\label{valra}
\rho_{M_1}(M)=\delta(M+M_1)-\delta(M+M_1-a)
\ee
Here $a$ is held fixed and the family is obtained by taking $M_1$ to have values in the range $M_1\in (a,\infty]$.
The corresponding solutions are  given by 
\be
\label{fpsima} 
\Psi_{M_1}(l,\phi)=e^{-il\sqrt{\phi^2+M_1}}-e^{-il\sqrt{\phi^2+M_1-a}}
\ee
Acting on a wave function of the form, eq.(\ref{sola}), the  operator ${\hat M}$, eq.\eqref{lconint}, takes the form
\be
\label{formM}
{\hat M}=\partial_l^2+\phi^2
\ee
and we see that $\Psi$ in eq.(\ref{sola}) is an eigenvector of ${\hat M}$ with eigenvalue $M$. 
It then follows that the two terms appearing on the RHS of eq.(\ref{fpsima}) are eignevectors of ${\hat M}$ with eigenvalues. 
 $-M_1$ and $ -(M_1-a)$ respectively. 
Different solutions in this family will therefore contain different eigenvectors of $M$ and therefore will be linearly independent, leading to an infinite number of linearly independent solutions. The solutions being considered here do not have a finite expectation value for the moments $\langle l^n \rangle$, however a similar construction using Gaussian wave functions which was discussed in section \ref{shifgaus} gives rise to an infinite family of normalisable solutions with finite expectation values for the moments as well.

\subsection{$\rho(M)$: A Theta  Function}
\label{condest}
Next we consider an example where the norm is   conserved but  $p(l) $ is not positive for all values of $l,\phi$. However, interestingly, once $\phi$ becomes large, we find that $p(l)$ does  becomes positive for all $l$. 
The coefficient function $\rho(M)$ is given by
%
\begin{align}
	\rho(M)=\begin{cases}
		1\quad -M_0\leq M\leq 0\\
		-1\quad -2M_0\leq M\leq -M_0
	\end{cases}
	\label{rhoman}
\end{align}
for some $M_0>0$. 
The other coefficient functions ${\tilde \rho}, \rho_1,\rho_2$ vanish. 
Note that 
\begin{align}
	\int dM \rho(M)=0\label{rhominze}
\end{align}
Since $\rho(M)$ is also real, with support only for $M<0$,  and $\rho_2$ vanishes,  it follows that the norm will be conserved, see section \ref{consnorm}. 
This can be checked explicitly by evaluating the wavefunction, which is given by 
\begin{align}
	\hat{\Psi}=\frac{e^{-i l \sqrt{M_0+\phi ^2}} \left(4+4 i l \sqrt{M_0+\phi ^2}\right)+e^{-i l \sqrt{2 M_0+\phi ^2}} \left(-2-2 i l \sqrt{2 M_0+\phi ^2}\right)+e^{-i l \phi } (-2-2 i l \phi )}{l^3}\label{wfexeph}.
\end{align}
(we see from eq.(\ref{wfexeph}) that by rescaling {$\phi\rightarrow \phi \sqrt{M_0}$, $l\rightarrow {l\over \sqrt{M_0}}$} we can set $M_0=1$, but we do not do so here  and continue to retain the $M_0$ dependence below). 

From eq.(\ref{wfexeph}) is follows that the quantity $\mathcal{C}_N$  has the limiting behaviour
\begin{align}
	\mathcal{C}_N=\begin{cases}
		\order(l),\quad l\rightarrow 0\\
		\order{(l^{-3})}\quad l\rightarrow \infty\label{cnurho}
	\end{cases}
\end{align}
and vanishes at both limits showing that the norm is conserved. 

The probability density $p(l)$ at $\phi=0$ is shown in Fig.\ref{CDSphi}, we see that it becomes negative, roughly  in the range  $l\in[15.2, 15.4]$, (in units of $M_0$=1).
More generally from eq.\eqref{plosc} we see that 
\be
\label{pltheta}
p(l,\phi)={1\over l^2} \int dM  dM' \rho(M) \rho(M') \{ (\sqrt{\phi^2-M} + \sqrt{\phi^2-M'} ) e^{-il \sqrt{\phi^2-M}}e^{il\sqrt{\phi^2-M'}}\}
\ee

For  $\phi^2\gg M_0$ this becomes 
\be
\label{pltb}
p(l,\phi)\simeq {2 \phi \over l^2} \abs{\int dM   \rho(M)   e^{-il \sqrt{\phi^2-M}}}^2
\ee
which is manifestly positive for all $l$. This is connected to the comments made in section \ref{pbtime}  that in such cases, where $\rho$ has compact support, the KG norm  becomes well approximately by the $L_2$ norm. 
In fact by the time $\phi^2=100 M_0$ we find   that $p(l)$ is positive for all $l$, see Fig \ref{CDSphi}. 
{\begin{figure}[h!]
		\centering
		\subfigure[]{\includegraphics[width=0.45\textwidth]{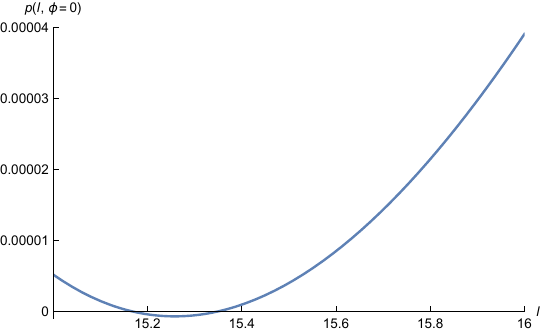}}\hfill
		\subfigure[]{\includegraphics[width=0.45\textwidth]{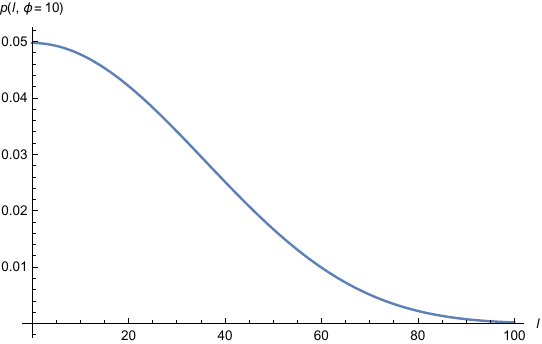}} 
		\caption{We have taken $M_0=1$ in both the figures.(a)$\phi=0$, (b) $\phi=10$.}
		\label{CDSphi}
\end{figure}}
We can also see from eq.(\ref{wfexeph}) that for $\phi^2\gg M_0$, {at large $l$, ${\hat \Psi}\sim {e^{-il\phi}\over l^2}$. Thus $\langle l\rangle $,$ \langle l^2\rangle $ are finite but higher moments will diverge. }

%


\section{Multiple Boundaries}
\label{dtprop}

Most of the  discussion in this paper deals with  a single universe.
 Here  we make some comments on wave functions involving two universes. Our discussion will involve  either two universes which are produced out of ``nothing" through tuneling, or decay into nothing, or a transition amplitude for a single universe in the far past to tunnel to  one  in  the far future. We will use path integral quantisation combined with what has been learnt about canonical quantisation in the single universe sector above,  in the discussion. Dealing with two, and more generally multi-universes, in the canonical framework requires us to ``third quantise" the theory - a task we will not attempt here.


When both universes are in the asymptotic dS limit, i.e. for  $\phi, l\rightarrow\infty $ with ${l\over \phi}$ kept fixed, 
one can use the path integral to calculate the wave function of interest or the transition amplitude, \cite{Maldacena:2019cbz,Cotler:2019nbi,Moitra:2022glw}
The result is given by, 
\begin{align}
	G^{\pm\pm}_{2}(\phi_1, l_1, \phi_2,l_2)&=\int_{0}^\infty b db \Psi^{\pm }_{T}( \phi_1,l_1, b)\Psi^{\pm }_{T}( \phi_2, l_2,b)\nonumber\\
		\Psi_{T}^\pm( \phi,l,b)&={
			\sqrt{\phi}\over \sqrt{\pm 16 i\pi^2 Gl } }e^{\pm i \left({\phi {l} \over 8 \pi G}  + { b^2 	\phi\over 16 \pi G l}\right)}\label{zht}
\end{align}
where the subscript ${\rm 2}$ on G indicates that the  geometry involves  two  boundaries and $\Psi_T$ is the trumpet answer obtained by doing a path integral with one boundary being a geodesic boundary of length $b$ and the other boundary being an asymptotic boundary.  The signs $\pm$ in indicate that the corresponding result is either for expanding (-) or contracting (+) branch of the universe.

  It seems reasonable to interpret $G^{--}_{2}$, which involves two factors of $\Psi_{T}^-$, as  giving the wave function for   two expanding universes which are produced from ``nothing" by quantum tunneling. This is the similar   to the single boundary case, where an analogous  path integral  gives the wave function for the 
  expanding branch of the Hartle-Hawking state.    The arguments $\phi_1,l_1,$ and $\phi_2,l_2,$ of the two wave functions  refer to the values the dilaton and the length  in the two  universes respectively. 
  The two wave functions $\Psi_T^-$ also depend on the parameter $b$ which is integrated over, which means  the wave functions of the two universes produced in this way are correlated with each other.   
  
  Similarly, $G^{++}_{2}$ which involves two factors of $\Psi_{T}^+$, can be interpreted as giving the wave function of two contracting universes which can  eventually  ``annihilate" each other and tunnel into nothing. 
  
  In contrast, it seems more reasonable to interpret  $G^{-+}_{2}(\phi_1,l_1, \phi_2,l_2)$ involving  a factor of $\Psi_T^-(\phi_1,l_1,b)$ and $ \Psi_T^+(\phi_2,l_2,b)$ 
  as a kind of propagator, giving  the transition amplitude for an initial universe  which has size $l_2$ when dilaton takes value $\phi_2$, to tunnel in the far future into a universe with size $l_1$ when the dilaton takes value $\phi_1$. 
  
  The single boundary wave functions we have dealt with above are  proportional to the factor $e^{S_0}$,  eq.\eqref{vals0}; in contrast $G_{2}^{\pm\pm}$ is $\order(1)$, see footnote near eq.\eqref{vals0} in section \ref{canquJT}.  
  
  The fact that the wave function $\Psi^{\pm}_T$ must satisfy the Klein-Gordon equation, eq.(\ref{newwdw}) away from the asymptotic  $dS$ limit, and match the result obtained above in the asymptotic case, fixes the form of   the wave function for the two universe sector for general values of $l_1,\phi_1, l_2, \phi_2$. 
   This form is  given by replacing $\Psi_T^{-}$ in eq.(\ref{zht}) with 
   \be
   \label{rep}
   \Psi_T^-(\phi,l,b)={\phi l \over \sqrt{l^2+b^2} }H^{(2)}_1(\phi \sqrt{l^2+b^2})
   \ee
      where the superscript ``$2$" in the Hankel function indicates that it is of the second kind, and the subscript ``$1$" indicates that the  order is unity. This leads to the more general expression for the expanding two universe wave function,
   \be
   \label{twoe}
   G_{2}^{--}=\int b db \Psi_T^-(\phi_1,l_1,b)\Psi_T^-(\phi_2,l_2,b).
   \ee
  
  Similarly for the two contracting universes the wave function is given by 
  \be
  \label{threee}
  G_{2}^{++}=\int b db \Psi_T^+(\phi_1,l_1,b)\Psi_T^+(\phi_2,l_2,b)
\ee
where 
\be
\label{valha}
\Psi_T^+(\phi,l,b)={l  \phi \over \sqrt{l^2+b^2}}  H^{(1)}_1(\phi\sqrt{l^2+b^2})
\ee
Note that since $H_1^{(1)}, H_1^{(2)}$ are complex conjugates,  $G_{2}^{--}, G_{2}^{++}$ are also complex conjugates of each other. 
Also, that to obtain 
  the wave function analogous to ${\hat \Psi}$ in the single universe case which solves the  KG equation, eq.(\ref{dshamcon}) we replace $\Psi_T^{\pm}$ by 
  ${\hat \Psi_T^{\pm}}={\Psi_T^{\pm}\over l}$ in $G_{2}^{++}, G_{2}^{--}$. Below we denote by ${\hat G}^{\pm\pm}$  the corresponding expressions involving ${\hat \Psi}_T^{\pm}$:
  \be
  \label{denG}
  {\hat G}_{2}^{\pm \pm}=\int b db {\hat \Psi}^{\pm}(\phi_1,l_1, b) {\hat \Psi}^{\pm}(\phi_2,l_2,b).
  \ee
  
  As noted above the dependence on  $b$, which is a modulus in the double trumpet  path integral, correlates the  wave functions of the two universes in $G_{2}^{--}, G_{2}^{++}$. To find the contribution arising from $G_{2}^{--}$ to the density matrix of the first universe, we need to integrate out the dependence on the second universe's variables. One natural way to try and do this is to use the KG inner product, eq.\eqref{kgnorm}. This gives the contribution to the density matrix for the first universe to be
  \be
  \label{dfu}
 {\hat  \rho}(l_1, {\tilde l}_1, \phi)=\int dl_2 ({\hat G}^{--}_{2})^*(\phi, l_1,\phi_2,l_2)i  \overleftrightarrow{\partial_{l_2}} {\hat G_{2}}^{--}(\phi,{\tilde l}_1,\phi_2, l_2)
  \ee
  where $A\overleftrightarrow{\del}B=A\del B-B\del A$.
  Note that the integral over $l_2$, the length of the second universe, is being carried out for a fixed value of the dilaton, $\phi_2$, in the second universe. 
  Also, $\phi$ denotes the value of the dilaton in the first universe at which its density matrix is being evaluated, and  accordingly we have taken the second argument in $({\hat G}_{2}^{--})^*$ and ${\hat G}_{2}^{--}$ to both equal $\phi$. 
  
  Recall that both factors of ${\hat G}_{2}^{--}$ above involve (independent)  integrals over the $b$ modulus. Denoting the integration variables as $b_1,b_2$, one 
  unfortunately finds that the RHS above diverges like 
  $\ln(b)$ as discussed in Appendix \ref{discondb}.  
  
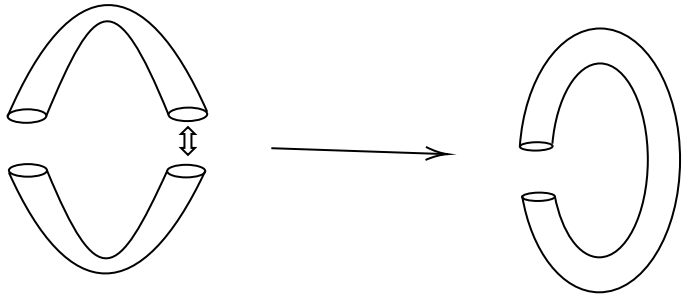
\begin{figure}[!h]
  	  \label{dens1u}
  	\centering
  	
  	\tikzset{every picture/.style={line width=0.75pt}} 
  	
  	\begin{tikzpicture}[x=0.75pt,y=0.75pt,yscale=-1,xscale=1]
  		
  		\draw    (100,114.22) .. controls (129.99,181.69) and (165.82,182.4) .. (198,113.51) ;
  		\draw    (119.01,113.16) .. controls (144.61,175.72) and (154.12,168.34) .. (178.99,113.51) ;
  		\draw   (100,113.16) .. controls (100,111.42) and (104.26,110) .. (109.51,110) .. controls (114.76,110) and (119.01,111.42) .. (119.01,113.16) .. controls (119.01,114.91) and (114.76,116.33) .. (109.51,116.33) .. controls (104.26,116.33) and (100,114.91) .. (100,113.16) -- cycle ;
  		\draw   (178.99,113.51) .. controls (178.99,111.77) and (183.24,110.35) .. (188.49,110.35) .. controls (193.74,110.35) and (198,111.77) .. (198,113.51) .. controls (198,115.26) and (193.74,116.68) .. (188.49,116.68) .. controls (183.24,116.68) and (178.99,115.26) .. (178.99,113.51) -- cycle ;
  		\draw    (198.98,83.69) .. controls (167.49,11.82) and (131.05,11.63) .. (99.36,85.98) ;
  		\draw    (179.67,85.12) .. controls (152.71,18.46) and (143.15,26.52) .. (118.69,85.68) ;
  		\draw   (199,84.82) .. controls (199.03,86.69) and (194.72,88.28) .. (189.38,88.36) .. controls (184.04,88.44) and (179.69,86.99) .. (179.67,85.12) .. controls (179.64,83.25) and (183.95,81.66) .. (189.29,81.58) .. controls (194.62,81.49) and (198.97,82.95) .. (199,84.82) -- cycle ;
  		\draw   (118.69,85.68) .. controls (118.72,87.56) and (114.41,89.14) .. (109.07,89.23) .. controls (103.73,89.31) and (99.38,87.86) .. (99.36,85.98) .. controls (99.33,84.11) and (103.64,82.52) .. (108.98,82.44) .. controls (114.31,82.36) and (118.66,83.81) .. (118.69,85.68) -- cycle ;
  		\draw    (231,102) -- (317,104.93) ;
  		\draw [shift={(319,105)}, rotate = 181.95] [color={rgb, 255:red, 0; green, 0; blue, 0 }  ][line width=0.75]    (10.93,-3.29) .. controls (6.95,-1.4) and (3.31,-0.3) .. (0,0) .. controls (3.31,0.3) and (6.95,1.4) .. (10.93,3.29)   ;
  		\draw   (185.88,94.86) -- (189.38,91.36) -- (192.88,94.86) -- (191.13,94.86) -- (191.13,101.86) -- (192.88,101.86) -- (189.38,105.36) -- (185.88,101.86) -- (187.63,101.86) -- (187.63,94.86) -- cycle ;
  		\draw  [draw opacity=0] (355.05,101.06) .. controls (357.31,67.65) and (374.52,41.67) .. (395.27,41.78) .. controls (417.42,41.9) and (435.22,71.7) .. (435.02,108.35) .. controls (434.82,144.99) and (416.7,174.6) .. (394.55,174.48) .. controls (376.15,174.38) and (360.76,153.8) .. (356.19,125.83) -- (394.91,108.13) -- cycle ; \draw   (355.05,101.06) .. controls (357.31,67.65) and (374.52,41.67) .. (395.27,41.78) .. controls (417.42,41.9) and (435.22,71.7) .. (435.02,108.35) .. controls (434.82,144.99) and (416.7,174.6) .. (394.55,174.48) .. controls (376.15,174.38) and (360.76,153.8) .. (356.19,125.83) ;  
  		\draw  [draw opacity=0] (371.29,99.75) .. controls (373.34,76.76) and (383.3,59.3) .. (395.18,59.37) .. controls (408.44,59.44) and (419.07,81.33) .. (418.93,108.26) .. controls (418.78,135.19) and (407.91,156.96) .. (394.65,156.89) .. controls (384.57,156.83) and (376.02,144.19) .. (372.55,126.31) -- (394.91,108.13) -- cycle ; \draw   (371.29,99.75) .. controls (373.34,76.76) and (383.3,59.3) .. (395.18,59.37) .. controls (408.44,59.44) and (419.07,81.33) .. (418.93,108.26) .. controls (418.78,135.19) and (407.91,156.96) .. (394.65,156.89) .. controls (384.57,156.83) and (376.02,144.19) .. (372.55,126.31) ;  
  		\draw   (371.4,100.96) .. controls (371.38,99.8) and (367.71,98.92) .. (363.2,98.99) .. controls (358.68,99.06) and (355.04,100.05) .. (355.06,101.21) .. controls (355.07,102.37) and (358.74,103.26) .. (363.26,103.19) .. controls (367.77,103.12) and (371.42,102.12) .. (371.4,100.96) -- cycle ;
  		\draw   (372.55,126.31) .. controls (372.53,125.15) and (368.86,124.27) .. (364.35,124.34) .. controls (359.83,124.41) and (356.19,125.4) .. (356.21,126.56) .. controls (356.22,127.72) and (359.89,128.61) .. (364.41,128.54) .. controls (368.92,128.47) and (372.57,127.47) .. (372.55,126.31) -- cycle ;

  	\end{tikzpicture}
  \caption{Contribution to the reduced density matrix of a single universe from the density matrix of the two-universe sector.}

  \end{figure}

  
  On the other hand from the path integral, at least schematically one would expect, as shown in Fig \ref{dens1u} below that the density matrix should be given by the propagator
  $G_{DT}^{-+}( \phi,l_1,  \phi,{\tilde l}_1)$, which is finite. We leave a better understanding of this issue for the future. 
  
  Before ending this section let us comment on some other issues which also need further investigation. 
   First, note  that 
  \cite{pagedensuni,Penington:2019kki}  discuss  contributions such as the double trumpet to the multi-boundary wave function and the resulting  density matrix for a single universe. It is  argued in \cite{Penington:2019kki} that when all components are included one gets a pure state for the single universe. We hope to return to this issue in the future. 
  
  Second,  let us also comment on the propagator $G_{2}^{-+}$, which gives the transition amplitude from an asymptotic dS in the far past to an asymptotic dS in the far future. 
 By convolving this propagator with the wave function of an initial state in the far past,  and the wave function of a final state in the far future, one might hope to compute the transition amplitude for  the initial state to tunnel to the final state. Following the line of thought above, one can  carry out this convolution using the KG norm. Convolving with an initial state ${\hat \Psi}_I$ in this way, gives a final state in the far future,  
 \be
 \label{ffs}
 {\hat \Psi}_f(\phi_f,l_f)=i\int dl_i {\hat G^{-+}}_{2}(\phi_f,l_f, \phi_i, l_i)\overleftrightarrow{\del_l}{\hat \Psi}_i(\phi_i,l_i)
 \ee
 where subscripts $i,f$ stand for initial and final respectively.
 Once again there is a  $b$ modulus integral present in ${\hat G}^{-+}_{2}$ which needs to be carried out, and one needs to be careful about potential divergences when $b\rightarrow 0$. An important check which should be  met is that  the final state wave function 
 does not depend on the instant of gluing, $\phi_2$.  We discuss some details of these in appendix \ref{discondb} but leave a more detailed investigation of this also for the future. 
 
Finally, note that the states given in eq.(\ref{rep}) and eq.(\ref{valha}) are obtained by replacing the ``no-boundary" condition in the  HH case with a geodeisc boundary of length $b$. It is interesting to ask whether other states discussed in this paper can also be obtained by suitable modifications in the boundary conditions. 

\section{Conclusions}
\label{concs}
In this paper we  canonically quantised JT gravity in de Sitter space,  following the procedure first discussed   by \cite{Hennauxjt}  and studied  its consequences with particular attention to 
 the problem of time. Choosing the dilaton as a clock we defined physical states, their norm and expectation values for operators. The number of gauge invariant states in the theory we found are, to begin with,  infinite. Requiring  that they have a have a finite and  conserved norm, and well-defined expectation values for operators, cuts down this number. Requiring that a good classical limit is obtained, where different branches corresponding to different expanding and contracting classical universes decohere, and fluctuations about the mean for expectation values become small, reduces this number even further. 

One conclusion from our study then is that while the problem of time can be solved satisfactorily, by using the dilaton as a physical clock, requiring that  this happens does impose   a significant constraint on  states. 
Another conclusion is that even after  meeting all the  requirements we mentioned above,  one is still left with  an infinite number of acceptable  states in the theory.
A precise discussion leading to this conclusion can be found in   sections \ref{shifgaus} and \ref{deldens}\footnote{
This infinity seems to disagree   with the deSitter entropy since in   JT theory the analogue of higher dimensional dS entropy  is $S_0$, eq.\eqref{vals0}, which is finite. 
However our analysis above was carried out in the single universe sector, without considering any topology change,  and therefore in-effect we have taken $S_0\rightarrow \infty$. 
It is in principle possible that a third quantised description leads to a finite number of states which agrees with the de Sitter entropy. For some discussion along these lines in the AdS context, see \cite{Marolf:2020xie,Post:2022dfi},
}.

The Hartle-Hawking state is a physical state - in fact requiring that it meets the gauge  constraints is an important consistency check on the canonical quantisation procedure we followed, \cite{Maldacena:2019cbz, Verlinde:2020zld}. However, this state is not the best behaved; its norm  diverges, due to a singularity at the turning point (where the universe has minimum length in the global dS classical solution). Several other states in comparison are better behaved, with a finite norm, and well defined expectation values, as mentioned above.

There are several open questions and further directions to pursue, some of which were also mentioned in the introduction. 

There are other physical  states, besides the Hartle -Hawking states,  which are also not satisfactory based on  their behaviour, as was discussed at various points in  the paper. 
Their  failure  to meet some of  the conditions we impose, for example of a finite and conserved norm,  can itself be instructive and potentially interesting from a more general point of view, since in these cases   the standard rules of Hamiltonian quantum mechanics do not apply, at least in some epochs. 
For example, if the coefficient functions $\rho$ and $\rho_1$ are non vanishing satisfying the condition eq.(\ref{condnna}), with these functions  being real and $\rho$ having compact support for $M<M_0$, where $M_0>0$, then as discussed in section \ref{consnorm}, around eq.(\ref{condlate}),  the norm is not conserved at early times for $\phi^2<M_0$ and only becomes conserved for $\phi^2>M_0$. A conventional quantum mechanical description then does not exist when the universe is young  and only holds  at a sufficiently later time. 
If the support for $\rho$ is non-compact and extends till $M\rightarrow \infty$ but decays exponentially at large $M$, as for the Gaussian coefficient case studied in 
section \ref{gausswf1}, then the norm is never conserved but its rate of change is exponentially small at late times. It would be interesting to explore the consequences of such departures from standard Hamiltonian quantum mechanics more generally. 

It would also be interesting to explore other possible physical clocks, besides the dilaton, to understand how different is the resulting description. One such possibility is to use the extrinsic curvature, $K$,  of a spatial slice.{ In ADM gauge, eq.\eqref{adm} this is given by 
	\begin{align}
		K=\frac{\dot{g}_1}{2Ng_1}\sim - \frac{\pi_{\phi}}{\sqrt{g_1}}\label{exkpig1}
	\end{align}

The HH state is defined using the  path integral and corresponds  to having  ``no-boundary". It would be  interesting to understand  other physical states in terms of suitable boundary conditions in the path integral description. It is worth mentioning in this context that the states $\Psi^{\pm}$ given in eq.(\ref{rep}), (\ref{valha}),
can be obtained by replacing the no-boundary condition with a geodesic of length $b$, as discussed in section \ref{dtprop}; perhaps this idea can be extended for  other states as well.

The canonical quantisation we have  carried out can be extended quite easily to a more general potential for the dilaton \cite{nanda:2023yta}, including potentials which are slowly varying and of interest from the point of view of inflation. Exploring  solutions of this type, especially from the point of view of eternal inflation, would be very interesting.

One important omission is that we have not included matter and dealt only with the pure JT theory. In the presence of  matter solving the gauge constraints becomes 
much more non-trivial. For the Hartle-Hawking state though the path integral could be carried out, \cite{Maldacena:2019cbz,Moitra:2022glw}, at least with  conformal matter, and we hope 
that progress can be made for other states as well, perhaps using a combination of canonical and path integral methods. 

We have mostly studied the single universe sector in this paper. The   path integral, for the no-boundary case,   leads to processes  with topology change corresponding to universes splitting or joining to form multiple universes. Understanding these in general  in the  canonical formulation would be interesting and would require us to ``third quantise" the theory. 

%


It is tempting to interpret the coefficient functions $\rho$ and ${\tilde \rho}$, which appear in the wave function, eq.(\ref{genwfinl}),  as being related to the  density of eigenstates of the operator 
${\hat P}$, which is the translation operator along the boundary, and the wave function then being related to $\Tr(e^{\pm i {\hat P}})$, see discussion after eq.(\ref{genwfinl}) and in appendix \ref{dsadmmass}. 
However it is not clear if such an interpretation is always possible. In the best behaved cases where eq.(\ref{condrr}) must be met, ${\rho}$ and ${\tilde \rho}$ cannot both be positive, for example. Also when ${\rho}$ and ${\tilde \rho}$ are independent functions how should one then interpret the presence of two densities of states?
If, we do not require  eq.(\ref{condrr}) to be met, one can take $\rho={\tilde \rho}$, with $\rho$ being positive. And then try to find the appropriate random matrix theory as was done for the HH wave function, \cite{Maldacena:2019cbz,Moitra:2022glw,Cotler:2019nbi}. Higher genus and multi boundary amplitudes can then be calculated using this matrix theory, as was done in \cite{Saad:2019lba,Maldacena:2019cbz,Cotler:2019nbi}, for the HH wave function. We leave such an investigation for the future.

Finally, in this paper, we had in mind restricting the range of the dilaton to lie in the region, 
\be
\label{rangedil}
\phi\ge 0.
\ee
 However, the quantisation procedure did not, per se, depend on this range, and in fact the basis, eq.(\ref{genwfinl}), in which we expanded states, consists of  even functions of $\phi$, which automatically extend to $\phi<0$. The exact value chosen for the lower limit of the   dilaton in eq.(\ref{rangedil}) was somewhat adhoc but in the classical theory  some lower limit seems warranted, both from the perspective of obtaining JT gravity from higher dimensions, and keeping in mind the behaviour of black holes where  a  singularity is taken to  arise  at the minimum allowed value of $\phi$.  After quantisation though, we found that  several states behave in a smooth manner, with no pathology even for arbitrarily negative values of $\phi$. One set of examples where this is most striking are 
 the shifted Gaussian states discussed in section \ref{shifgaus}, with $\langle{\hat M}\rangle$ being negative. These  correspond to  classical solutions discussed in section (\ref{cllimiwfa}), eq.(\ref{dsmilne}), with $m<0$, for which the geometry has no singularity  and can be extended from the far past to the far future.  By taking the width of the Gaussian to be sufficiently narrow, eq.(\ref{delsign}),  we found that the quantum states behave classically for all values of $\phi$.  
 The smooth and near-classical behaviour of these quantum state suggests that in such cases, at least, the range of the  dilaton could also be extended, with  the corresponding quantum states starting as a  contracting universe in the far past, $\phi\rightarrow -\infty$,  and smoothly evolving into an expanding universe  in the far future, $\phi\rightarrow \infty$.  It would  be interesting  to add  matter for  such states  too and study the resulting behaviour. 
 We leave a more detailed analysis of this and related issues also for the future.


%
%
%
%
%
%
%
%
%
%
%

\section{Acknowledgements}
We thank Sumit Das, Abhijit Gadde, Alok Laddha, Andrei Linde, Gautam Mandal, Shiraz Minwalla, Onkar Parrikar and Amitabh Virmani for discussions. 
SPT acknowledges support from the organisers of the conference, ``Entanglement, Large N and Blackholes" at APCTP, Korea, and the  
``KIAS-APCTP Frontiers of Theoretical Physics 2023" conference in Jeju, Korea. He also acknowledges support from the ICTP, Trieste,  for 
attending the ``Huddle on Entanglement, Black Holes and Spacetime". SKS acknowledges support from the organisers of the conference ``Quantum Black Holes, Quantum
Information and Quantum Strings" at APCTP, Korea and the  ``17th Kavli Asian Winter School" at IBS, Daejeon, Korea. KKN and SPT acknowledge support from Government of India, 
Department of Atomic Energy, under Project  Identification No. RTI 4002 and from the Quantum Space-Time Endowment of the Infosys Science Foundation.   The work of SKS is supported by MEXT KAKENHI Grant-in-Aid for Transformative Research Areas A “Extreme Universe” No. 21H05184. Most of all, we are grateful to the people of India and Japan for generously supporting research in String Theory.
\newpage

\appendix

\section{Covariant phase-space quantization}
\label{symcur}
In this appendix, we quantise the classical system of JT gravity by following the route of covariant phase space quantization. The main idea is to compute the symplectic two form on the space of classical solutions. The next step is to convert the classical phase variables to quantum operators with the appropriate commutation relations determined by this symplectic two form.

Let us begin by computing the symplectic two form in JT gravity. The first step is to compute the pre-symplectic one form, which is the set of boundary terms that arises when deriving the equations of motion. The next step is to construct a two-form current by taking a second variation of this pre-symplectic one-form. This current is then integrated on a Cauchy slice to give the required symplectic two-form. The JT action, in units of $8\pi G=1$ is given by,
\begin{equation}
	I_{\text{bulk}}= \frac{1}{2} \int \sqrt{-g} \phi (R- \Lambda)\label{jtin8g1}
\end{equation}
Taking the variation we obtain,
\begin{equation}
	\delta I_{\text{bulk}} = \frac{1}{2} \int \sqrt{-g} \delta\phi (R- \Lambda) + \frac{1}{2}\int \phi \delta(\sqrt{-g}) (R-\Lambda)+ \frac{1}{2} \int \sqrt{-g}\phi \,\delta g^{\alpha \beta}\, R_{\alpha \beta}  + \frac{1}{2} \int \phi \sqrt{-g} g^{\alpha \beta} \delta R_{\alpha \beta} \label{actionvar1}
\end{equation}
The first three terms do not give boundary terms and so we can ignore them.  Only the fourth term will give us the required boundary terms since it involves two derivatives acting on the variation of the metric.
The variation of the Ricci tensor is given by 
\begin{align}
	\delta R_{\alpha \beta}&=\nabla_{\mu} \delta \Gamma ^{\mu}_{\alpha \beta} - \nabla_{\alpha} \delta \Gamma^{\mu}_{\mu \beta},\nonumber\\
	\delta\Gamma^{\mu}_{\alpha\beta}&=\half g^{\mu \nu}(\nabla_\alpha \delta g_{\nu\beta}+\nabla_\beta \delta g_{\alpha\nu}-\nabla_\nu \delta g_{\alpha\beta})\label{delrindelg}
\end{align}
Using this, the last term in eq.\eqref{actionvar1} can be written as
\begin{align}
	\delta I_{\text{bulk}}=\half \int \nabla_\mu \left(\phi \sqrt{-g} \,(g^{\alpha\beta}\delta\Gamma^\mu_{\alpha\beta}-g^{\mu \beta}\delta\Gamma^{\nu}_{\nu\beta}) \right)-\half \int \sqrt{-g}\,\nabla_\mu\phi \,(g^{\alpha\beta}\delta\Gamma^\mu_{\alpha\beta}-g^{\mu \beta}\delta\Gamma^{\nu}_{\nu\beta})\label{oninbpar}
\end{align}
We need to do one more integration by parts since the second term above involves derivatives acting on variations. Using the variation of Christoffel symbol in terms of variations of metric components, explicitly given in eq.\eqref{delrindelg}, it can be easily checked that,
\begin{align}
	g^{\alpha\beta}\delta\Gamma^\mu_{\alpha\beta}-g^{\mu \beta}\delta\Gamma^{\nu}_{\nu\beta}= \nabla_\nu (g^{\mu\alpha} g^{\nu\beta} \delta g_{\alpha\beta}- g^{\mu \nu} g^{\alpha \beta} \delta g_{\alpha \beta})\label{chrisid}
\end{align}
Using this in eq.\eqref{oninbpar}, and dropping the bulk term that we get after doing an integration by parts, we see that the total set of boundary terms are given by 
\begin{align}
	\delta I_{\text{bulk},\del}&=	\half \int \left[\nabla_\mu \left(\phi \sqrt{-g} \,(g^{\alpha\beta}\delta\Gamma^\mu_{\alpha\beta}-g^{\mu \beta}\delta\Gamma^{\nu}_{\nu\beta}) \right)-\nabla_\nu\left(\sqrt{-g}\nabla_\mu\phi\, (g^{\mu\alpha} g^{\nu\beta} \delta g_{\alpha\beta}- g^{\mu \nu} g^{\alpha \beta} \delta g_{\alpha \beta}) \right) \right]\nonumber\\
	&=	\half \int \nabla_\mu\left[\phi \sqrt{-g} \,(g^{\alpha\beta}\delta\Gamma^\mu_{\alpha\beta}-g^{\mu \beta}\delta\Gamma^{\nu}_{\nu\beta}) -\sqrt{-g}\nabla_\nu\phi\, (g^{\nu\alpha} g^{\mu\beta} \delta g_{\alpha\beta}- g^{\mu \nu} g^{\alpha \beta} \delta g_{\alpha \beta}) \right]
	\label{bdtdels}
\end{align}
So, the pre-symplectic one-form is given by 
\begin{align}
	\Theta^\mu&=\half \left(\phi \sqrt{-g} \,(g^{\alpha\beta}\delta\Gamma^\mu_{\alpha\beta}-g^{\mu \beta}\delta\Gamma^{\nu}_{\nu\beta}) -\sqrt{-g}\nabla_\nu\phi\, (g^{\nu\alpha} g^{\mu\beta} \delta g_{\alpha\beta}- g^{\mu \nu} g^{\alpha \beta} \delta g_{\alpha \beta})\right)\nonumber\\
	&=\half \sqrt{-g} (g^{\nu\alpha} g^{\mu\beta} - g^{\mu \nu} g^{\alpha \beta} )\left(\phi \, \nabla_\nu \delta g_{\alpha \beta}-\nabla_\nu\phi\, \delta g_{\alpha \beta}\right)
	\label{presymp}
\end{align}
The symplectic current is  given in terms of the pre-symplectic form  by 
\begin{equation}
	\sqrt{-g} J^{\mu} =-\delta \Theta^\mu\label{sycurpresymp}
\end{equation}
A straightforward but tedious algebra also shows that the symplectic current defined above is conserved and hence can be evaluated on any Cauchy slice. Instead, we will evaluate this current in a particular gauge and show that it is conserved.  The symplectic two-form is then given by 
\begin{align}
	\omega=\int d\Sigma_\mu \sqrt{-g} J^\mu\label{sympcur}
\end{align}
\subsection{Feffermann-Graham Gauge}
\label{fgforcpq}
Let us now evaluate it on a solution shown in eq.\eqref{dsmilne} which we write here for convenience, 
\begin{equation}
	ds^2 = -\frac{dr^2}{r^2-m} + (r^2-m) dx^2, \hspace{1cm}  x \sim x+1 ,\hspace{1cm} \phi = A r.\label{milsolin}
\end{equation}
It will be convenient to transform the above solution to a FG gauge, as that will make easier the comparison with Brown-York stress tensor calculation that we later carry out. We can transform it to FG gauge by doing the coordinate transformation

\begin{equation}
	r = \frac{1}{ z} + \frac{m z}{4}.\label{miltofg}
\end{equation}

following which the solution becomes
\begin{equation}
	ds^2 = -\frac{dz^2}{z^2} + \frac{1}{z^2} \left(1 - \frac{m z^2}{4}\right)^2 dx^2, \hspace{2cm} \phi = \frac{A}{z} +  \frac{A m}{4} z \label{solJT}
\end{equation}

It turns out that in this gauge, only $J^z$ component of the symplectic current is non-zero  and is given by 
\begin{equation}
	\sqrt{-g} J^{z} = \frac{1}{2} \delta A \wedge \delta m.  \label{symcurr}
\end{equation}
Thus, it is easy to verify that the current is conserved as follows
\begin{align}
	\nabla_\mu J^\mu =\frac{1}{\sqrt{-g}}\del_z( \sqrt{-g}J^z)=0\label{crntcons}
\end{align}
Thus, the symplectic two-form on the phase space is given by 
\begin{equation}
	\omega = \int dx \sqrt{-g} J^{z} = \half \delta A \wedge \delta m \label{current}
\end{equation}
Thus, the quantization of the classical variables $(A, m)$ that parametrize the classical solutions leads to
\begin{equation}
	[m , A] = -2{i}.\label{mAcom}
\end{equation}
To compare with the canonical quantization and the corresponding wavefunction discussed in section \ref{cllimiwf}, we need to use $M$,eq.\eqref{mas}, as one of the phase-space variables. A straightforward calculation using eq.\eqref{solJT},\eqref{mAcom} yields
\begin{equation}
	M=  m A^2. \label{Mcos}
\end{equation}
It then follows from eq.\eqref{mAcom} that 
\begin{equation}
	\left[M, \frac{1}{A}\right]= 2i. \label{comm}
\end{equation}
Thus, an eigenstate of the operator $M$ with eigenvalue $\frac{1}{A}$ is given by 
\begin{align}
	\Psi_M=e^{\frac{iM}{2A}}\label{psimA}
\end{align}
Thus, an eigenfunction of $\hat{M}$ with an eigenvalue $M$ is of the form shown above, in the basis of its conjugate momentum $\frac{1}{2A}$. We can compare this with a eigenfunction of $\hat{M}$ obtained using the canonical quantization procedure in section \ref{cllimiwf}. Considering the expanding branch wavefunction $\Psi=e^{-il\sqrt{\phi^2-M}}$ and expanding it for large $\phi$ gives
\begin{align}
	\Psi=e^{-il\phi}e^{\frac{ilM}{2\phi}}\label{psiexexp}
\end{align}
Upto to the phase factor $e^{il\phi}$ this is the same as eq.\eqref{psimA} if we identify $\frac{1}{A}=\frac{l}{\phi}$. Indeed this is case for on-shell configurations mentioned in eq.\eqref{milsolin}, when $\phi$ is large, 
\begin{equation}
	l =  \sqrt{r^2-m} =  \sqrt{\frac{\phi^2}{A^2}- \frac{M}{A^2}} \implies \frac{1}{A} =\frac{l}{\sqrt{\phi^2-M}}\sim \frac{l}{\phi} \label{rel}
\end{equation}
\subsubsection{Calculation of ADM Momentum}
\label{dsadmmass}
The ADM momentum can be calculated using the holographic renomalization procedure. In a ADM like gauge,
\begin{align}
	ds^2=-\frac{dz^2}{z^2}+\gamma_{xx}(z,x) dx^2\label{ds2gage}
\end{align}
 It is given by (in units of $8\pi G=1$)
\begin{align}
	\label{ADMdefP}
	P_{\text{ADM}}= -\frac{2z}{\sqrt{\gamma}} \frac{\delta I_{OS}}{\delta \gamma^{xx} }
\end{align}
where $I_{OS}$ is the on-shell value of the full JT action including the boundary terms and counterterms. As we will see, we will require a counter term to render $P_{\text{ADM}}$ finite. The full action is then given by 
\begin{align}
	I_{\text{OS}} =I_{\text{JT}}+I_{\text{ct}}\label{itot}
\end{align}
where $I_{\text{JT}}$ is given in eq.\eqref{jtact} and $I_{\text{ct}}$ is given by 
\begin{align}
	I_{\text{ct}}=\int \sqrt{\gamma}\,\phi \,dx \label{ictact}
\end{align}
 The variation of each of the terms in eq.\eqref{itot} can be carried out as follows.  
The term $I_{\text{JT}}$ eq.\eqref{jtact}, consists of two pieces, the bulk term and the Gibbons-Hawking term. The variation of the bulk part is already computed earlier which when evaluated on a solution, just gives the boundary terms mentioned in eq.\eqref{bdtdels},
 gives\footnote{Note that the Stokes theorem for a spacelike boundary has the form $\int \sqrt{-g}\nabla_\mu J^\mu=-\int_\del \sqrt{\gamma} n_\mu J^\mu$, where $n^\mu$ is the outward-directed normal vector. See chapter 3 of \cite{poisson2004relativist}.}
\begin{align}
	\delta I_{\text{bulk},\del} &= -\half \int \sqrt{\gamma}\,n_\mu\left[\phi  \,(g^{\alpha\beta}\delta\Gamma^\mu_{\alpha\beta}-g^{\mu \beta}\delta\Gamma^{\nu}_{\nu\beta}) -\nabla_\nu\phi\, (g^{\nu\alpha} g^{\mu\beta} \delta g_{\alpha\beta}- g^{\mu \nu} g^{\alpha \beta} \delta g_{\alpha \beta}) \right]\nonumber \\
	&= \frac{1}{2} \int {\sqrt{\gamma_{xx}}} \left(\frac{\delta \gamma_{xx}}{\gamma_{xx}} \,z\partial_z \phi +{z\,}\phi \left(- \frac{ \partial_z \delta \gamma_{xx}}{\gamma_{xx}} +  \frac{\partial_z \gamma_{xx}}{2\gamma_{xx}}\frac{\delta \gamma_{xx}}{\gamma_{xx}}\right)\right)
	\label{varbulkjt}
\end{align}
In obtaining the second line above, we used the fact that the normal vector has the non-zero component $n_z=z^{-1}$. 
The Gibbons-Hawking boundary term, in the gauge eq.\eqref{ds2gage}, reads
\begin{align}
	I_{\text{GH}}= - \int dx\sqrt{\gamma}\, \phi \,K=\int dx \sqrt{\gamma_{xx}}\phi \left(\frac{z\del_z \gamma_{xx}}{2\gamma_{xx}}\right)
	\label{actghting}
\end{align}
whose variation with respect to the boundary metric is then given by 
\begin{align}
	\delta I_{\text{GH}}= \frac{1}{2}\int \sqrt{ \gamma_{xx}} \,z\,\phi \left(\frac{\partial_z \delta \gamma_{xx}}{\gamma_{xx}} - \frac{\delta \gamma_{xx} \partial_z \gamma_{xx}}{2 \gamma_{xx}^2}\right)
	\label{varghing}
\end{align}
Combining eq.\eqref{varbulkjt} and eq.\eqref{varghing} we get,
\begin{align}
	\label{varJTfull}
	\delta I_{\text{JT}} = \frac{1}{2}\int {z\sqrt{\gamma_{xx}}}{\gamma^{xx}} \delta \gamma_{xx} \partial_z \phi = - \frac{1}{2}\int {z\sqrt{\gamma_{xx}}}{\gamma_{xx}} \delta \gamma^{xx} \partial_z \phi 
\end{align}
Here we have used the fact that $\delta \gamma_{xx}= - \delta \gamma^{xx} \gamma_{xx}^2$. The variation of the counterterm with respect to the boundary metric is given by 
\begin{align}
	\delta I_{\text{ct}}= -\frac{1}{2}\int \sqrt{\gamma_{xx}} \phi \gamma_{xx} \delta \gamma^{xx}.
	\label{varcting}
\end{align}
Combining eq.\eqref{varJTfull} and eq.\eqref{varcting}, we get
\begin{align}
	\delta I_{\text{OS}}= - \frac{1}{2}\int {\sqrt{\gamma_{xx}}}{\gamma_{xx}} \delta \gamma^{xx} (z\partial_z \phi +\phi)\label{iosvar}
\end{align}
and using this in eq.\eqref{ADMdefP} we get,
\begin{equation}
	P_{\text{ADM}} = \lim_{z \rightarrow \epsilon} z \gamma_{xx} (z \partial_z \phi + \phi).\label{admmofo}
\end{equation}
Substituting the values for $\phi$ and $\gamma_{xx}$ given in \eqref{solJT} we get,
\begin{equation}
	P_{\text{ADM}} = \lim_{z \rightarrow \epsilon} z \frac{1}{z^2} \left(1 - \frac{m z^2}{4}\right)^2 \frac{m A}{2} z = m \frac{A}{2}.\label{admmval}
\end{equation}
Hence, comparing the above result with \eqref{Mcos}, we get,
\begin{equation}
P_{\text{ADM}}={M\over 2A}
	\label{admmvsm}
\end{equation}

It is also worth noting that from eq.(\ref{admmvsm}), eq.(\ref{psiexexp}), and eq.(\ref{rel}) we see that 
a general expanding branch wave function of the form 
\be
\label{wffax}
\Psi=\int dM \rho(M) e^{-il\sqrt{\phi^2-M}}
\ee
asymptotically, for $l,\phi$  large,  has the form\footnote{More precisely this asymptotic form is valid for large $\phi$ if  $\rho$ has compact support, 
or more generally when the WKB approximation is valid, see appendix \ref{allnfwf}.}
\be
\label{assformwf}
\Psi\simeq e^{-il\phi} \int dM \rho(M) e^{i P(M)}\sim \int dM \rho(M) e^{iP(M)}
\ee
where in the last expression we have ignored  the overall phase.
The last expression on the RHS is  suggestive  of  being a trace, 
\be
\label{wft}
\Psi\sim \Tr(e^{i{\hat P}}),  
\ee
with $ \rho(M)$ being the density of eigenstates of   the operator ${\hat P}$, with eigenvalues  lying in the interval $[M, M+dM]$.
Similarly, a  contracting branch wave function would be related to 
\be
\label{wftc}
\Psi\sim \Tr(e^{-i{\hat P}})
\ee
Note in this context  that for a constant  value of $A$, $P$ and $M$ are proportional to each other, eq.(\ref{admmvsm}),\eqref{psiexexp}.
However, as is noted in section (\ref{cllimiwf}),  states  often do not have a positive definite value for $\rho$, and ${\tilde \rho}$, and these functions in fact do not  even have to be real -  for such states a direct interpretation  of the wave function in terms of a density of states is not possible.
\section{Wavefunction and momentum constraint}
\label{momcowf}

In this appendix, we shall show that the wavefunction given in eq.\eqref{wfsinw} satisfies the momentum constraint equation. In particular, we need to show that 

\begin{align}
	\frac{\delta \mathcal{S}}{\delta\phi}=\frac{\sqrt{g_1}}{\mathcal{Q}}\pqty{\left(\frac{\phi'}{\sqrt{g_1}}\right)'+{\sqrt{g_1}\phi}}\label{momcseq	}
\end{align}

To evaluate the variation of $\mathcal{S}$ with respect to $\phi$, we first note that 
\begin{align}
	\frac{\delta \mathcal{S}}{\delta \phi}=\frac{\del \mathcal{S}}{\del\phi}-\pqty{\frac{\del \mathcal{S}}{\del\phi'}}'\label{sphiel}
\end{align}
where as before primes denote derivative with respect to $x$. 
From eq.\eqref{wfsinw}, it is easy to show that 
\begin{align}
	\frac{\del \mathcal{S}}{\del\phi}&=-\frac{\mathcal{Q}\phi g_1}{\phi'^2-\mathcal{Q}^2},\nonumber\\
	\frac{\del \mathcal{S}}{\del\phi'}&=-\tanh^{-1}\pqty{\frac{\mathcal{Q}}{\phi'}}\label{sphphdd}
\end{align}
where in obtaining the second equation above we used the explicit expression for $Q$ in terms of $\phi,\phi'$ in eq.\eqref{wfsinw} from which we get
\begin{align}
	\frac{\del \mathcal{Q}}{\del\phi}=\frac{\phi g_1}{\mathcal{Q}},\quad \frac{\del \mathcal{Q}}{\del \phi'}=\frac{\phi'}{\mathcal{Q}},\quad \frac{\del\mathcal{Q}}{\del g_1}=\frac{\mathcal{Q}^2-\phi'^2}{2 \mathcal{Q} g_1}\label{Qphphd}
\end{align}
Using eq.\eqref{sphphdd} in eq.\eqref{sphiel} we find
\begin{align}
	\frac{\delta \mathcal{S}}{\delta \phi}=\frac{\mathcal{Q} \left(g_1 \phi+\phi''\right)-\mathcal{Q}' \phi'}{\mathcal{Q}^2-\phi'^2}\label{Qintch}
\end{align}
Inserting the value of $\mathcal{Q}'$ using the chain rule
\begin{align}
	\mathcal{Q}'(x)=\frac{\del\mathcal{Q}}{\del\phi}\phi'+\frac{\del\mathcal{Q}}{\del\phi'}\phi''+\frac{\del\mathcal{Q}}{\del g_1}g_1'\label{Qchrule}
\end{align}
and using the results of eq.\eqref{Qphphd}, it is straightforward to obtain  eq.\eqref{momcseq	}.
\section{WKB Approximation}
\label{allnfwf}

In this appendix we revisit the WDW equation and try to solve it in the WKB approximation. Let us start with a wavefunction of the form 
\begin{align}
	\hat{\Psi}=e^{S}\label{wkbpsi}
\end{align}
The WKB expansion in our case corresponds to a $G$ expansion\footnote{For related work on WKB solutions of WDW equation in higher dimensions see \cite{kiefer1991quantum} and \cite{singh1989notes}}. of the exponent $S$. Let us consider the following expansion
\begin{align}
	S=\frac{S_{-1}}{G}+S_0+G S_1+G^2S_2\dots\label{Sexpa}
\end{align}
The WDW equation with explicit factors of $G$ reads (in $8\pi =1$ units,)
\begin{align}
	G^2\del_\phi\del_l\hat{\Psi}+l\phi\hat{\Psi}=0\label{gnexp}
\end{align}
Using the ansatz eq.\eqref{wkbpsi} and the expansion eq.\eqref{Sexpa} we get, to leading order
\begin{align}
	\del_l S_{-1}\del_\phi S_{-1}+l\phi =0\label{S0eq}
\end{align}
the solution for which is easily seen to be
\begin{align}
	S_{-1}=\pm i l\phi\label{sm1}
\end{align}
Expanding eq.\eqref{gnexp} to higher orders in $G$, we get the following equations for $S_0, S_1$ and $S_2$
\begin{align}
	\del_l S_0 \del_\phi S_{-1}+	\del_l S_{-1} \del_\phi S_0+\del_l\del_\phi S_{-1}&=0\nonumber\\
	\del_l S_1\del_\phi S_{-1}+	\del_l S_{-1}\del_\phi S_1+\del_lS_0\del_\phi S_0+\del_l\del_\phi S_0&=0\nonumber\\
	\del_lS_2\del_\phi S_{-1}+		\del_\phi S_2\del_l S_{-1}+\del_l\del_\phi S_1+\del_lS_0\del_\phi S_1+\del_lS_1\del_\phi S_0&=0\label{S0S1S2eq}.
\end{align}
We can solve these equations iteratively, first solving for $S_0$ using $S_{-1}$ in eq.\eqref{sm1}, then using $S_{0},S_{-1}$ to obtain $S_1$ and so on. 
The corresponding solutions for $S_0,S_1, S_2$ are then given by 
\begin{align}
	\eta &=\frac{\phi}{l}\nonumber\\
	S_0&=-\half \ln (l\phi)+\ln g_0(\eta)\nonumber\\
	S_1&=\frac{1}{g_0(\eta)}\frac{i}{2\phi l }\left(-\eta^2g_0''(\eta)-\eta g_0'(\eta)+\frac{1}{4}g_0(\eta)\right)\nonumber\\
	S_2&=\frac{1}{8l^2\phi^2}
	\left(-\frac{\eta^4 g_0^{(4)}(\eta)}{ g_0(\eta)}-\frac{6 \eta^3 g_0^{(3)}(\eta)}{ g_0(\eta)}+\frac{\eta^4 g_0''(\eta)^2}{ g_0(\eta)^2}-\frac{5 \eta^2 g_0''(\eta)}{ g_0(\eta)}+\frac{\eta^2 g_0'(\eta)^2}{ g_0(\eta)^2}+\frac{\eta g_0'(\eta)}{ g_0(\eta)}+\frac{2\eta^3 g_0'(\eta) g_0''(\eta)}{ g_0(\eta)^2}-\frac{1}{2}\right)\label{s1s2s3}
\end{align}
where $g_0(\eta)$ is an arbitrary function of $\eta$. Note that at each order, a homogeneous solution can be added. For example, in the equation for $S_1$, apart from the response to the source terms $S_0,S_{-1}$, a homogenous solution to the differential equation for $S_1$ can always be added. This can then act as a source term for the next iteration, $S_2$. The same arguments can be repeated for every iteration. This can effectively be thought of as an expansion in $G$ of the function $g_{0}(\eta)$. We do not consider such an expansion and only restrict to evaluating the response to the source terms at each order except in $S_0$ where we introduced the homogeneous solution $g_0(\eta)$.
So, the wavefunction can be written as
\begin{align}
	\hat{\Psi}&\simeq \frac{e^{-il\phi}}{\sqrt{l\phi}}g_0(\eta)e^{GS_1+G^2S_2}\nonumber\\
	&\simeq \frac{e^{-il\phi}}{\sqrt{l\phi}}g_0(\eta)	\left(1+G S_1+G^2\left(S_2+\half S_1^2\right)\right)\nonumber\\
	&\simeq \frac{e^{-il\phi}}{\sqrt{l\phi}}	\left(g_0- \frac{iG}{2\phi l }\left(\eta^2g_0''+\eta g_0'-\frac{1}{4}g_0\right)-\frac{G^2}{8l^2\phi^2}\left(\eta^4 g_0^{(4)}+6 \eta^3 g_0^{(3)}+\frac{9}{2} \eta^2 g_0''-\frac{3}{2} \eta g_0'+\frac{9 g_0}{16}\right)\right)
	\label{psihgn}
\end{align}
We see from eq.\eqref{s1s2s3} that for any bounded function $g_0(\eta)$ the small $G$ expansion is the same as the expansion in large $l \phi$, with the ratio $\frac{l}{\phi}$ held fixed. Thus, the WKB limit corresponds to 
\begin{align}
	l\phi\gg G,\,\quad \frac{l}{\phi}=\text{fixed}\label{wkb}
\end{align}
An alternate way to understand the above WKB solution is as follows. Consider a general solution to the WDW equation in eq.\eqref{genwfinl}. Without loss of generality let us restrict to the case where we have only expanding branch, {\it i.e.} $\rho\neq 0, \tilde{\rho}=0$. Assuming $\rho$ has a compact support we can expand the square root in the exponent for large $\phi$ satisfying 
\begin{align}
	M\ll \phi^2\label{sqrtexpcon}
\end{align}
as 
\begin{align}
	\hat{\Psi}=\frac{1}{l}\int dM \,\rho(M)e^{-il\phi+\frac{ilM}{2\phi}+\frac{ilM^2}{8\phi^3}+\frac{ilM^3}{16\phi^5}+\frac{5ilM^4}{128\phi^7}+\dots}\label{psisqexp}
\end{align}
where dots denote higher order terms which are suppressed by higher powers of $\frac{M}{\phi^2}$ (Note that eq.\eqref{psisqexp} is in units of $8\pi G=1$). We are interested only in comparing with the expansion in eq.\eqref{psihgn} for which it suffices to consider the expansion till $M^4$ in the exponent and  drop the terms indicated by dots above.  In the limit where 
\begin{align}
	\frac{lM}{\phi}\sim\order(1),\,\, \frac{lM^2}{\phi^3}\ll 1\label{lphcond}
\end{align}
we  can consistently drop the terms denoted by dots in eq.\eqref{psisqexp}. 
Combining eq.\eqref{sqrtexpcon} and eq.\eqref{lphcond}, we obtain
\begin{align}
	1\ll l\phi\sim\frac{\phi^2}{M}\ll \left(\frac{\phi^2}{M}\right)^2\label{lphicond}
\end{align}
Thus in this region, we have
\begin{align}
	\hat{\Psi}\simeq&\frac{e^{-il\phi}}{l}\int dM \,\rho(M)e^{\frac{ilM}{2\phi}}\left(1+\frac{ilM^2}{8\phi^3}+\frac{ilM^3}{16\phi^5}- \frac{lM^4 (-5 i + l \phi)}{128\phi^7}\right)\nonumber\\
	=&\frac{e^{-il\phi}}{l}\left(1-\frac{il}{2\phi}\del_l^2- \frac{l}{2\phi^2}\del_l^3 - \frac{l  (-5 i + l \phi) }{8\phi^3}\del_l^4\right)\int dM\,\rho(M)e^{\frac{ilM}{2\phi}}
	\label{psitoled}
\end{align}
Note that the term proportional to $M^4$ has contribution from the two terms in the expansion of the exponent. One of them is from expanding $e^{\frac{ilM^2}{8\phi^3}}$ to quadratic order and another from $e^{\frac{ilM^4}{\phi^7}}$ to linear order. 
Let
\begin{align}
	\hat{g}_0\left(\frac{\phi}{l}\right)=\sqrt{\frac{\phi}{l}}\int dM \,\rho(M)e^{\frac{ilM}{2\phi}},\quad \eta=\frac{\phi}{l}\label{godef}
\end{align}
We find that the wavefunction is given by 
\begin{align}
	\hat{\Psi}&\simeq \frac{e^{-il\phi}}{\sqrt{l\phi}}\left(\hat{g}_0+\frac{i}{2l\phi}\left(-\eta^2\hat{g}_0''-\eta\,\hat{g}_0'+\frac{1}{4}\hat{g}_0\right) \right)\nonumber\\
	&- \frac{e^{-il\phi}}{\sqrt{l\phi}} \frac{1}{8 l^2 \phi^2}
	\left(\eta^4\hat{g}_0^{(4)}+6\,\eta^3\hat{g}_0 ^{(3)}+ \frac{9}{2}\,\eta^2\hat{g}_0''- \frac{3}{2}\, \eta\hat{g}_0'+ \frac{9}{16}\hat{g}_0\right)  \nonumber \\
	&+ \frac{e^{-il\phi}}{\sqrt{l\phi}} \frac{i}{16 l^3 \phi^3} 	\left(10 \,\eta^4\hat{g}_0^{(4)}+100\, \eta^3\hat{g}_0 ^{(3)}+ 225\, \eta^2\,\hat{g}_0''- 75\, \eta\,\hat{g}_0'+ \frac{75}{8}\hat{g}_0\right)\label{psihex}
\end{align}
The second line is from the quadratic expansion of $e^{\frac{ilM^2}{8\phi^3}}$ and the third line is from the linear term in $e^{5ilM^4\over 128\phi^7}$. 
In the limit of large $l\phi$, we see from eq.\eqref{psihex} that the third line is smaller compared to the second. This is essentially just the fact that 
\begin{align}
	\frac{l^2M^4}{\phi^6} \gg \frac{lM^4}{\phi^7} \quad\iff\quad l\phi\gg 1\label{lphg1}
\end{align}
We can then write the wavefunction as
\begin{align}
	\hat{\Psi}&\simeq \frac{e^{-il\phi}}{\sqrt{l\phi}}\left(\hat{g}_0+\frac{i}{2l\phi}\left(-\eta^2\hat{g}_0''-\eta\,\hat{g}_0'+\frac{1}{4}\hat{g}_0\right) \right)\nonumber\\
	&- \frac{e^{-il\phi}}{\sqrt{l\phi}} \frac{1}{8 l^2 \phi^2}
	\left(\eta^4\hat{g}_0^{(4)}+6\,\eta^3\hat{g}_0 ^{(3)}+ \frac{9}{2}\,\eta^2\hat{g}_0''- \frac{3}{2}\, \eta\hat{g}_0'+ \frac{9}{16}\hat{g}_0\right) + \order\left({{(l \phi)^{-\frac{7}{2}}}}\right) \label{psitolgo}
\end{align}
where derivatives of the function $\hat{g}_0$ is with respect to its argument. 
Comparing with eq.\eqref{psihgn} we see that we indeed reproduce the leading terms in the WKB expansion with the identification that $g_0(\eta)=\hat{g}_0(\eta)$. 

\section{Norm of the Hartle-Hawking state}
\label{nrmhh}

In this appendix, we give an explanation of the divergence of the Hartle-Hawking wavefunction from the point of view of the path integral. The Hartle-Hawking wavefunction  corresponds to evaluating the path integral along a complex contour in the space of metrics, starting from north/south pole of the euclidean sphere and continuing at the equator  to  Lorentzian dS \cite{Maldacena:2019cbz,Moitra:2021uiv}. 
The norm of the state is then obtained by calculating the sphere partition function, since the Lorentzian sections which contribute an imaginary term in the exponent cancel out of $|\Psi|^2$. Alternatively, in JT gravity,  we can also  start with a disk in $-AdS_2$, see \cite{Moitra:2021uiv},  but we will not do so here and  consider the more conventional contour involving the euclidean sphere below. 

The resulting norm is then given by 
\be
\label{normh1a}
|\Psi|^2=\int {D[g] D[\phi] \over {\it \text{Vol}(\text{diff})}} e^{-S}
\ee
where the integral is over metrics of $S^2$ topology, ${\it \text{Vol}(\text{diff})}$ is the volume of diffeomorphism and $S$ is the JT action given in eq.\eqref{jtact}.
The path integral for JT gravity was analyzed in the second order formalism see \cite{Moitra:2021uiv}, we will use the analysis  of that paper to base  our discussion below and adopt mostly  the same notation here.

The measure for summing over metrics is related to the inner product in the space of metric deformations, this  is  of the ultra-local form and given by
\be
\label{inprods}
(\delta _1g, \delta_2 g)=\int d^2x \sqrt{g} g^{ac}g^{bd}\delta_1g_{ab} \delta_2g_{cd}
\ee
A general metric deformation can be decomposed as 
\be
\label{decoma}
\delta g_{ab}=\delta \sigma g_{ab} \oplus {\it range} P \oplus {\it Ker} P^\dagger
\ee
where $\sigma$ is the conformal factor and $P$ is an operator which maps vector fields $V_a$ to metric deformations and is given by 
\be
\label{defpp}
(PV)_{ab}=\nabla_a V_b+\nabla_b V_a -\nabla \cdot V g_{ab}
\ee
${\it Ker} P ^\dagger$ refers to the kernel of $P^\dagger$ which vanishes on $S^2$ due to absence of moduli. 
 
 Rotating the contour for the dilaton to lie along the imaginary axis, in the fashion which is now customary in the study of JT gravity,
 we get, upto finite normalisation factors which we will not be careful about  here, 
 \be
 \label{normb}
 |\Psi|^2= \int D[\sigma] {D[PV]\over {\it \text{Vol}(\text{diff})}}\delta (R-2) e^{-S}
 \ee
 Expanding the metric as 
 \be
 \label{expm}
 g_{ab}=e^{2\sigma} {\hat g}_{ab}
 \ee
 where ${\hat g}_{ab}$ is the round metric on the sphere with $R=2$
 we get 
 \be
 \label{expdel}
 \delta (R-2)=\delta(e^{-2\sigma}(2-2 {\hat \nabla}^2\sigma)-2)=\delta((-2)({\hat \nabla}^2+2)\delta \sigma)
 \ee
 where in the last step we have linearised in $\delta \sigma$. 
 
  The diffeomorphisms whose  volume appears in the denominator in eq.(\ref{normb})  include those generated by the Conformal Killing Vectors (CKVs) which are annihilated by $P$.
  The integral $\int D[PV]$ in contrast, is over   metrics which lie in the range of $P$, and therefore does not involve the CKVs. The two sets of diffeomorphisms are in this sense mismatched. 
  Carrying out the  integral $\int D[PV]$ now  gives
  \be
  \label{carrint}
  |\Psi|^2=\int D[\sigma]{\sqrt{\text{det}'({\hat P}^\dagger {\hat P})}\over \text{Vol}(\text{CKV}) } \delta((-2)({\hat \nabla}^2+2)\delta \sigma)
  \ee
  where the remaining factor of $\text{Vol}(\text{CKV})$ in the denominator arises because of the  mismatch mentioned above. 
  The hats over $P, P^\dagger$ and $\nabla^2$ indicate that the operators and determinants are defined with respect to the fiducial round metric ${\hat g}_{ab}$. 
  And the prime in $\text{det}'({\hat P}^\dagger {\hat P})$ indicates that only non-zero modes are to be included. 
  
 Finally the  last  integral over the Liouville mode can  be carried out next using the delta function.  Expanding $\delta \sigma$  in spherical harmonics
  \be
  \label{expdelsig}
  \delta \sigma = \sum_{l,m}c_{lm}Y_{lm}
  \ee
   we see that the delta function sets all the coefficients $c_{lm}=0$ except for the three with $l=1$. These three are related to conformal Killing vectors on the sphere. 
  There are six such CKVs, three of them are isometries and do not change $\sigma$, the remaining three correspond to the  $l=1$ zero modes 
  of $({\hat \nabla}^2+2)$.
  
  Carrying out the integrals over all the other $c_{lm}$ coefficients, besides the $l=1$ ones, then gives 
  \be
  \label{carrp}
  |\Psi|^2={\sqrt{\text{det}'({\hat P}^\dagger {\hat P})} \over \sqrt{\text{det}'({\hat \nabla}^2+2)}} {1\over \text{Vol}(\text{CKV})}
  \int \prod_{m=\pm1,0} dc_{1,m} \delta(0)
\ee
where the primes over the two determinants indicate that we are only including non-zero modes of ${\hat P}^\dagger {\hat P}$ and ${\hat \nabla}^2+2$. These determinants depend on the background metric ${\hat g}_{ab}$ but not on the CKVs.
The $\delta(0)$ factor stands for   the three delta  functions with zero arguments which arise for the three $l=1$ zero modes.
The integral over these zero modes, done more correctly  away from the linearised approximation in $\sigma$ we considered above, is an integral over the manifold 
$SL(2,C)/SU(2)$. This integral therefore gives a factor of $\text{Vol}(SL(2,C)/SU(2))$ in the numerator of eq.(\ref{carrp}) which cancels the divergence in the denominator due to $\text{Vol}(\text{CKV})$,  with the ratio
$\text{Vol}(SL(2,C)/SU(2))/\text{Vol}(\text{CKV})$  now being  finite and equal to 
\be
\label{rataaha}
{\text{Vol}(SL(2,C)/SU(2)) \over \text{Vol}(\text{CKV})}= {1\over \text{Vol}(SU(2))}
\ee
Eq.(\ref{carrp}) then takes the form 
\be
\label{formfi}
|\Psi|^2={\sqrt{\text{det}'({\hat P}^\dagger {\hat P})} \over \sqrt{\text{det}'({\hat \nabla}^2+2)}} {1\over \text{Vol}(SU(2))}
  \delta(0)
  \ee

We see finally that the three delta functions with zero argument  still remain in the numerator. These are divergent.
 As a result  the norm $|\Psi|^2$ diverges.

}

\section{More on Gaussian example}
\label{sadapp}
In this appendix, we provide some more details of the Gaussian example discussed in subsection \ref{gausscf} where the saddle point analysis was presented for the region $l\sigma\ll \phi^3$. Here we show the saddle point analysis for the region $\l\sigma\gtrsim\phi^3$.  First consider the region $l\sigma\gg \phi^3$. In this region, we can self-consistently find a saddle point by  assuming $M\gg \phi^2$. Also, $x_0$ can be dropped in eq.\eqref{spt1} when eq.\eqref{condaga} is satisfied. Doing so, the saddle point equation becomes
\begin{align}
	M=-  \sigma \left(  {l\over 2 \sqrt{M}}\right)\label{sadptlpl}
\end{align}
and we get the saddle point to be
\begin{align}
	M_0\simeq re^{\frac{2i\pi}{3}},\quad r=\left(\frac{l\sigma}{2}\right)^{\frac{2}{3}}\label{llaphsad}
\end{align}
As can be seen from eq.\eqref{sadptlpl}, the solutions involves cubic roots of (-1). The solution in eq.\eqref{llaphsad} is only one of the solutions. In general, in doing the saddle point analysis, we should consider all saddle points that are compatible with the given contour for the integration variable, {\it i.e.} all the saddle points through which the contour can be deformed smoothly to pass through without encountering singularties or branch cuts. In our case, the contour is as shown in fig.\ref{mbranch} and it can be verified that the only saddle point that this contour can be deformed to go through is the one shown above in eq.\eqref{llaphsad}.

 The justification for dropping $x_0$ in eq.\eqref{spt1} can now be understood by noting that for the saddle point  in eq.\eqref{llaphsad}, we find $\abs{\frac{l\sigma}{\sqrt{M_0}}}\sim\abs{M_0}\gg \phi^2 $.  Note that $x_0$ can be dropped in finding the saddle point value eq.\eqref{sadptlpl} if
 \begin{align}
 \abs{\frac{l\sigma}{\sqrt{M_0}}}\sim\abs{M_0}\gg \sigma x_0\Rightarrow	l\gg \sqrt{\sigma} \label{lsgsqs}
 \end{align}
is met. It is easy to see that the above condition is less restrictive than $\phi^2\gg \sigma$ when $l\sigma\gg \phi^3$. If the condition eq.\eqref{condaga} is met then the condition in eq.\eqref{lsgsqs} is automatically met justifying the dropping of $x_0$ in obtaining the saddle point value. However, if we ignore $x_0$ totally, the wavefunction will be zero due to equal and opposite contributions from the two terms in eq.\eqref{psi2g}. This means that we have to work to leading order in $x_0$. Since, in the region under consideration $l\sigma\gg \phi^3$, the leading contribution involving $x_0$ comes from the on-shell where we use the saddle point value for coefficient of $x_0$ in eq.\eqref{psig}.  In this approximation, the wavefunction is given by 
\begin{align}
	\hat{\Psi}=\frac{1}{l}e^{S_0}\sin(M_0x_0),\quad S_0\simeq-\frac{M_0^2}{2\sigma}-il\sqrt{-M_0}\label{psilllphi3}
\end{align}
For the saddle point in eq.\eqref{llaphsad}, with the convention eq.\eqref{defa},\eqref{defs}, the value of the saddle point exponent is given by 
\begin{align}
	S_0=-\frac{3}{2\sigma}\left(\frac{l\sigma}{2}\right)^{\frac{4}{3}}e^{i\frac{\pi}{3}}
	\label{sadptval}
\end{align}

\subsection{Validity of the saddle point approximation}
\label{valdsadap}
We shall now show that the saddle point analysis carried out in the example in subsection \ref{gausscf} is indeed a good approximation. Consider a wavefunction of the form  
\begin{align}
	\Psi\sim \int dM e^{S(M)}\label{psiinSexp}
\end{align}
Around a saddle point $M_0$, we have the expansion
\begin{align}
	S(M)=S(M_0)+\frac{1}{2}S''(M_0)(M-M_0)^2+\frac{1}{6}S'''(M_0)(M-M_0)^3+\dots\label{sadexp}
\end{align}
where dots denote the higher order terms. To justify the saddle point analysis where we neglect terms starting from $S'''(M_0)$, we need to show that
\begin{align}
	\abs{\frac{S'''(M)(M-M_0)}{S''(M_0)}}\simeq\abs{\frac{S'''(M)}{(S''(M_0))^{\frac{3}{2}}}}\ll 1\label{sadjustcond}
\end{align}

We shall show this is the case in various regions of $l$. For the wavefunction considered in subsection \ref{gausscf}, the exponent is given by 
\begin{align}
	S=-\frac{M^2}{2\sigma}\pm i M x_0 -il \sqrt{\phi^2-M}\label{expgaus}
\end{align}
Let us analyze for the $+$ sign above as the other case is similar. 
 The second third derivatives are  then given by 
\begin{align}
	S''(M)=-\frac{1}{\sigma}+\frac{il}{4(\sqrt{{\phi}^2-M})^3},\quad S'''(M)=\frac{3i l}{8(\sqrt{{\phi}^2-M})^5}\label{s2s3p}
\end{align}
First, consider the region $l\sigma\gg \phi^3$. 
The
 saddle point value in this region is given by eq.\eqref{llaphsad} using which we get 
\begin{align}
	S''(M_0)=-\frac{3}{2\sigma}, \quad S'''(M_0)=\frac{3il}{8}\left(\frac{l\sigma}{2}\right)^{-\frac{5}{3}}e^{5i\pi\over 6}\label{s3s2lgph3}
\end{align}
and thus we get
\begin{align}
	\abs{\frac{S'''(M)}{(S''(M_0))^{\frac{3}{2}}}}\simeq \order{\left(\frac{\sqrt{\sigma}}{(l\sigma)^{\frac{2}{3}}}\right)}\ll 1\label{lpggphsadco}
\end{align}

For the region $l\sigma\sim\order{( \phi^3)}$,   $x_0$ can still be treated as  perturbation. In this regime the saddle point value is $M_0\sim\order{( \phi^2)}$. The fact that $x_0$ can be ignored  in finding the saddle point is again related to the condition eq.\eqref{condaga}. So, we have
\begin{align}
	S''(M_0)=\frac{1}{\sigma}\left(-1+\order{\left( \frac{l\sigma}{\phi^3}\right)}\right)\sim\order{\left(\frac{1}{
			\sigma}\right)},\quad S'''(M_0)\sim\order{\left(\frac{1}{\sigma\phi^2}\right)}\label{s2s3leqph3}
\end{align}
and thus we get
\begin{align}
	\abs{\frac{S'''(M)}{(S''(M_0))^{\frac{3}{2}}}}\simeq \order{\left(\frac{\sqrt{\sigma}}{{\phi^2}}\right)}\ll 1\label{lporphsadco}
\end{align}
where the last inequality above follows from the condition eq.\eqref{condaga}.
The reader may be worried at this point since there can be a cancellation of two term in $S''(M_0)$ in eq.\eqref{s2s3leqph3}. We can show that this cannot happen. If this were to happen, then both $S'(M_0)=S''(M_0)=0$. From the condition that $S''(M_0)=0$,we find that 
\begin{align}
	\sqrt{\phi^2-M_0}=\left(\frac{il\sigma}{4}\right)^{1/3}\label{phims2}
\end{align}
 Substituting this in the condition $S'(M)=0$,  eq.\eqref{spt1}, neglecting $x_0$ we get that 
.\begin{align}
	M_0= 2\left(\frac{il\sigma}{4}\right)^{2/3} \label{zeta0sdzer}
\end{align}
The two conditions eq.\eqref{phims2} and eq.\eqref{zeta0sdzer} are incompatible since if both are true we can use eq.\eqref{zeta0sdzer} in eq.\eqref{phims2} to find
\begin{align}
	M_0=\frac{2}{3}\phi^2
\end{align}	
which is incompatible with eq.\eqref{zeta0sdzer} since $i^\frac{2}{3}$ is complex.  Thus, we can safely assume that $S''(M_0)\neq 0$.

Let us now analyze the last remaining region in $l$, $l\sigma\ll \phi^3$. In this region, we had expanded ${\phi}^2-M$ in a series in $\frac{M}{{\phi}^2}$ and found a self-consistent saddle point by keeping only the linear term in $M$. It is easy to see that had we kept higher order terms in this expansion, we would get a contribution to $S'''$ given by 
\begin{align}
	S'''(M_0)\sim\order{\left(\frac{l}{{\phi}^5}\right)},\quad S''(M_0)\sim-\frac{1}{\sigma}\label{s3s2lllphi3}
\end{align}
Thus in this region again we get
\begin{align}
	\frac{S'''(M_0)}{S''(M_0)^\frac{3}{2}}\sim\order{\left(\frac{l\sigma}{{\phi}^3}\frac{\sqrt{\sigma}}{{\phi}^2}\right)}\ll 1\label{s3s2ll1lllph}
\end{align}
Thus we find that in all the regions of $l$, the condition eq.\eqref{sadjustcond} is satisfied justifying our use of the saddle point approximations.

	\subsection{Norm Conservation}
	\label{normconss}
	Consider the wave function given in eq.\eqref{pshrhsh}. Let us rewrite it here for convenience.
	\be
	\hat{\Psi}={1\over l}  \int_{-\infty}^{\phi^2} dM \,\rho(M)\,\, e^{-il\sqrt{\phi^2-M}} + \frac{1}{l} \int_{\phi^2}^\infty dM \rho_1(M) e^{-l \sqrt{M-\phi^2}} \label{pshrhsh2}
	\ee
	with $\rho(M), \rho_1(M) $ given by 
	\begin{eqnarray}
		\rho(M) & = & {2  \sin (M x_0)}\,e^{-\frac{M^2}{2\sigma} }, \ M\in [-\infty,\infty] \\
		\rho_1(M) & = & {2  \sin (M x_0)}\,e^{-\frac{M^2}{2\sigma} }, \ M \in [0,\infty]
		\label{rhomgsa2}
	\end{eqnarray}
	For norm to be conserved at all values of $\phi$, $\mathcal{C}_N$, eq.\eqref{normccond}, reproduced below, must vanish at $l=0,\infty$ for all values of $\phi$.
	\begin{equation}
		\mathcal{C}_N= i ( \hat{\Psi}\del_{\phi}\hat{\Psi}^*{}\,-\hat{\Psi}^*{\del_{\phi}}\hat{\Psi} )=0 \quad \text{at}\,\,{ l=0,\infty}
	\end{equation}
	Since,
	\begin{equation}
		\int_{-\infty}^{\infty} \rho(M) dM=0 \label{rhocond}
	\end{equation}
	at $l=0$ we get the following term,
	\begin{equation*}
		\mathcal{C}_N=  2 \phi \int_{\phi^2}^{\infty} \rho(M) \frac{1}{\sqrt{M-\phi^2}}  \int_{-\infty}^{\phi^2} \rho(M_1) \sqrt{\phi^2-M_1}  + 2\phi \int_{\phi^2}^{\infty} \rho(M) \sqrt{M-\phi^2} \int_{-\infty}^{\phi^2} \rho(M_1) \frac{1}{\sqrt{\phi^2-M_1}}
	\end{equation*}
	This term in general does not vanish. However in the limit $\phi^2 \gg \sigma$, in the integrals of the form $\int_{-\infty} ^{\phi^2}$, we can expand the square root since $\rho(M)$ has support only around $M=0$ as can be seen from eq.\eqref{rhomgsa2} and we then have,
	\begin{equation*}
		\mathcal{C}_N \approx  2 \phi^2 \int_{\phi^2}^{\infty} \rho(M) \frac{1}{\sqrt{M-\phi^2}}  \int_{-\infty}^{\phi^2} \rho(M_1) + 2 \int_{\phi^2}^{\infty} \rho(M) \sqrt{M-\phi^2} \int_{-\infty}^{\phi^2} \rho(M_1) 
	\end{equation*}
	Then, to leading order,
	\begin{equation*}
		\int_{-\infty}^{\phi^2} \rho(M_1) \sim 2 \sin(\phi^2 x_0) e^{-\frac{\phi^4}{2 \sigma}} \frac{\sigma}{\phi^2}
	\end{equation*}
	Thus we have,
	\begin{equation}
		\mathcal{C}_N \approx  4 \sigma \sin(\phi^2 x_0) e^{-\frac{\phi^4}{2 \sigma}}\int_{\phi^2}^{\infty} \rho(M) \frac{1}{\sqrt{M-\phi^2}} 
	\end{equation}
	
	Using the fact that $\rho(M)$ within the integral will be bounded by $|\rho(M)|<\sim e^{-\phi^4\over 2\sigma}$ we get that 
\be
\label{bcna}
{\cal C}_N\lesssim e^{-\phi^4\over \sigma}
\ee
and therefore the rate of change of the norm is also exponentially supressed going like
${\partial {\cal N}\over \partial \phi}\sim e^{-\phi^4\over \sigma}$.

	Furthermore using some reasonable values of $\sigma, x_0, \phi$ such that the limits $\phi^2 \gg \sigma, \sigma \gg 1, x_0 \sim \order{(1)}$ are obeyed one can also numerically verify the above conclusions. More generally, as can be seen from the above analysis, for $\rho$ that decay sufficiently fast at large $\phi$, the norm is conserved approximately for such large values of $\phi$.

\section{Rindler basis expansion}
\label{rindlerapp}

In this appendix, we give more details regarding the Rindler basis expansion of the wavefunction mentioned in section \ref{gensols}. We rewrite general solution for the wavefunction written in the Rindler basis eq.\eqref{exppsi} here for convenience.
\be
{\hat \Psi}=\int_{-\infty}^\infty dk \bigl[a(k) e^{i k \theta}J_{-i |k|}( \xi)+ b(k) e^{-i k \theta}J_{i |k|}( \xi)\bigr ]
\ee
At both the horizons, $\mathcal{H}_1,\mathcal{H}_2$, the coordinate $\xi\rightarrow 0$ and so we can expand the Bessel functions using the small argument expansion
\begin{align}
	J_\alpha(\xi)\sim \xi^\alpha+\dots\label{bessmar}
\end{align}
Using this, the wave function near $\xi\rightarrow 0$ becomes, 
\begin{align}
	\hat{\Psi}&=\int_{-\infty}^{\infty} dk \left(a(k) e^{ik\theta} \xi^{-i\abs{k}}+b(k) e^{-ik\theta}\xi^{i\abs{k}}\right)\nonumber\\
	&=\int_{-\infty}^0 dk\left(a(k)e^{iky_+}+b(k)e^{-iky_+}\right)+\int_{0}^{\infty} dk\left(a(k)e^{-iky_-}+b(k)e^{iky_-}\right) \nonumber \\
	&=\int_{-\infty}^{\infty} dk \Theta(-k) \left(a(k)e^{iky_+}+b(k)e^{-iky_+}\right)+\int_{-\infty}^{\infty} dk \Theta(k)\left(a(k)e^{-iky_-}+b(k)e^{iky_-}\right) \label{psiypym}
\end{align}
where 
\begin{align}
	y_+=\ln \xi\,+\theta=\ln v=2\ln\phi\nonumber\\
	y_-=\ln \xi\,-\theta=\ln u=2\ln l\label{ypm}
\end{align}
The operators $a(k), b(k)$ can be obtained by the usual inverse Fourier transform. Doing so, near $\xi\rightarrow 0$, we get
\begin{align}
	\int_{-\infty}^{\infty}dy_+ \, e^{iky_+}\,\del_{y^+}\hat{\Psi}=-ik\,\Theta(k)a({-k})-ik\,\Theta(-k)b(k)\nonumber\\
	\int_{-\infty}^{\infty}dy_- \, e^{iky_-}\,\del_{y^-}\hat{\Psi}=-ik\,\Theta(k)a({k})-ik\,\Theta(-k)b({-k})\label{psiakbk}
\end{align}
Let us now consider a wavefunction of the form eq.\eqref{mpsisol}. We have
\begin{align}
	&\del_{y^+}\hat{\Psi}\bigg\vert_{\mathcal{H}_1}=\frac{\phi}{2}\del_\phi\hat{\Psi}\bigg\vert_{l\rightarrow 0}=-\frac{i\phi^2}{2\sqrt{\phi^2-M}}\nonumber\\
	&\del_{y^-}\hat{\Psi}\bigg\vert_{\mathcal{H}_2}=\frac{l}{2}\del_l\hat{\Psi}\bigg\vert_{\phi\rightarrow 0}=-\frac{1}{2l}\left({il\sqrt{-M}}+1\right)e^{-il\sqrt{-M}}\label{delypymph}
\end{align}
Let us take $M<0$ for a moment. Then 
\begin{align}
	\int_{-\infty}^{\infty}dy_+e^{iky_+}\del_{y^+}\hat{\Psi}=-i\int_{-\infty}^{\infty}d\phi' \frac{e^{2ik\phi'+2\phi'}}{\sqrt{e^{2\phi'}-M}}\label{delyp}
\end{align}
where we have done a variable change $\phi'=\ln\phi$. This integral can be done exactly on mathematica and gives a closed form expression in terms of $\beta$ function and ${}_2F_1$. For the $y_-$ integral, we have
\begin{align}
	\int_{-\infty}^{\infty}dy_- \, e^{iky_-}\,\del_{y^-}\hat{\Psi}&=-\int_0^{\infty}dl \,e^{-il\sqrt{-M}}l^{2ik-2}(1+il\sqrt{-M}) \nonumber\\
	&=-\frac{1}{(i\sqrt{-M})^{2ik-1}}\left(\Gamma(2ik)+\Gamma(2ik-1)\right)
	\label{delymin}
\end{align}
The results eq.\eqref{delyp}, eq.\eqref{delymin} combined with eq.\eqref{psiakbk} can be used to read off $a(k),b(k)$. The analysis of a more general wavefunction eq.\eqref{gnthefn} can also be done in a similar manner to obtain $a(k),b(k)$ in terms of $\rho,\tilde{\rho},\rho_1,\rho_2$.

\section{Disk with  matter}
\label{diswmat}
In this appendix, we consider addition of local bulk matter to JT gravity and comment on the quantization of the combined system.
The full action is then given by 
\begin{align}
	I=I_{\text{JT}}+I_\text{M}\label{JT+matca}
\end{align}
with $I_{\text{JT}}$ given by eq.\eqref{jtact}.  
For convenience, let us consider bosonic conformal matter fields with the action
\begin{align}
	I_\text{M}=\frac{1}{2}\int d^2x \,\sqrt{-g}(\del\varphi)^2\label{scmt}
\end{align}
To handle matter fields, we find it convenient to work in the conformal gauge as opposed to the ADM gauge used in previous sections. 

Consider the metric in the conformal gauge, 
\begin{align}
	ds^2=-e^{2\rho}dx^+ dx^-=-e^{2\rho}(dt^2-dx^2)\label{confg}
\end{align}
where $x^\pm=t\pm x$. The JT action, eq.\eqref{jtact}, in this gauge becomes,
\begin{align}
	I_{\text{JT}}&=\frac{1}{8\pi G}\pqty{\frac{1}{2}\int d^2x \sqrt{-g}\phi(R-2)-\int_{\del}\phi(K-1)}\nonumber\\
	&=\frac{-1}{8\pi G}\int d^2x \pqty{\dot{\rho}  \dot{\phi}-\rho' \phi'+\phi e^{2\rho}}\label{jtinconfg}
\end{align}
where overdots and primes denote derivatives with respect to $t$ and $x$ respectively. 
The canonical momentum for the $\rho$ and $\phi$ are given by 
\begin{align}
	\pi_\rho=-\frac{\dot{\phi}}{8\pi G},\,\,\pi_\phi=-\frac{\dot{\rho}}{8\pi G}\label{piconfs}
\end{align}

The constraint equations in the conformal gauge are given by $T_{++},T_{--}$. The remaining component $T_{+-}$ is a dynamical equation. For the gravitational part these are given by 
\begin{align}
	&T_{++}^G= \nabla_{+} \partial_+ \phi = \partial_+ ^2 \phi - 2 \partial_+ \rho \partial_+ \phi \nonumber\\
	&T_{--}^G= \nabla_{-} \partial_- \phi = \partial_- ^2 \phi - 2 \partial_- \rho \partial_- \phi\nonumber\\
	&T_{+-}^G= -\partial_+ \partial_- \phi + \frac{1}{2} e^{2 \rho} \phi
	\label{Tcomps}
\end{align}
We can also choose to impose a linear combination of these. The particular linear combination that corresponds to the Hamiltonian constraint in the ADM gauge is the $T_{tt}^G$ given by
\begin{align}
	T_{tt}^G= T_{++}^G + T_{--}^G + 2 T_{+-}^G = \phi'' + e^{2 \rho} \phi - \dot{\phi} \dot{\rho} - \rho' \phi' \label{Tttgrav}
\end{align}
where superscript $G$ indicates that it corresponds to the gravitational part of the theory. Note that in the equation above we changed $\partial_{\pm}$ into $\partial_{t,x}$ while writing the final equality.

Similarly, for the matter fields corresponding to the action eq.\eqref{scmt}, in the classical theory  it is given by, the conjugate momentum for the matter field is given by 
\begin{align}
	\pi_\varphi=\dot{\varphi}\label{matconjmom}
\end{align}
and the $T_{tt}$ component is given by 
\begin{align}
T_{tt}^M=T_{++}^M + T_{--}^M + 2 T_{+-}^M=\frac{1}{2}\pqty{\dot{\varphi}^2+\varphi'^2}=\frac{1}{2}\pqty{\pi_\varphi^2+\varphi'^2}\label{mathden}
\end{align}
However, the corresponding expression for the quantum problem in the presence of dynamical gravity will be different. 
The combined $T_{tt}$ is given by ,
\begin{align}
T_{tt}=-\frac{1}{8\pi G}\pqty{(8\pi G)^2\pi_\phi\pi_\rho+\rho'\phi'-\phi e^{2\rho} - \phi''}+T_{tt}^M\label{fullhden}
\end{align}
So, the constraint that needs to be satisfied by the physical states is given by the operator equation
\begin{align}
	\left(-\frac{1}{8\pi G}\pqty{(8\pi G)^2\pi_\phi\pi_\rho+\rho'\phi'-\phi e^{2\rho} - \phi''}+T_{tt}^M\right)|\hat{\Psi}\rangle =0\label{psiop}
\end{align}
The above equation is now non-trivial to solve. Some simplification can be made by making a judicious choice of slice where $\phi'=0=\rho'$. The possibility of such a choice is discussed below. However, we leave a more detailed study of the solutions of the states that satisfy the constraint eq.\eqref{psiop} for the future \cite{nanda:2023yta}.

\subsection{Consistency of conformal gauge and constant dilaton slice}
\label{confmatgc}

In this appendix, we try to show that the choice of $\phi'=0=\rho'$ is consistent with the conformal gauge choice used in the analysis of WDW wavefunction in the presence of matter. In other words, we show that any one particular time slice can be chosen in such a way that both the dilaton and the conformal factor are constant on the chosen time slice. This, in particular, will be useful if we know the operator form of the matter stress tensor precisely. It will reduce the complexity of the WDW equation in the presence of matter as we can choose a nice slice on which $\phi'=0=\rho'$, classically and then turn the resulting equation into an operator form.

Suppose, initally, $\phi$ is not constant on the chosen time slice, say $t=t_0$. This means that $\del_x\phi(t_0,x)\neq 0$. Let us, without any loss of generality, assume that $\del_x\phi(t_0,x)\ll 1$  so that we can work with infinitesimal coordinate transformation. We show that we can always do an infinitesimal coordinate transformation which keep us in conformal gauge but alter the time slice slightly so that on the modified time slice, $\phi$ is a constant. 

Let us do an infinitesimal coordinate transformation,

\begin{align}
	&x^\pm=\tilde{x}^\pm + \epsilon^\pm(\tilde{x}^\pm)\nonumber\\
	\Rightarrow \,\,&t=\tilde{t}+\epsilon^{t}(\tilde{t},\tilde{x}), \,\,x=\tilde{x}+\epsilon^{x}(\tilde{t},\tilde{x})\nonumber\\	\text{where}\,\,& \epsilon^{t}=\frac{1}{2}(\epsilon^++\epsilon^-),\,\,\quad \epsilon^{x}=\frac{1}{2}(\epsilon^+-\epsilon^-)	\nonumber\\
	&\tilde{x}^\pm=\tilde{t}\pm \tilde{x}\label{txtil}
\end{align}
Now, on the time slice $\tilde{t}=t_0$, we would like to make $\phi$ to be constant by choosing $\epsilon^\pm$ appropriately. In other words, we want to impose,
\begin{align}
	\del_{\tilde{x}}\phi(\tilde{t},\tilde{x})\big\vert_{\tilde{t}=t_0}=0\label{codicon}
\end{align}
Using the chain rule of partial derivatives and expanding the above in terms of derivatives of $t,x$, we get
\begin{align}
	\pqty{\frac{\del t}{\del \tilde{x}}\del_t\phi+\frac{\del x}{\del \tilde{x}}\del_x\phi}\bigg\vert_{\tilde{t}=t_0}=0\label{dilconcon}
\end{align}
From eq.\eqref{txtil}, we get,
\begin{align}
	&\frac{\del t}{\del \tilde{x}}=\frac{1}{2}\del_{\tilde{x}}(\epsilon^+(\tilde{x}^+)+\epsilon^-(\tilde{x}^-))=\frac{1}{2}(\epsilon^+{}'- \epsilon^-{}')\nonumber\\
	&\frac{\del x}{\del \tilde{x}}=1+\frac{1}{2}\del_{\tilde{x}}(\epsilon^+(\tilde{x}^+)-\epsilon^-(\tilde{x}^-))=1+\frac{1}{2}(\epsilon^+{}'+\epsilon^-{}')	\label{dtdtil}
\end{align}
where the primes above denote derivatives with respect to the arguments. The condition eq.\eqref{dtdtil}, to leading order in $\epsilon$, then becomes
\begin{align}
	\pqty{\epsilon^+{}'-\epsilon^-{}'+2\frac{\del_x\phi}{\del_t\phi}}\bigg\vert_{\tilde{t}=t_0}=0\label{epsphcon}
\end{align}
So, by choosing $\epsilon^\pm$ to  satisfy the condition above, we can set dilaton to be constant on the time slice $\tilde{t}=t_0$. Having done this, we now show that we can do a further infinitesimal coordinate transformation to now achieve $\del_x\rho=0$. Let us first relabel the coordinates $(\tilde{t},\tilde{x})$ to $(t,x)$.  Suppose, to begin with $\del_x\rho\neq 0$ but small. Having set $\del_x\phi=0$, we want to do another coordinate of the form eq.\eqref{txtil} but with the condition that $\del_x\phi$ remains zero. This means that the new coordinate transformation should satisfy the analog of eq.\eqref{epsphcon} with $\del_x\phi=0$ so that the condition becomes,
\begin{align}
	\pqty{\epsilon^+{}'(\tilde{x}^+)-\epsilon^-{}'(\tilde{x}^-)}\bigg\vert_{\tilde{t}=t_0}=0\label{seepscon}
\end{align}
To begin with, the metric was given by 
\begin{align}
	ds^2=-e^{2\rho}dx^+ dx^-\label{confg22}
\end{align}
Doing a coordinate transformation of the form eq.\eqref{txtil} then gives the metric to be
\begin{align}
	&ds^2= - e^{2\tilde{\rho}}d\tilde{x}^+d\tilde{x}^-\nonumber\\
	\text{where}&\quad \tilde{\rho}(\tilde{x}^+,\tilde{x}^-)=\rho(x^+(\tilde{x}^+),x^-(\tilde{x}^+))+\epsilon^+{}'(\tilde{x}^+)+\epsilon^-{}'(\tilde{x}^-)
	\label{tilmiet}
\end{align}
Thus, the condition that $\del_{\tilde{x}}\tilde{\rho}\big\vert_{\tilde{t}=t_0}=0$ becomes
\begin{align}
	\del_{\tilde{x}}\tilde{\rho}\big\vert_{\tilde{t}=t_0}=0\Rightarrow \pqty{\frac{\del t}{\del \tilde{x}}\del_t\rho+\frac{\del x}{\del \tilde{x}}\del_x\rho+\epsilon^+{}''-\epsilon^-{}''}\bigg\vert_{\tilde{t}=t_0}=0\label{rhopzco}
\end{align}
From eq.\eqref{seepscon}, it follows that 
\begin{align}
	&\epsilon^-{}'(\tilde{x}^-)=\epsilon^+{}'(2\tilde{t}_0-\tilde{x}^-)\nonumber\\
	\Rightarrow & \epsilon^-{}''(\tilde{x}^-)=-\epsilon^+{}''(2\tilde{t}_0-\tilde{x}^-)=-\epsilon^+{}''(x^+)\big\vert_{\tilde{t}=t_0}\label{epssecdc}
\end{align}
Using eq.\eqref{dtdtil},eq.\eqref{seepscon} and eq.\eqref{epssecdc} in eq.\eqref{rhopzco}, we get
\begin{align}
	\pqty{\epsilon^+{}''(\tilde{x}^+)+\frac{1}{2}\del_x\rho}\bigg\vert_{\tilde{t}=t_0}=0\label{epiddc}
\end{align}
Thus, we see that we can choose an appropriate $\epsilon^+$ so that above condition is satisfied which in turn means that we can set the conformal factor to be constant on the appropriate time slice. So, in effect, by choosing $\epsilon^+,\epsilon^-$ appropriately, we see that we can set $\phi'=0=\rho'$ on any particular time slice. 
\section{Double trumpet}
\label{discondb}

In this appendix, we give some details regarding the evaluation of the density matrix for the single universe sector from the two universe density matrix discussed in section \ref{dtprop}. Following this, we also show how the gluing of a state to the propagator can only be carried out in the far past. 

\subsection{Evaluation of the reduced density matrix}
{ The contribution to the reduced density matrix for the single universe,{ by tracing out over one universe in a two-universe density matrix is given by, eq.\eqref{dfu},}}

\be
\label{dfu1}
{\hat  \rho}(l_1, {\tilde l}_1, \phi)=\int dl_2 ({\hat G}^{--}_{2} (l_1,\phi, l_2,\phi_2))^* i  \overleftrightarrow{\partial_{l_2}} {\hat G_{2}}^{--}({\tilde l}_1,\phi, l_2,\phi_2)
\ee
where
\begin{align}
	\hat{G}_{2}^{--}(l_1, \phi_1, l_2,\phi_2)&=\int b db\hat{\Psi}^-_T(l_1, \phi_1, b) \hat{\Psi}^-_T(l_2, \phi_2, b)\nonumber\\
	\hat{\Psi}_T^-(l,\phi,b)&={ \phi \over \sqrt{l^2+b^2}}  H^{(2)}_1\left(\phi\sqrt{l^2+b^2}\right) \label{g2mm}
\end{align}
So, the reduced density matrix is given by 
\begin{align}
	{\hat  \rho}(l_1, {\tilde l}_1, \phi)=i\int b_1 db_1 \,b_2db_2\,\hat{\Psi}^+_T(l_1, \phi, b_1)\hat{\Psi}^-_T(\tilde{l}_1, \phi, b_2)\int dl_2\left(\hat{\Psi}^+_T(l_2, \phi_2, b_1)\overleftrightarrow{\del_{l_2}}\hat{\Psi}^-_T(l_2, \phi_2, b_2)\right)\label{rinpsih}
\end{align}
The region of large $l_2$ in the above integral does not have a divergence. This is because in this region, using the large argument expansion of the Hankel functions,
\begin{align}
	H_{m}^{(1)}(T)&\sim \sqrt{\frac{2}{\pi T }}e^{i(T-\frac{m\pi}{2}-\frac{\pi}{4})}\quad \quad \text{for}\quad -\pi<\text{arg}(T)<2\pi\nonumber\\
	H_{m}^{(2)}(T)&\sim \sqrt{\frac{2}{\pi T }}e^{-i(T-\frac{m\pi}{2}-\frac{\pi}{4})}\quad \quad \text{for}\quad -2\pi<\text{arg}(T)<\pi\label{hankasym}
\end{align}
it can be seen that the $l_2$ integral has the form $\int \frac{dl_2}{l_2^4}$ which is finite.

Since the integral is regular in the large $l_2$ limit, to check for the finiteness of eq.\eqref{rinpsih}, we then focus on the regions where $l,b\ll 1$. More precisely, the region of our interest is
\begin{align}
	\phi_2\sqrt{l_2^2+b_1^2}\ll 1,\, 	\phi_2\sqrt{l_2^2+b_2^2}\ll 1\label{lphsmcond}
\end{align} 
In this region, the $l_2$ integral in eq.\eqref{rinpsih} can be approximated as
\begin{align}
	\int_0 dl_2\left(\hat{\Psi}^+_T(l_2, \phi_2, b_1)\overleftrightarrow{\del_{l_2}}\hat{\Psi}^-_T(l_2, \phi_2, b_2)\right)\sim \frac{4 \left(b_1^4+2 b_1^2 b_2^2 \left(\ln \left(b_2^2\right)-\ln \left(b_1^2\right)\right)-b_2^4\right)}{\pi ^2 b_1^2 b_2^2 \left(b_1^2-b_2^2\right)^2}+\order\left(l_2^2\right)\label{l2int}
\end{align}


Now, depending on the order of the evaluation of the integral corresponding to $b_1,b_2$ for finite values of $l_1,\tilde{l}_1,\phi$, we get divergences in $\hat{\rho}$ of the form
\begin{align}
	a_1 \ln(b_1)^2+	a_2 \ln(b_2)^2+a_3  \ln(b_1)\ln(b_2)\label{b1b2int}
\end{align}
where $a_1,a_2,a_3$ are some $\order{(1)}$ constants.

\subsection{Dependence on instant of gluing}
\label{depinsgl}
Let us now mention some details about gluing an initial state to a propagator as mentioned in eq.\eqref{ffs}.
The final state wavefunction defined by eq.\eqref{ffs} should satisfy certain criteria to be considered as a reasonable state of the universe at late times.
First of all, a natural requirement for the above wavefunction is that it be independent of the time of gluing of the initial wavefunction with the propagator. Mathematically, this means that
\begin{align}
	\del_{\phi_i}\hat{\Psi}_f(\phi_f,l_f)=0\label{phiidc}
\end{align}
Further, we require that this final state wavefunction should have a finite and conserved norm as in the case of single boundary wavefunction. Let us consider each of these requirements in turn. To reduce clutter, we  denote the propagator $\hat{G}_2^{-+}$ as just $\hat{G}$.
To check eq.\eqref{phiidc}, 
using the fact that both $\hat{G}$ and $\hat{\Psi}_i$ satisfy WDW equation, we find
\begin{align}
	\del_{\phi_i}\hat{\Psi}_f(\phi_f,l_f)&=i\int dl_i\,( \del_{\phi_i}\hat{G}{\del_{l_i}}\hat{\Psi}_i+\hat{G}{\del_{\phi_i}\del_{l_i}}\hat{\Psi}_i- \del_{\phi_i}\del_{l_i}\hat{G}\hat{\Psi}_i-\del_{l_i}\hat{G}\,{\del_{\phi_i}}\hat{\Psi}_i )\nonumber\\
	&=i\int dl_i ( \del_{\phi_i}\hat{G}{\del_{l_i}}\hat{\Psi}_i+{{\phi_i}{l_i}}\hat{G}\hat{\Psi}_i- {\phi_i}{l_i}\hat{G}\hat{\Psi}_i-\del_{l_i}\hat{G}{\del_{\phi_i}}\hat{\Psi}_i )\nonumber\\
	&=i ( \hat{\Psi}_i\del_{\phi_i}\hat{G}{}\,-\hat{G}{\del_{\phi_i}}\hat{\Psi}_i )\bigg\vert_{l_i=0}^{\infty}\label{glucondi}
\end{align}
where in obtaining the first line from the second we used the fact that $\hat{G},\hat{\Psi}$ satisfies the WDW equation and the third line is obtained from the second by an integration by parts and using WDW equation again. 
Thus, the requirement that the final wavefunction be independent of the instant of gluing translates into the condition that the above boundary term vanishes. We shall see that this boundary term generally does not vanish for arbitrary $\phi_i$ but only vanishes in the limit of $\phi_i\rightarrow \infty$. Thus, we shall be forced to glue the initial wavefunction to the propagator only at $\phi_i\rightarrow\infty$. First consider the upper limit of $l_i\rightarrow \infty$. We only consider initial state wavefunctions that die sufficiently fast in this limit so that the contribution to the boundary term from $l_i\rightarrow \infty$ vanishes. Then the contribution from the $l_i=0$ end remains to be evaluated.

To analyze the boundary term in $l_i\rightarrow 0$ limit is more non-trivial. For the propagator, we get
\begin{align}
	\hat{G}&=\int_0^\infty bdb \pqty{\frac{\phi_i}{\sqrt{l_i^2+b^2}}}\pqty{\frac{\phi_f}{\sqrt{l_f^2+b^2}}} H^{(1)}_1(\phi_i\sqrt{l_i^2+b^2})H^{(2)}_1(\phi_f\sqrt{l_f^2+b^2})\nonumber\\
	&\xrightarrow{l_i\rightarrow 0}\phi_i\phi_f\int_0^\infty \frac{db}{\sqrt{l_f^2+b^2}}\, H^{(1)}_1(\phi_ib)H^{(2)}_1(\phi_f\sqrt{l_f^2+b^2})\label{ligex}
\end{align}
Noting the Hankel expansions in the small argument limit as
\begin{align}
	H_1^{(1)}(x)\xrightarrow{x\rightarrow 0}-\frac{2 i}{\pi  x}+\dots\label{h1smargli}
\end{align}
we see that the $b$ integral in eq.\eqref{ligex} diverges logarithmically at small $b$ for any fixed value of $\phi_i$. Thus, we cannot glue at finite value of $\phi_i$ and get a wavefunction which is independent of the instant of gluing. On the other hand, in the limit $\phi_i\rightarrow\infty$, we see that the initial wavefunction goes as, using eq.\eqref{genexwf} and eq.\eqref{condrr} with $\tilde{\rho},\rho_{1},\rho_2=0$,
\begin{align}
	\partial_{\phi_i} \hat{\Psi}_i\sim 	-\frac{i}{2\phi_i^2}\int dM \rho(M) M \implies \hat{\Psi}_i\sim  \frac{i}{2\phi_i} \int dM \rho(M) M \label{iniwfinlp}
\end{align}
Also, from eq.\eqref{ligex}, we get
\begin{align}
	\hat{G}\xrightarrow{l_i\rightarrow 0,\phi_i\rightarrow\infty}\sqrt{\frac{2}{\pi\phi_i}}\phi_f e^{-\frac{3i \pi}{4}}\int \frac{db}{\sqrt{l_f^2+b^2}\sqrt{b}} \,\,e^{i\phi_i b}	H^{(2)}_1(\phi_f\sqrt{l_f^2+b^2})\label{ginphzf}
\end{align}
Using eq.\eqref{iniwfinlp} and eq.\eqref{ginphzf} it is easy to convince oneself that the boundary term vanishes in the limit $\phi_i\rightarrow\infty$ as both $\hat{\Psi}_i$ and $\hat{G}$ vanish as $\phi_i\rightarrow\infty$.  
\subsection{Norm of final state}
\label{normfinst}
Let us now investigate the normalizability of the final state wavefunction. As in the case of single boundary wavefunction, we explore the possibility of Klein-Gordon norm.  The norm is given by 
\begin{align}
	\langle \hat{\Psi}_f,\hat{\Psi}_f\rangle \sim i\int dl_f\, \hat{\Psi}_f^*(\phi_f,l_f)\overleftrightarrow{\del_{l_f}}\hat{\Psi}_f(l_f,\phi_f)\label{psifnorm}
\end{align}
Again, the natural question would be if the norm is conserved and finite. Let us first look at the issue of conservation which amounts to 
\begin{align}
	\del_{\phi_f}\langle \hat{\Psi}_f(\phi_f,l_f),\hat{\Psi}_f(\phi_f,l_f)\rangle=0\label{finwfnorm}
\end{align}
The discussion for the conservation of norm will parallel that of the single boundary wavefunction in subsection \ref{consnorm}. 
A series of steps similar to those in eq.\eqref{glucond} will give
\begin{align}
	\del_{\phi_f}\langle \hat{\Psi}_f(\phi_f,l_f),\hat{\Psi}_f(\phi_f,l_f)\rangle=i ( \hat{\Psi}_f\del_{\phi_f}\hat{\Psi}_f^*{}\,-\hat{\Psi}_f^*{\del_{\phi_f}}\hat{\Psi}_f )\bigg\vert_{l_f=0}^{\infty}\label{bdtenor}
\end{align}
The dependence of the late time wavefunction on $l_f$ comes entirely from the propagator eq.\eqref{ligex}. It can be written as
\begin{align}
	\hat{\Psi}_f&=i \int  b \,db \hat{\Psi}^+_{T}(\phi_f,l_f, b)\hat{f}(b)\nonumber\\
	\hat{f}(b)&=\int dl_i \left({\hat{\Psi}^-_{T}(\phi_i,l_i, b)\del_{l_i}\hat{\Psi}(\phi_i, l_i)-\hat{\Psi}_i(\phi_i, l_i) \del_{l_i}\hat{\Psi}^-_{T}(\phi_i,l_i,b)}\right)\label{psifinht}
\end{align}

The vanishing of the boundary term for $l_f\rightarrow\infty$ can be argued as follows.  In the limit $l_f\gg b$, assuming the $b$ integral in eq.\eqref{psifinht} has most of its support for $b<\infty$, we can write eq.\eqref{ligex} to leading order as
\begin{align}
	\hat{G}\simeq -\frac{i\sqrt{2}}{\sqrt{\pi}}e^{-3i\pi\over 4}\int_0^\infty bdb  \pqty{\frac{\phi_i}{\sqrt{l_i^2+b^2}}}H^{(1)}_1(\phi_i\sqrt{l_i^2+b^2})\pqty{\frac{\sqrt{\phi_f}}{\sqrt{(l_f^2+b^2)^{3\over 2}}}} e^{-i\phi_f\sqrt{l_f^2+b^2}	}\label{lfpgglar}
\end{align}
It is easy to see from the above expression that $\hat{G}\sim \frac{1}{l_f^{3\over 2}}$ which then means that $\hat{\Psi}_f$ also has a similar behaviour at large $l_f$. In this limit of large $l_f$, the derivatives of $\phi_f$ can be dealt with in a WKB approximation which act on $e^{-i\phi_fl_f}$ to give a factor of $l_f$. So, the full boundary term is suppressed by a factor of $l_f^{-2}$ in this limit and hence vanishes. 

Now coming to the small $l_f$ limit,  we get
\begin{align}
	\hat{\Psi}_f\simeq i\int \,db \hat{f}(b){\phi_f}H_1^{(2)}(\phi_f b)\label{psifsmlf}
\end{align}
If $\hat{f}(b)$ does not vanish in the limit of small $b$, we see that the wavefunction will have a divergence in the $b$ integral from the region of small $b$, which can be seen using the small argument limit of $H_1^{(2)}$.  It is easy to see from the expression for $\hat{f}(b)$ in eq.\eqref{psifinht} that it does not vanish generally as $b\rightarrow 0$. This then means that the wavefunction has a divergence in the $l_f\rightarrow 0$ limit. This divergence is easy to understand as follows.  Since the late time wavefunction also satisfies WDW equation, it can be written in the form eq.\eqref{nearl}  for some $\rho,\tilde{\rho},\rho_1,\rho_2$.  However, the analog of the condition eq.\eqref{condrr} is not satisfied now and hence the divergence. Thus, the boundary term in eq.\eqref{bdtenor} may not vanish and so the norm need not necessarily be conserved.  Moreover this divergence of the wavefunction in the limit $l_f\rightarrow 0$ would also mean that its norm will not be finite. 

Thus, we are led to the disturbing conclusion that even if we start with the initial wavefunction with all the good properties such as finite and conserved norm, the propagator will propagate this `good' initial wavefunction to one whose norm is neither conserved nor finite at late times.

\section{Moments of M}
\label{expvalm}
 In this appendix, we give a more general analysis for computing the expectation value of moments of $M$. Consider a wavefunction with only an expanding branch, {\it i.e.,$\rho\neq0$} and all other coefficient functions to vanish ($\tilde{\rho}=\rho_1=\rho_2=0$). The wavefunction is then given by 
 \begin{align}
 	\hat{\Psi}=\frac{1}{l}\int dM \rho(M) e^{-il\sqrt{\phi^2-M}}\label{psirhoexp}
 \end{align}
From the general analysis in subsection \ref{consnorm}, we see that $\rho$ has to be real and should have support only over $M<0$ for the norm to be conserved. We take these requirements to be satisfied for the $\rho(M)$ in eq.\eqref{psirhoexp} and thus belongs to the case 1), eq.\eqref{rhogaus} studied there. 
Let us first compute the first moment of $\hat{M}$. This is given in terms of the conjugate momentum $\pi_l$ as
\begin{align}
	\langle \hat{M}\rangle=\mathcal{N}^{-1}(\langle \phi^2\rangle  -\langle \pi_l^2\rangle)\label{minpil2}
\end{align}
where $\mathcal{N}$ is the normalization of the state under consideration. 
The higher moments can be similarly computed in terms of higher moments of $\pi_l$ as
\begin{align}
	\langle \hat{M}^n\rangle =\mathcal{N}^{-1}\langle (\phi^2-\pi_l^2)^n\rangle \label{himinhipil}
\end{align}

But even before computing the first moment $\langle \hat{M}\rangle$, let us evaluate the norm. Inserting the wavefunction in eq.\eqref{psirhoexp} in eq.\eqref{kgnorm}, we get
\begin{align}
	\langle \hat{\Psi},\hat{\Psi}\rangle =i(-i)\int dM_a dM_b\rho(M_a)\rho(M_b) (\sqrt{\phi^2-M_a}+\sqrt{\phi^2-M_b}) \int_0^\infty \frac{dl}{l^2} e^{-il \sqrt{\phi^2-M_a}+il\sqrt{\phi^2-M_b}}\label{normeq1}
\end{align}
For brevity of notation from now on, let us denote
\begin{align}
	q_i=\sqrt{\phi^2-M_i},\quad i={a,b}\label{qidf}
\end{align}
The $l$ integral in eq.\eqref{normeq1} naively diverges at the lower limit $l=0$. One way we had taken care of these divergences is to take $\rho$ satisfy the condition eq.\eqref{condrr}, {i.e}
\begin{align}
	\int dM \rho(M)=0\label{intrhom}
\end{align}
So, we introduce a small regulator, $\epsilon$ in the integral and finally take the limit $\epsilon\rightarrow 0$. The norm is then given by
\begin{align}
		\langle \hat{\Psi},\hat{\Psi}\rangle =\lim_{\epsilon\rightarrow 0^+}i(-i)\int dM_a dM_b\rho(M_a)\rho(M_b) (q_a+q_b) \int_\epsilon^\infty \frac{dl}{l^2} e^{-il (q_a-q_b)}\label{normeq2}
\end{align}
The value of the $l$ integral in the limit $\epsilon\rightarrow 0$ is given by 
\begin{align}
	\int_\epsilon^\infty \frac{dl}{l^2} e^{il q}\simeq \frac{1}{\epsilon } - i q (\ln ( q)+\ln (-i\epsilon )+\gamma -1)+\order{(\epsilon)},\qquad \text{if} \quad \Im(q)>0\label{l2inteze}
\end{align}
Using the above result in eq.\eqref{normeq2}, we easily see that the terms proportional to $\epsilon^{-1}$ vanish due to the condition eq.\eqref{intrhom}. For the terms of $\order{(\epsilon^0)}$, we see that all the terms except the $\ln(q)$ again vanish due to the antisymmetry of $q=q_b-q_a$ under exchange of $M_1, M_2$ where as the rest of the integrand is symmetric. So, the only term that gives a non-vanishing result  is $\ln(q)$ and so the norm becomes
\begin{align}
		\langle \hat{\Psi},\hat{\Psi}\rangle &=\lim_{\epsilon\rightarrow 0^+}i(-i)^2\int dM_a dM_b\rho(M_a)\rho(M_b) (q_a+q_b)(q_b-q_a)\ln(q_b-q_a)\nonumber\\
		&=\lim_{\epsilon\rightarrow 0^+}-i\int dM_a dM_b\rho(M_a)\rho(M_b) (M_a-M_b)\ln(\sqrt{\phi^2-M_b}-\sqrt{\phi^2-M_a})
		\label{normeq3}
\end{align}
This will give a non-zero result due to the presence of log term as it gives a phase factor under exchange of $M_a, M_b$. Let us now compute the first moment. As can be seen from eq.\eqref{minpil2}, we need to compute the second moment of $\pi_l$. Using the definition for the moments in eq.\eqref{mommom} it is given by 
\begin{align}
	\langle \pi_l^2\rangle = \int dM_a dM_b\rho(M_a)\rho(M_b) (q_a+q_b) \int_\epsilon^\infty \frac{dl}{l^2} e^{-il (q_a-q_b)}\left(\frac{(q_a+q_b)^2}{4}-\frac{3}{2 l^2}\right)\label{pil2}
\end{align}
Now we need one more integral $\int \frac{dl}{l^4}e^{ilq}$ which in the small $\epsilon$ limit is given by 
\begin{align}
	\int\frac{dl}{l^4}e^{ilq}=\frac{1}{3 \epsilon ^3}+\frac{i q}{2 \epsilon ^2}-\frac{q^2}{2 \epsilon }+\frac{1}{36} q^3 (6 i \ln (q)+6 i \ln (\epsilon )-11 i+6 i \gamma +3 \pi )\label{l4int}
\end{align}
Using this and eq.\eqref{l2inteze} to compute the $l$ integral in eq.\eqref{pil2}, it will turn out that all the terms except the $\ln q$ term vanishes for reasons mentioned above eq.\eqref{normeq3}. We then get
\begin{align}
		\langle \pi_l^2\rangle &= \frac{1}{2} i\int dM_a dM_b\rho(M_a)\rho(M_b)  \left(q_a^4-q_b^4\right) \ln (q_b-q_a) \nonumber\\
		&=-\frac{1}{2} i\int dM_a dM_b\rho(M_a)\rho(M_b) (M_b-M_a)(M_a + M_b -2 \phi^2)\ln(\sqrt{\phi^2-M_b}-\sqrt{\phi^2-M_a})\label{mambexp}
\end{align}
So, on the whole using the results eq.\eqref{normeq3} and eq.\eqref{mambexp} in eq.\eqref{minpil2}, we get
\begin{align}
	\langle \hat{M}\rangle =\frac{i}{2 \mathcal{N}}\int dM_a dM_b\rho(M_a)\rho(M_b) (M_b-M_a)(M_a + M_b )\ln(\sqrt{\phi^2-M_b}-\sqrt{\phi^2-M_a})\label{mexp}
\end{align}
 The arguments above eq.\eqref{normeq3} are essentially the same for higher moments. We expect in the case of higher moments, it is again the $\ln$ term that gives a non-zero result which can be evaluated straightforwardly. We expect the result for any $n$ to be,
 \begin{equation}
 		\langle \hat{M}^n\rangle =\frac{i}{\mathcal{N}(n+1)}\int dM_a dM_b\rho(M_a)\rho(M_b) (M_b^{n+1}-M_{a} ^{n+1}) \ln(\sqrt{\phi^2-M_b}-\sqrt{\phi^2-M_a}) \label{mexpn}
 \end{equation}
\subsection{An example: Delta function}
\label{dosexofm}
Let's apply the formulas eq.\eqref{normeq3} and eq.\eqref{mexpn} to the example considered in \ref{deldens}. For convenience we rewrite eq.\eqref{rhomdel} here.
\begin{align}
	\rho(M)=\delta(M+M_1)-\delta(M+M_2),\quad M_1,M_2>0 \label{rhomdel2}
\end{align}
Using the above in equation in eq.\eqref{normeq3} yields,
\begin{equation}
	\mathcal{N} = \pi \abs{M_1-M_2}
\end{equation}
which agrees with eq.\eqref{normdefwf}. Similarly, the mean $\langle \hat{M}\rangle$ and variance $\sqrt{\langle\hat{M}^2\rangle - \langle \hat{M}\rangle ^2}$ are given by,
\begin{align}
	& \langle \hat{M}\rangle = -\frac{1}{2} (M_1+M_2) \label{meanM}\\
	& \Delta M = \sqrt{\langle\hat{M}^2\rangle - \langle \hat{M}\rangle ^2} = \frac{\abs{M_1-M_2}}{2 \sqrt{3}} \label{varM}.
\end{align}
The value of mean as given in eq.\eqref{meanM} is to be expected since it is the average of two values that $M$ can take given eq.\eqref{rhomdel2}.{ One final thing to note is that the values calculated above for various moments of $\hat{M}$ can be checked directly using the wave function computed in \ref{deldens}, eq.\eqref{wfdel} and are found to match}.

		\vspace{-0.6cm}
		
		\newpage
		\bibliographystyle{JHEP}
		\bibliography{refs}

\providecommand{\href}[2]{#2}\begingroup\raggedright\begin{thebibliography}{10}

\bibitem{Spradlin:2001pw}
M.~Spradlin, A.~Strominger and A.~Volovich, \emph{{Les Houches lectures on de
  Sitter space}},  in \emph{{Les Houches Summer School: Session 76: Euro Summer
  School on Unity of Fundamental Physics: Gravity, Gauge Theory and Strings}},
  pp.~423--453, 10, 2001.
\newblock \href{http://arxiv.org/abs/hep-th/0110007}{{\tt hep-th/0110007}}.

\bibitem{Halliwell:1989myn}
J.~J. Halliwell, \emph{{INTRODUCTORY LECTURES ON QUANTUM COSMOLOGY}},  in
  \emph{{7th Jerusalem Winter School for Theoretical Physics: Quantum Cosmology
  and Baby Universes}}, 1989.
\newblock \href{http://arxiv.org/abs/0909.2566}{{\tt 0909.2566}}.

\bibitem{Hartle:1992as}
J.~B. Hartle, \emph{{Space-time quantum mechanics and the quantum mechanics of
  space-time}},  in \emph{{Les Houches Summer School on Gravitation and
  Quantizations, Session 57}}, pp.~0285--480, 7, 1992.
\newblock \href{http://arxiv.org/abs/gr-qc/9304006}{{\tt gr-qc/9304006}}.

\bibitem{Isham:1992ms}
C.~J. Isham, \emph{{Canonical quantum gravity and the problem of time}},
  {\emph{NATO Sci. Ser. C} {\bf 409} (1993) 157--287},
  [\href{http://arxiv.org/abs/gr-qc/9210011}{{\tt gr-qc/9210011}}].

\bibitem{Witten:2001kn}
E.~Witten, \emph{{Quantum gravity in de Sitter space}},  in \emph{{Strings
  2001: International Conference}}, 6, 2001.
\newblock \href{http://arxiv.org/abs/hep-th/0106109}{{\tt hep-th/0106109}}.

\bibitem{Linde:1990flp}
A.~D. Linde, \emph{{Particle physics and inflationary cosmology}}, vol.~5.
\newblock 1990.

\bibitem{Gell-Mann:1989lly}
M.~Gell-Mann and J.~B. Hartle, \emph{{QUANTUM MECHANICS IN THE LIGHT OF QUANTUM
  COSMOLOGY}},  8, 1989.

\bibitem{Hartletasi}
J.~B. Hartle, \emph{{Quantum Cosmology, TASI Lectures 1985}}, .

\bibitem{Vilenkin:1986cy}
A.~Vilenkin, \emph{{Boundary Conditions in Quantum Cosmology}},
  \href{http://dx.doi.org/10.1103/PhysRevD.33.3560}{\emph{Phys. Rev. D} {\bf
  33} (1986) 3560}.

\bibitem{PhysRevD.28.2960}
J.~B. Hartle and S.~W. Hawking, \emph{Wave function of the universe},
  \href{http://dx.doi.org/10.1103/PhysRevD.28.2960}{\emph{Phys. Rev. D} {\bf
  28} (Dec, 1983) 2960--2975}.

\bibitem{Banks:1984cw}
T.~Banks, \emph{{T C P, Quantum Gravity, the Cosmological Constant and All
  That...}}, \href{http://dx.doi.org/10.1016/0550-3213(85)90020-3}{\emph{Nucl.
  Phys. B} {\bf 249} (1985) 332--360}.

\bibitem{JACKIW1985343}
R.~Jackiw, \emph{Lower dimensional gravity},
  \href{http://dx.doi.org/https://doi.org/10.1016/0550-3213(85)90448-1}{\emph{Nuclear
  Physics B} {\bf 252} (1985) 343 -- 356}.

\bibitem{Teitelboim:1983ux}
C.~Teitelboim, \emph{{Gravitation and Hamiltonian Structure in Two Space-Time
  Dimensions}},
  \href{http://dx.doi.org/10.1016/0370-2693(83)90012-6}{\emph{Phys. Lett.} {\bf
  126B} (1983) 41--45}.

\bibitem{Hennauxjt}
M.~Henneaux, \emph{Quantum gravity in two dimensions: Exact solution of the
  jackiw model},
  \href{http://dx.doi.org/10.1103/PhysRevLett.54.959}{\emph{Phys. Rev. Lett.}
  {\bf 54} (Mar, 1985) 959--962}.

\bibitem{Maldacena:2019cbz}
J.~Maldacena, G.~J. Turiaci and Z.~Yang, \emph{{Two dimensional Nearly de
  Sitter gravity}},  \href{http://arxiv.org/abs/1904.01911}{{\tt 1904.01911}}.

\bibitem{Verlinde:2020zld}
L.~V. Iliesiu, J.~Kruthoff, G.~J. Turiaci and H.~Verlinde, \emph{{JT gravity at
  finite cutoff}},
  \href{http://dx.doi.org/10.21468/SciPostPhys.9.2.023}{\emph{SciPost Phys.}
  {\bf 9} (2020) 023}, [\href{http://arxiv.org/abs/2004.07242}{{\tt
  2004.07242}}].

\bibitem{Moitra:2021uiv}
U.~Moitra, S.~K. Sake and S.~P. Trivedi, \emph{{Jackiw-Teitelboim Gravity in
  the Second Order Formalism}},  \href{http://arxiv.org/abs/2101.00596}{{\tt
  2101.00596}}.

\bibitem{Moitra:2022glw}
U.~Moitra, S.~K. Sake and S.~P. Trivedi, \emph{{Aspects of Jackiw-Teitelboim
  gravity in Anti-de Sitter and de Sitter spacetime}},
  \href{http://dx.doi.org/10.1007/JHEP06(2022)138}{\emph{JHEP} {\bf 06} (2022)
  138}, [\href{http://arxiv.org/abs/2202.03130}{{\tt 2202.03130}}].

\bibitem{Vilenkin:2021awm}
G.~Fanaras and A.~Vilenkin, \emph{{Jackiw-Teitelboim and Kantowski-Sachs
  quantum cosmology}},  \href{http://arxiv.org/abs/2112.00919}{{\tt
  2112.00919}}.

\bibitem{Narayan:2020pyj}
K.~Narayan, \emph{{Aspects of two-dimensional dilaton gravity, dimensional
  reduction, and holography}},
  \href{http://dx.doi.org/10.1103/PhysRevD.104.026007}{\emph{Phys. Rev. D} {\bf
  104} (2021) 026007}, [\href{http://arxiv.org/abs/2010.12955}{{\tt
  2010.12955}}].

\bibitem{Witten:2020ert}
E.~Witten, \emph{{Deformations of JT Gravity and Phase Transitions}},
  \href{http://arxiv.org/abs/2006.03494}{{\tt 2006.03494}}.

\bibitem{Blommaert:2020tht}
A.~Blommaert, \emph{{Searching for butterflies in dS JT gravity}},
  \href{http://arxiv.org/abs/2010.14539}{{\tt 2010.14539}}.

\bibitem{Chakraborty:2023los}
T.~Chakraborty, J.~Chakravarty, V.~Godet, P.~Paul and S.~Raju,
  \emph{{Holography of information in de Sitter space}},
  \href{http://arxiv.org/abs/2303.16316}{{\tt 2303.16316}}.

\bibitem{Chakraborty:2023yed}
T.~Chakraborty, J.~Chakravarty, V.~Godet, P.~Paul and S.~Raju, \emph{{The
  Hilbert space of de Sitter quantum gravity}},
  \href{http://arxiv.org/abs/2303.16315}{{\tt 2303.16315}}.

\bibitem{Constantinidis:2008ty}
C.~P. Constantinidis, O.~Piguet and A.~Perez, \emph{{Quantization of the
  Jackiw-Teitelboim model}},
  \href{http://dx.doi.org/10.1103/PhysRevD.79.084007}{\emph{Phys. Rev. D} {\bf
  79} (2009) 084007}, [\href{http://arxiv.org/abs/0812.0577}{{\tt 0812.0577}}].

\bibitem{Stanford:2019vob}
D.~Stanford and E.~Witten, \emph{{JT Gravity and the Ensembles of Random Matrix
  Theory}},  \href{http://arxiv.org/abs/1907.03363}{{\tt 1907.03363}}.

\bibitem{Stanford:2020qhm}
D.~Stanford and Z.~Yang, \emph{{Finite-cutoff JT gravity and self-avoiding
  loops}},  \href{http://arxiv.org/abs/2004.08005}{{\tt 2004.08005}}.

\bibitem{Strobl:1993yn}
T.~Strobl, \emph{{Quantization and the issue of time for various
  two-dimensional models of gravity}},
  \href{http://dx.doi.org/10.1142/S0218271894000460}{\emph{Int. J. Mod. Phys.
  D} {\bf 3} (1994) 281--284}, [\href{http://arxiv.org/abs/hep-th/9308155}{{\tt
  hep-th/9308155}}].

\bibitem{Anegawa:2023wrk}
T.~Anegawa, N.~Iizuka, S.~K. Sake and N.~Zenoni, \emph{{Is action complexity
  better for de Sitter space in Jackiw-Teitelboim gravity?}},
  \href{http://dx.doi.org/10.1007/JHEP06(2023)213}{\emph{JHEP} {\bf 06} (2023)
  213}, [\href{http://arxiv.org/abs/2303.05025}{{\tt 2303.05025}}].

\bibitem{Aguilar-Gutierrez:2021bns}
S.~E. Aguilar-Gutierrez, A.~Chatwin-Davies, T.~Hertog, N.~Pinzani-Fokeeva and
  B.~Robinson, \emph{{Islands in Multiverse Models}},
  \href{http://dx.doi.org/10.1007/JHEP11(2021)212}{\emph{JHEP} {\bf 11} (2021)
  212}, [\href{http://arxiv.org/abs/2108.01278}{{\tt 2108.01278}}].

\bibitem{Jensen:2023eza}
J.~Cotler and K.~Jensen, \emph{{Isometric evolution in de Sitter quantum
  gravity}},  \href{http://arxiv.org/abs/2302.06603}{{\tt 2302.06603}}.

\bibitem{Baek:2022ozg}
J.-H. Baek and K.-S. Choi, \emph{{Islands in proliferating de Sitter spaces}},
  \href{http://dx.doi.org/10.1007/JHEP05(2023)098}{\emph{JHEP} {\bf 05} (2023)
  098}, [\href{http://arxiv.org/abs/2212.14753}{{\tt 2212.14753}}].

\bibitem{Castro:2022cuo}
A.~Castro, F.~Mariani and C.~Toldo, \emph{{Near-extremal limits of de Sitter
  black holes}}, \href{http://dx.doi.org/10.1007/JHEP07(2023)131}{\emph{JHEP}
  {\bf 07} (2023) 131}, [\href{http://arxiv.org/abs/2212.14356}{{\tt
  2212.14356}}].

\bibitem{Aalsma:2022swk}
L.~Aalsma, S.~E. Aguilar-Gutierrez and W.~Sybesma, \emph{{An
  outsider\textquoteright{}s perspective on information recovery in de Sitter
  space}}, \href{http://dx.doi.org/10.1007/JHEP01(2023)129}{\emph{JHEP} {\bf
  01} (2023) 129}, [\href{http://arxiv.org/abs/2210.12176}{{\tt 2210.12176}}].

\bibitem{Rahman:2022jsf}
A.~A. Rahman, \emph{{dS JT Gravity and Double-Scaled SYK}},
  \href{http://arxiv.org/abs/2209.09997}{{\tt 2209.09997}}.

\bibitem{Svesko:2022txo}
A.~Svesko, E.~Verheijden, E.~P. Verlinde and M.~R. Visser, \emph{{Quasi-local
  energy and microcanonical entropy in two-dimensional nearly de Sitter
  gravity}}, \href{http://dx.doi.org/10.1007/JHEP08(2022)075}{\emph{JHEP} {\bf
  08} (2022) 075}, [\href{http://arxiv.org/abs/2203.00700}{{\tt 2203.00700}}].

\bibitem{Teresi:2021qff}
D.~Teresi, \emph{{Islands and the de Sitter entropy bound}},
  \href{http://dx.doi.org/10.1007/JHEP10(2022)179}{\emph{JHEP} {\bf 10} (2022)
  179}, [\href{http://arxiv.org/abs/2112.03922}{{\tt 2112.03922}}].

\bibitem{Kames-King:2021etp}
J.~Kames-King, E.~M.~H. Verheijden and E.~P. Verlinde, \emph{{No Page curves
  for the de Sitter horizon}},
  \href{http://dx.doi.org/10.1007/JHEP03(2022)040}{\emph{JHEP} {\bf 03} (2022)
  040}, [\href{http://arxiv.org/abs/2108.09318}{{\tt 2108.09318}}].

\bibitem{Addazi:2021vvy}
A.~Addazi, J.~Bilski, Q.~Gan and A.~Marciano, \emph{{New Massive JT
  Multi-Gravity and N-Replica of SYK Models}},
  \href{http://arxiv.org/abs/2106.02384}{{\tt 2106.02384}}.

\bibitem{Niermann:2021wco}
L.~Niermann and T.~J. Osborne, \emph{{Holographic networks for
  (1+1)-dimensional de~Sitter space-time}},
  \href{http://dx.doi.org/10.1103/PhysRevD.105.125009}{\emph{Phys. Rev. D} {\bf
  105} (2022) 125009}, [\href{http://arxiv.org/abs/2102.09223}{{\tt
  2102.09223}}].

\bibitem{Balasubramanian:2020xqf}
V.~Balasubramanian, A.~Kar and T.~Ugajin, \emph{{Islands in de Sitter space}},
  \href{http://dx.doi.org/10.1007/JHEP02(2021)072}{\emph{JHEP} {\bf 02} (2021)
  072}, [\href{http://arxiv.org/abs/2008.05275}{{\tt 2008.05275}}].

\bibitem{Hartman:2020khs}
T.~Hartman, Y.~Jiang and E.~Shaghoulian, \emph{{Islands in cosmology}},
  \href{http://dx.doi.org/10.1007/JHEP11(2020)111}{\emph{JHEP} {\bf 11} (2020)
  111}, [\href{http://arxiv.org/abs/2008.01022}{{\tt 2008.01022}}].

\bibitem{Chen:2020tes}
Y.~Chen, V.~Gorbenko and J.~Maldacena, \emph{{Bra-ket wormholes in
  gravitationally prepared states}},
  \href{http://dx.doi.org/10.1007/JHEP02(2021)009}{\emph{JHEP} {\bf 02} (2021)
  009}, [\href{http://arxiv.org/abs/2007.16091}{{\tt 2007.16091}}].

\bibitem{Mirbabayi:2020grb}
M.~Mirbabayi, \emph{{Uptunneling to de Sitter}},
  \href{http://dx.doi.org/10.1007/JHEP09(2020)070}{\emph{JHEP} {\bf 09} (2020)
  070}, [\href{http://arxiv.org/abs/2003.05460}{{\tt 2003.05460}}].

\bibitem{Cotler:2019dcj}
J.~Cotler and K.~Jensen, \emph{{Emergent unitarity in de Sitter from matrix
  integrals}}, \href{http://dx.doi.org/10.1007/JHEP12(2021)089}{\emph{JHEP}
  {\bf 12} (2021) 089}, [\href{http://arxiv.org/abs/1911.12358}{{\tt
  1911.12358}}].

\bibitem{witten2020matrix}
E.~Witten, \emph{Matrix models and deformations of jt gravity},
  {\emph{Proceedings of the Royal Society A} {\bf 476} (2020) 20200582}.

\bibitem{eberhardt20232d}
L.~Eberhardt and G.~J. Turiaci, \emph{2d dilaton gravity and the weil-petersson
  volumes with conical defects}, {\emph{arXiv preprint arXiv:2304.14948} (2023)
  }.

\bibitem{mertens2023solvable}
T.~G. Mertens and G.~J. Turiaci, \emph{Solvable models of quantum black holes:
  a review on jackiw--teitelboim gravity}, {\emph{Living Reviews in Relativity}
  {\bf 26} (2023) 4}.

\bibitem{Bousso:2001mw}
R.~Bousso, A.~Maloney and A.~Strominger, \emph{{Conformal vacua and entropy in
  de Sitter space}},
  \href{http://dx.doi.org/10.1103/PhysRevD.65.104039}{\emph{Phys. Rev. D} {\bf
  65} (2002) 104039}, [\href{http://arxiv.org/abs/hep-th/0112218}{{\tt
  hep-th/0112218}}].

\bibitem{Crnkovic:1986ex}
C.~Crnkovic and E.~Witten, \emph{{COVARIANT DESCRIPTION OF CANONICAL FORMALISM
  IN GEOMETRICAL THEORIES}}, .

\bibitem{nanda:2023yta}
K.~K. Nanda, S.~K. Sake and S.~P. Trivedi, \emph{{Quantization of 2d Dilaton
  gravity with a general potential}}, {\emph{Work in Progress} }.

\bibitem{Stanford:2017thb}
D.~Stanford and E.~Witten, \emph{{Fermionic Localization of the Schwarzian
  Theory}}, \href{http://dx.doi.org/10.1007/JHEP10(2017)008}{\emph{JHEP} {\bf
  10} (2017) 008}, [\href{http://arxiv.org/abs/1703.04612}{{\tt 1703.04612}}].

\bibitem{Mahajan:2021nsd}
R.~Mahajan, D.~Stanford and C.~Yan, \emph{{Sphere and disk partition functions
  in Liouville and in matrix integrals}},
  \href{http://dx.doi.org/10.1007/JHEP07(2022)132}{\emph{JHEP} {\bf 07} (2022)
  132}, [\href{http://arxiv.org/abs/2107.01172}{{\tt 2107.01172}}].

\bibitem{Cotler:2019nbi}
J.~Cotler, K.~Jensen and A.~Maloney, \emph{{Low-dimensional de Sitter quantum
  gravity}},  \href{http://arxiv.org/abs/1905.03780}{{\tt 1905.03780}}.

\bibitem{pagedensuni}
D.~N. Page, \emph{Density matrix of the universe},
  \href{http://dx.doi.org/10.1103/PhysRevD.34.2267}{\emph{Phys. Rev. D} {\bf
  34} (Oct, 1986) 2267--2271}.

\bibitem{Penington:2019kki}
G.~Penington, S.~H. Shenker, D.~Stanford and Z.~Yang, \emph{{Replica wormholes
  and the black hole interior}},
  \href{http://dx.doi.org/10.1007/JHEP03(2022)205}{\emph{JHEP} {\bf 03} (2022)
  205}, [\href{http://arxiv.org/abs/1911.11977}{{\tt 1911.11977}}].

\bibitem{Marolf:2020xie}
D.~Marolf and H.~Maxfield, \emph{{Transcending the ensemble: baby universes,
  spacetime wormholes, and the order and disorder of black hole information}},
  \href{http://dx.doi.org/10.1007/JHEP08(2020)044}{\emph{JHEP} {\bf 08} (2020)
  044}, [\href{http://arxiv.org/abs/2002.08950}{{\tt 2002.08950}}].

\bibitem{Post:2022dfi}
B.~Post, J.~van~der Heijden and E.~Verlinde, \emph{{A universe field theory for
  JT gravity}}, \href{http://dx.doi.org/10.1007/JHEP05(2022)118}{\emph{JHEP}
  {\bf 05} (2022) 118}, [\href{http://arxiv.org/abs/2201.08859}{{\tt
  2201.08859}}].

\bibitem{Saad:2019lba}
P.~Saad, S.~H. Shenker and D.~Stanford, \emph{{JT gravity as a matrix
  integral}},  \href{http://arxiv.org/abs/1903.11115}{{\tt 1903.11115}}.

\bibitem{poisson2004relativist}
E.~Poisson, \emph{A relativist's toolkit: the mathematics of black-hole
  mechanics}.
\newblock Cambridge university press, 2004.

\bibitem{kiefer1991quantum}
C.~Kiefer and T.~P. Singh, \emph{Quantum gravitational corrections to the
  functional schr{\"o}dinger equation}, {\emph{Physical Review D} {\bf 44}
  (1991) 1067}.

\bibitem{singh1989notes}
T.~Singh and T.~Padmanabhan, \emph{Notes on semiclassical gravity},
  {\emph{Annals of Physics} {\bf 196} (1989) 296--344}.

\end{thebibliography}\endgroup

	\end{document}